\documentclass[journal]{IEEEtran}
\usepackage{amsmath,amsfonts}
\usepackage{algorithmic}
\usepackage{algorithm}
\usepackage{array}
\usepackage[caption=false,font=normalsize,labelfont=sf,textfont=sf]{subfig}
\usepackage{textcomp}
\usepackage{stfloats}
\usepackage{url}
\usepackage{verbatim}
\usepackage{graphicx}
\usepackage{cite}
\usepackage{color}
\usepackage[nolist,printonlyused]{acronym}      

\begin{document}

\begin{acronym}
  \acro{2G}{Second Generation}
  \acro{3G}{3$^\text{rd}$~Generation}
  \acro{3GPP}{3$^\text{rd}$~Generation Partnership Project}
  \acro{4G}{4th~Generation}
  \acro{5G}{5th~Generation}
  \acro{6G}{6th~Generation}
  \acro{AA}{Antenna Array}
  \acro{AC}{Admission Control}
  \acro{ACL}{adjacent channel leakage}
  \acro{AD}{Attack-Decay}
  \acro{ADC}{analog-to-digital converter}
  \acro{ADSL}{Asymmetric Digital Subscriber Line}
  \acro{AHW}{Alternate Hop-and-Wait}
  \acro{AMC}{Adaptive Modulation and Coding}
	\acro{AP}{Access Point}
  \acro{APA}{Adaptive Power Allocation}
  \acro{AR}{autoregressive}
  \acro{ARMA}{Autoregressive Moving Average}
  \acro{ATES}{Adaptive Throughput-based Efficiency-Satisfaction Trade-Off}
  \acro{AWGN}{additive white Gaussian noise}
  \acro{BB}{branch and bound}
  \acro{BD}{block diagonalization}
  \acro{BER}{bit error rate}
  \acro{BF}{Best Fit}
  \acro{BLER}{BLock Error Rate}
  \acro{BPC}{Binary power control}
  \acro{BPSK}{binary phase-shift keying}
  \acro{BPA}{Best \ac{PDPR} Algorithm}
  \acro{BRA}{Balanced Random Allocation}
  \acro{BS}{base station}
  \acro{CAP}{Combinatorial Allocation Problem}
  \acro{CAPEX}{Capital Expenditure}
  \acro{CBF}{Coordinated Beamforming}
  \acro{CBR}{Constant Bit Rate}
  \acro{CBS}{Class Based Scheduling}
  \acro{CC}{Congestion Control}
  \acro{CDF}{Cumulative Distribution Function}
  \acro{CDMA}{Code-Division Multiple Access}
  \acro{CL}{Closed Loop}
  \acro{CLI}{cross-link interference}
  \acro{CLPC}{Closed Loop Power Control}
  \acro{CNR}{Channel-to-Noise Ratio}
  \acro{CPA}{Cellular Protection Algorithm}
  \acro{CPICH}{Common Pilot Channel}
  \acro{CoMP}{Coordinated Multi-Point}
  \acro{CQI}{Channel Quality Indicator}
  \acro{CRM}{Constrained Rate Maximization}
	\acro{CRN}{Cognitive Radio Network}
  \acro{CS}{Coordinated Scheduling}
  \acro{CSI}{channel state information}
  \acro{CSIR}{channel state information at the receiver}
  \acro{CSIT}{channel state information at the transmitter}
  \acro{CUE}{cellular user equipment}
  \acro{DAC}{digital-to-analog converter}
  \acro{D2D}{device-to-device}
  \acro{DCA}{Dynamic Channel Allocation}
  \acro{DE}{Differential Evolution}
  \acro{DFT}{Discrete Fourier Transform}
  \acro{DIST}{Distance}
  \acro{DL}{downlink}
  \acro{DMA}{Double Moving Average}
	\acro{DMRS}{Demodulation Reference Signal}
  \acro{D2DM}{D2D Mode}
  \acro{DMS}{D2D Mode Selection}
  \acro{DNN}{Deep Neural Network} 
  \acro{DoA}{direction-of-arrival}
  \acro{DOCSIS}{Data Over Cable Service Interface Specification}
  \acro{DPC}{Dirty Paper Coding}
  \acro{DRA}{Dynamic Resource Assignment}
  \acro{DSA}{Dynamic Spectrum Access}
  \acro{DSM}{Delay-based Satisfaction Maximization}
  \acro{EBD}{electrical balance duplexer}
  \acro{ECC}{Electronic Communications Committee}
  \acro{EFLC}{Error Feedback Based Load Control}
  \acro{EI}{Efficiency Indicator}
  \acro{eNB}{Evolved Node B}
  \acro{EPA}{Equal Power Allocation}
  \acro{EPC}{Evolved Packet Core}
  \acro{EPS}{Evolved Packet System}
  \acro{E-UTRAN}{Evolved Universal Terrestrial Radio Access Network}
  \acro{ES}{Exhaustive Search}
  \acro{FD}{full-duplex}
  \acro{FDD}{frequency division duplexing}
  \acro{FDM}{Frequency Division Multiplexing}
  \acro{FER}{Frame Erasure Rate}
  \acro{FF}{Fast Fading}
  \acro{FSB}{Fixed Switched Beamforming}
  \acro{FST}{Fixed SNR Target}
  \acro{FTP}{File Transfer Protocol}
  \acro{GA}{Genetic Algorithm}
  \acro{GBR}{Guaranteed Bit Rate}
  \acro{GLR}{Gain to Leakage Ratio}
  \acro{GOS}{Generated Orthogonal Sequence}
  \acro{GPL}{GNU General Public License}
  \acro{GRP}{Grouping}
  \acro{HARQ}{Hybrid Automatic Repeat Request}
  \acro{HD}{half-duplex}
  \acro{HMS}{Harmonic Mode Selection}
  \acro{HOL}{Head Of Line}
  \acro{HSDPA}{High-Speed Downlink Packet Access}
  \acro{HSPA}{High Speed Packet Access}
  \acro{HTTP}{HyperText Transfer Protocol}
  \acro{ICMP}{Internet Control Message Protocol}
  \acro{ICI}{Intercell Interference}
  \acro{ID}{Identification}
  \acro{IETF}{Internet Engineering Task Force}
  \acro{ILP}{Integer Linear Program}
  \acro{JRAPAP}{Joint RB Assignment and Power Allocation Problem}
  \acro{MEC}{Multi-Access edge computing}
  \acro{UID}{Unique Identification}
  \acro{IAB}{integrated access and backhaul}
  \acro{IBFD}{in-band full-duplex}
  \acro{IID}{Independent and Identically Distributed}
  \acro{IIR}{Infinite Impulse Response}
  \acro{ILP}{Integer Linear Problem}
  \acro{IMT}{International Mobile Telecommunications}
  \acro{INV}{Inverted Norm-based Grouping}
  \acro{IoT}{Internet of Things}
  \acro{IP}{Internet Protocol}
  \acro{IPv6}{Internet Protocol Version 6}
  \acro{I/Q}{in-phase/quadrature}
  \acro{IRS}{Intelligent Reflective Surface}
  \acro{ISAC}{integrated sensing and communications}
  \acro{ISD}{Inter-Site Distance}
  \acro{ISI}{Inter Symbol Interference}
  \acro{ITU}{International Telecommunication Union}
  \acro{JCAS}{joint communications and sensing}
  \acro{JOAS}{Joint Opportunistic Assignment and Scheduling}
  \acro{JOS}{Joint Opportunistic Scheduling}
  \acro{JP}{Joint Processing}
	\acro{JS}{Jump-Stay}
    \acro{KF}{Kalman filter}
  \acro{KKT}{Karush-Kuhn-Tucker}
  \acro{L3}{Layer-3}
  \acro{LAC}{Link Admission Control}
  \acro{LA}{Link Adaptation}
  \acro{LC}{Load Control}
  \acro{LNA}{low-noise amplifier}
  \acro{LOS}{Line of Sight}
  \acro{LP}{Linear Programming}
  \acro{LS}{least squares}
  \acro{LTE}{Long Term Evolution}
  \acro{LTE-A}{LTE-Advanced}
  \acro{LTE-Advanced}{Long Term Evolution Advanced}
  \acro{M2M}{Machine-to-Machine}
  \acro{MAC}{Medium Access Control}
  \acro{MANET}{Mobile Ad hoc Network}
  \acro{MC}{Modular Clock}
  \acro{MCS}{Modulation and Coding Scheme}
  \acro{MDB}{Measured Delay Based}
  \acro{MDI}{Minimum D2D Interference}
  \acro{MF}{Matched Filter}
  \acro{MG}{Maximum Gain}
  \acro{MH}{Multi-Hop}
  \acro{MIMO}{multiple-input multiple-output}
  \acro{MINLP}{Mixed Integer Nonlinear Programming}
  \acro{MIP}{Mixed Integer Programming}
  \acro{MISO}{Multiple Input Single Output}
  \acro{ML}{maximum likelihood}
  \acro{MLWDF}{Modified Largest Weighted Delay First}
  \acro{mmWave}{millimeter wave}
  \acro{MME}{Mobility Management Entity}
  \acro{MMSE}{minimum mean squared error}
  \acro{MOS}{Mean Opinion Score}
  \acro{MPF}{Multicarrier Proportional Fair}
  \acro{MRA}{Maximum Rate Allocation}
  \acro{MR}{Maximum Rate}
  \acro{MRC}{maximum ratio combining}
  \acro{MRT}{Maximum Ratio Transmission}
  \acro{MRUS}{Maximum Rate with User Satisfaction}
  \acro{MS}{mobile station}
  \acro{MSE}{mean squared error}
  \acro{MSI}{Multi-Stream Interference}
  \acro{MTC}{Machine-Type Communication}
  \acro{MTSI}{Multimedia Telephony Services over IMS}
  \acro{MTSM}{Modified Throughput-based Satisfaction Maximization}
  \acro{MU-MIMO}{multiuser multiple input multiple output}
  \acro{MU}{multi-user}
  \acro{NAS}{Non-Access Stratum}
  \acro{NB}{Node B}
  \acro{NE}{Nash equilibrium}
  \acro{NCL}{Neighbor Cell List}
  \acro{NLP}{Nonlinear Programming}
  \acro{NLOS}{Non-Line of Sight}
  \acro{NMSE}{Normalized Mean Square Error}
  \acro{NORM}{Normalized Projection-based Grouping}
  \acro{NP}{Non-Polynomial Time}
  \acro{NR}{New Radio}
  \acro{NRT}{Non-Real Time}
  \acro{NSPS}{National Security and Public Safety Services}
  \acro{NTNs}{non-terrestrial networks}
  \acro{O2I}{Outdoor to Indoor}
  \acro{OFDMA}{orthogonal frequency division multiple access}
  \acro{OFDM}{orthogonal frequency division multiplexing}
  \acro{OFPC}{Open Loop with Fractional Path Loss Compensation}
  \acro{O2I}{Outdoor-to-Indoor}
  \acro{OL}{Open Loop}
  \acro{OLPC}{Open-Loop Power Control}
  \acro{OL-PC}{Open-Loop Power Control}
  \acro{OPEX}{Operational Expenditure}
  \acro{ORB}{Orthogonal Random Beamforming}
  \acro{JO-PF}{Joint Opportunistic Proportional Fair}
  \acro{OSI}{Open Systems Interconnection}
  \acro{PA}{power amplifier}
  \acro{PAIR}{D2D Pair Gain-based Grouping}
  \acro{PAPR}{Peak-to-Average Power Ratio}
  \acro{PBCH}{physical broadcast channel}
  \acro{P2P}{Peer-to-Peer}
  \acro{PC}{Power Control}
  \acro{PCI}{Physical Cell ID}
  \acro{PDF}{Probability Density Function}
  \acro{PDPR}{pilot-to-data power ratio}
  \acro{PER}{Packet Error Rate}
  \acro{PF}{Proportional Fair}
  \acro{P-GW}{Packet Data Network Gateway}
  \acro{PL}{Pathloss}
  \acro{PMN}{Perceptive Mobile Network}
  \acro{PPR}{pilot power ratio}
  \acro{PRB}{physical resource block}
  \acro{PROJ}{Projection-based Grouping}
  \acro{ProSe}{Proximity Services}
  \acro{PS}{Packet Scheduling}
  \acro{PSAM}{pilot symbol assisted modulation}
  \acro{PSO}{Particle Swarm Optimization}
  \acro{PZF}{Projected Zero-Forcing}
  \acro{QAM}{Quadrature Amplitude Modulation}
  \acro{QoS}{Quality of Service}
  \acro{QPSK}{Quadri-Phase Shift Keying}
  \acro{RAISES}{Reallocation-based Assignment for Improved Spectral Efficiency and Satisfaction}
  \acro{RAN}{Radio Access Network}
  \acro{RA}{Resource Allocation}
  \acro{RAT}{Radio Access Technology}
  \acro{RATE}{Rate-based}
  \acro{RB}{resource block}
  \acro{RBG}{Resource Block Group}
  \acro{REF}{Reference Grouping}
  \acro{RF}{radio frequency}
  \acro{RIS}{reconfigurable intelligent surface}
  \acro{RLC}{Radio Link Control}
  \acro{RM}{Rate Maximization}
  \acro{RNC}{Radio Network Controller}
  \acro{RND}{Random Grouping}
  \acro{RRA}{Radio Resource Allocation}
  \acro{RRM}{Radio Resource Management}
  \acro{RSCP}{Received Signal Code Power}
  \acro{RSRP}{Reference Signal Receive Power}
  \acro{RSRQ}{Reference Signal Receive Quality}
  \acro{RR}{Round Robin}
  \acro{RRC}{Radio Resource Control}
  \acro{RSSI}{Received Signal Strength Indicator}
  \acro{RT}{Real Time}
  \acro{RU}{Resource Unit}
  \acro{RUNE}{RUdimentary Network Emulator}
  \acro{RV}{Random Variable}
  \acro{RX}{receiver}
  \acro{SAC}{Session Admission Control}
  \acro{SBFD}{sub-band full-duplex}
  \acro{SCM}{Spatial Channel Model}
  \acro{SC-FDMA}{Single Carrier - Frequency Division Multiple Access}
  \acro{SD}{Soft Dropping}
  \acro{S-D}{Source-Destination}
  \acro{SDPC}{Soft Dropping Power Control}
  \acro{SDMA}{Space-Division Multiple Access}
  \acro{SER}{Symbol Error Rate}
  \acro{SES}{Simple Exponential Smoothing}
  \acro{S-GW}{Serving Gateway}
  \acro{SI}{self-interference}
  \acro{SINR}{signal-to-interference-plus-noise ratio}
  \acro{SIC}{self-interference cancellation}
  \acro{SIP}{Session Initiation Protocol}
  \acro{SISO}{single-input single-output}
  \acro{SIMO}{Single Input Multiple Output}
  \acro{SIR}{signal-to-interference ratio}
  \acro{SLNR}{Signal-to-Leakage-plus-Noise Ratio}
  \acro{SMA}{Simple Moving Average}
  \acro{SNR}{signal-to-noise ratio}
  \acro{SORA}{Satisfaction Oriented Resource Allocation}
  \acro{SORA-NRT}{Satisfaction-Oriented Resource Allocation for Non-Real Time Services}
  \acro{SORA-RT}{Satisfaction-Oriented Resource Allocation for Real Time Services}
  \acro{SPF}{Single-Carrier Proportional Fair}
  \acro{SRA}{Sequential Removal Algorithm}
  \acro{SRS}{Sounding Reference Signal}
  \acro{STAR}{simultaneous transmit-and-receive}
  \acro{SU-MIMO}{single-user multiple input multiple output}
  \acro{SU}{Single-User}
  \acro{SVD}{Singular Value Decomposition}
  \acro{TCP}{Transmission Control Protocol}
  \acro{TDD}{time division duplexing}
  \acro{TDMA}{Time Division Multiple Access}
  \acro{TETRA}{Terrestrial Trunked Radio}
  \acro{TP}{Transmit Power}
  \acro{TPC}{Transmit Power Control}
  \acro{TTI}{Transmission Time Interval}
  \acro{TTR}{Time-To-Rendezvous}
  \acro{TSM}{Throughput-based Satisfaction Maximization}
  \acro{TU}{Typical Urban}
  \acro{TX}{transmitter}
  \acro{UE}{user equipment}
  \acro{UEPS}{Urgency and Efficiency-based Packet Scheduling}
  \acro{UL}{uplink}
  \acro{UMTS}{Universal Mobile Telecommunications System}
  \acro{URI}{Uniform Resource Identifier}
  \acro{URM}{Unconstrained Rate Maximization}
  \acro{UT}{user terminal}
  \acro{VR}{Virtual Resource}
  \acro{VoIP}{Voice over IP}
  \acro{WAN}{Wireless Access Network}
  \acro{WCDMA}{Wideband Code Division Multiple Access}
  \acro{WF}{Water-filling}
  \acro{WiMAX}{Worldwide Interoperability for Microwave Access}
  \acro{WINNER}{Wireless World Initiative New Radio}
  \acro{WLAN}{Wireless Local Area Network}
  \acro{WMPF}{Weighted Multicarrier Proportional Fair}
  \acro{WPF}{Weighted Proportional Fair}
  \acro{WSN}{Wireless Sensor Network}
  \acro{WWW}{World Wide Web}
  \acro{XIXO}{(Single or Multiple) Input (Single or Multiple) Output}
  \acro{XDD}{cross-division duplex}
  \acro{ZF}{zero-forcing}
  \acro{ZMCSCG}{Zero Mean Circularly Symmetric Complex Gaussian}
\end{acronym}

\title{Full-Duplex Wireless for 6G: \\Progress Brings New Opportunities and Challenges }

\author{Besma Smida,~\IEEEmembership{Senior Member,~IEEE}, Ashutosh Sabharwal,~\IEEEmembership{Fellow,~IEEE},\\ Gabor Fodor,~\IEEEmembership{Senior Member,~IEEE}, George C. Alexandropoulos,~\IEEEmembership{Senior Member,~IEEE},\\ Himal A. Suraweera,~\IEEEmembership{Senior Member,~IEEE}, and Chan-Byoung Chae,~\IEEEmembership{Fellow,~IEEE}

\thanks{The work by S. Besma was in part supported by NSF CAREER \#1620902. The work by A. Sabharwal was in part supported by NSF Grants 2215082, 1956297 and 2148313. The work of G. C. Alexandropoulos was partially supported by the SNS JU TERRAMETA project under EU's Horizon Europe research and innovation programme under Grant Agreement No 101097101. The work by C.-B. Chae was in part supported by NRF and IITP grants funded by the Korea Government (NRF-2020R1A2C4001941, 2022-0-00704).}

\thanks{B. Smida is with the Department of Electrical and Computer Engineering, University of Illinois Chicago, Chicago, IL 77005, USA (e-mail: smida@uic.edu).} 
\thanks{A. Sabharwal is with the Department of Electrical and Computer Engineering, Rice University, Houston, TX 77005, USA (e-mail: ashu@rice.edu).}
\thanks{G. Fodor is with Ericsson Research and KTH Royal Institute of Technology, Stockholm, Sweden (e-mail: gabor.fodor@ericsson.com).}
\thanks{G. C. Alexandropoulos is with the Department of Informatics and Telecommunications, National and Kapodistrian University of Athens, 15784 Athens, Greece (e-mail: alexandg@di.uoa.gr).}
\thanks{H. A. Suraweera is the Department of Electrical and Electronic Engineering, University of Peradeniya, Peradeniya 20400, Sri Lanka (e-mail: himal@eng.pdn.ac.lk).}
\thanks{C.-B. Chae is with the School of Integrated Technology, Yonsei University, 03722 Seoul, Republic of Korea (e-mail: cbchae@yonsei.ac.kr).}
\thanks{\textit{Corresponding authors are B. Smida and C.-B. Chae.}}}

\maketitle

\begin{abstract}
The use of in-band full-duplex (FD) enables nodes to simultaneously transmit and receive on the same frequency band, which challenges the traditional assumption in wireless network design. The full-duplex capability enhances spectral efficiency and decreases latency, which are two key drivers pushing the performance expectations of next-generation mobile networks. In less than ten years, in-band FD has advanced from being demonstrated in research labs to being implemented in standards, presenting new opportunities to utilize its foundational concepts. Some of the most significant opportunities include using FD to enable wireless networks to sense the physical environment, integrate sensing and communication applications, develop integrated access and backhaul solutions, and work with smart signal propagation environments powered by reconfigurable intelligent surfaces. However, these new opportunities also come with new challenges for large-scale commercial deployment of FD technology, such as managing self-interference, combating cross-link interference in multi-cell networks,  and coexistence of dynamic time division duplex, subband FD and FD networks.
\end{abstract}



\section{Introduction}

The Mobile Internet is a crucial worldwide infrastructure that supports countless services and has exhibited a remarkable growth trajectory that continues to surpass expectations. Over the last five years, wireless broadband subscriptions have experienced an annual increase of 7\%, with the total number recently surpassing $7.5$ billion~\cite{Ericsson:2022}. In the period between the third quarters of $2021$ and $2022$, mobile network data traffic grew by 38\%, reaching a usage rate of $108$ Exabytes (i.e., $100$$\times$10$^{18}$ bytes) per month. To meet the future growth of diverse applications, next-generation broadband networks must be ever-more spectrally efficient, and achieve lower latency while reducing infrastructure costs and energy consumption per bit~\cite{Chowdhury_2020,Tataria_2021}. For wireless innovators, these expectations necessitate questioning current network design principles and encouraging the introduction of disruptive technological solutions~\cite{TFaisal_2020,Wanshi_2023,Fodor:2021,chae_mole}.

A fundamental assumption in traditional wireless network design has been the \ac{HD} communication, according to which a node either transmits or receives a signal in a single channel usage. The \ac{HD} design assumption manifests itself in the design of all current networks, with \ac{FDD} and \ac{TDD} being the two most commonly used techniques. While conceptually simple, in-band \ac{FD}, which enables simultaneous transmissions and receptions in the same frequency band, was considered practically infeasible till its initial experimental results back in $2010$ using off-the-shelf radios~\cite{rice_conf1,stan_conf1}. Those pioneering proof-of-concepts demonstrated that \ac{SI} can be sufficiently mitigated, paving the way for practical implementations of the technology. In fact, \ac{FD} communications were independently discovered and experimentally confirmed by many research groups worldwide~\cite{old_fd_mit,Khan2, rice_conf2, stan_conf2,Khan1,rice_exp,passive_symmetric,rice_journal1,Righini_2022}, ushering a new theoretical and practical research avenue for the wireless communications community \cite{Riihonen_2011,full-tap_MIMO,Krikidis_TWC_2012, Hong_2014, Sabharwal2014IBFD,Heino_2015,Zhang:16,Shende_2018,M_Mohammadi_2019,Kolodziej:2019,Vaibhav:2020,Roberts_BookCh_2022}. 
This \ac{STAR} technology has the potential to increase the spectral efficiency and reduce the latency of wireless systems, which constitute two of the key performance  drivers for the design of next generation broadband networking. To this end, the \ac{FD} technology can enable the convergence of future networks towards a unified communication, sensing, and backhaul platform. 


Between 2010 and 2020, there was a significant research and development dedicated to in-band \ac{FD} wireless communication \cite{Lingyang_Song_Book, Alves_FD_Book, Kolodziej_Book}. In fact, by 2015, the cable modem industry had already adopted in-band full-duplex to create the \ac{DOCSIS}~4.0 standard, which enabled next-generation cable modems to operate in \ac{FD} mode~\cite{Berscheid:2019}. By 2020, \ac{FD} wireless products had begun to appear. 
Thus, within the span of just one decade, the concept of in-band \ac{FD} communication has moved from being a laboratory idea to being incorporated into telecommunications standards.


The evolution of envisioned applications and use cases is a defining characteristic of successful technological ideas, which drives technical progress. The same trend is observed in the development of in-band \ac{FD} technology, which is the central focus of this overview paper. The paper begins by outlining the fundamental characteristics and current progress in \ac{SI} cancellation (SIC), the core component of in-band \ac{FD} radios. The paper then explores the potential impact of this technology on the future design of wireless networks, underlining the motivation for this paper.

Section~\ref{sec:applications} of the paper provides an overview of the major current and future applications of \ac{FD} wireless, such as wireless sensing, \ac{ISAC}, and \ac{IAB}. The section also considers the technology's potential for networking latency reduction, its inclusion in 5G-Advanced systems, and its integration with emerging \ac{6G} technologies such as \acp{RIS} and \ac{NTNs}. For each application, the paper highlights important open questions and challenges for future research, with the aim of inspiring further investigation and research within the wireless communications community.

\section{Self-interference Cancellation Techniques \label{sec:foundations}}
\begin{figure}[!t]
	\begin{center}
		\centerline{\resizebox{0.9\columnwidth}{!}{\includegraphics{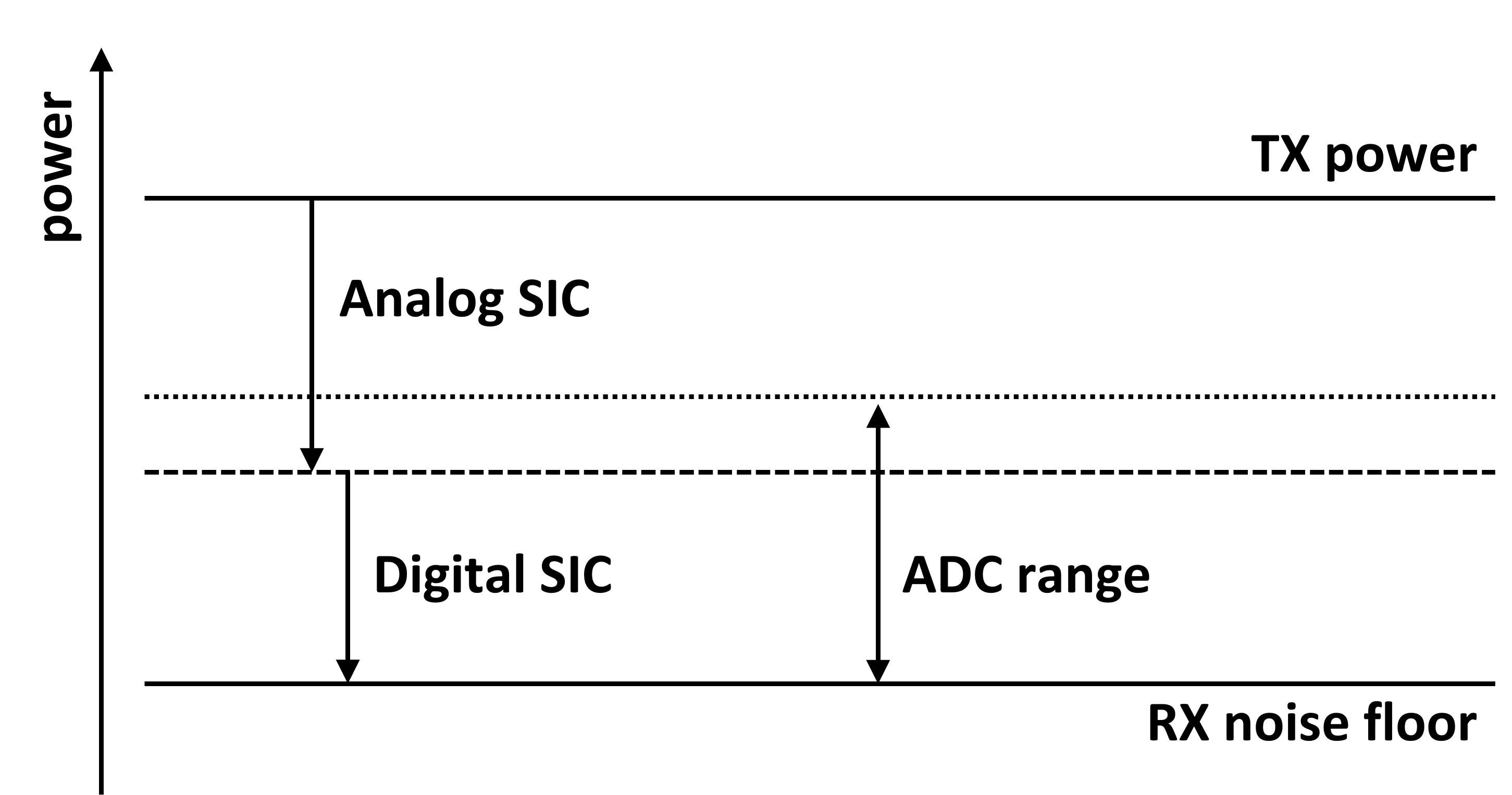}}}
		\caption{An illustrative example of the signal power levels at the outputs of the different modules comprising the \ac{SIC} unit in an in-band \ac{FD} transceiver.}	
		\label{fig.ADC}
	\end{center}
\end{figure}

For successful \ac{FD} communications, it is important to suppress \ac{SI} to a level below the receiver's thermal noise floor. In a system with multiple antennas, passive analog cancellation can be achieved using a directive antenna or an antenna cancellation technique. However, if the transmitter and receiver share a single antenna, an isolator is required to separate the SI from the signal of interest. The remaining SI after passive analog cancellation can be divided into linear and nonlinear components. Linear components are typically caused by multipath propagation between the transmitter and receiver, and in a single-antenna system with a circulator, leakage from the circulator is modeled as linear. On the other hand, nonlinear components mostly come from power amplifier nonlinearity, \ac{ADC} quantization noise, transmitter noise, and other \ac{RF} imperfections.

\begin{figure}[!t]
	\begin{center}
		\centerline{\resizebox{1.0\columnwidth}{!}{\includegraphics{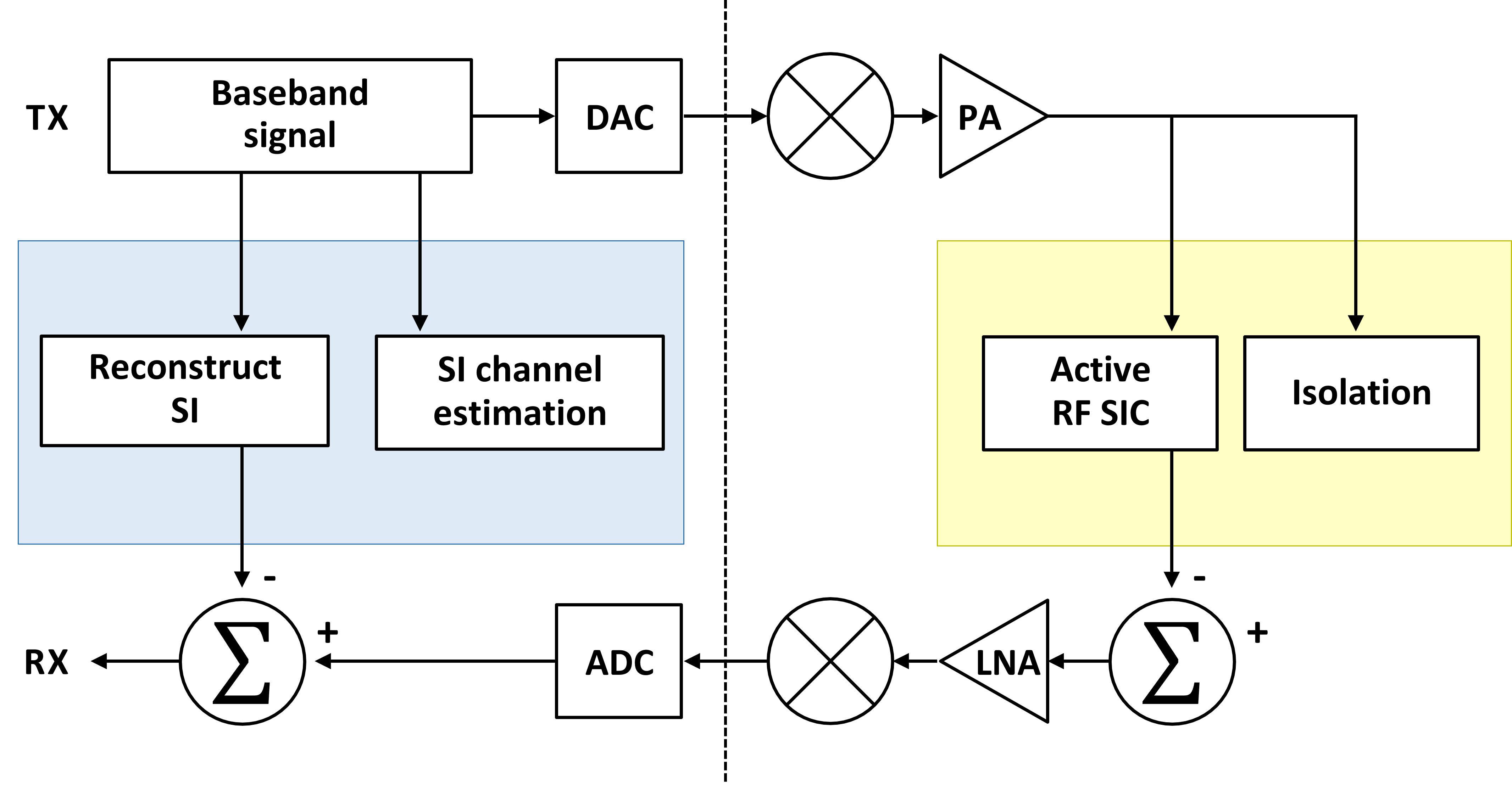}}}
		\caption{A block diagram of a single-antenna \ac{FD} system with generic analog- and digital-domain \ac{SIC}; DAC: digital-to-analog converter and LNA: low-noise amplifier.}
	\label{fig.blockdiagram}       
	\end{center}
\end{figure}

Channel estimation-based cancellation can be carried out in both analog and digital domains. The cancellation procedure consists of three steps: (1) modeling the SI channel, (2) estimating the \ac{SI} channel using perfect knowledge of the transmit signal, and (3) reconstructing and subtracting the SI from the received signal. Linear components of SI can be easily mitigated by existing channel estimation methods. Nonlinear components such as power amplifier nonlinearity and I/Q imbalance can be modeled as described in references~\cite{nonlinear_digital,nonlinear_digital2}. The primary challenge of SIC is the required dynamic range of the \ac{ADC} to acquire both the SI and signal-of-interest simultaneously. For instance, with a 14-bit \ac{ADC}, the dynamic range is 86 dB,\footnote{A dynamic range of a $q$-bit \ac{ADC} is calculated as $ 6.02\times q+1.76 \ [\text{dB}].$} which implies that the SI power after analog cancellation must not exceed the receiver noise floor by 86~dB. Fig.~\ref{fig.ADC} demonstrates the power levels in a successful \ac{SIC} scenario. The \ac{ADC}'s dynamic range determines the amount of required analog cancellation, which is typically not achievable with intrinsic cancellation alone. As a result, most full-duplex systems use additional active cancellation in the analog domain, which is also based on SI channel estimation. A full-duplex system that combines analog and digital domain \ac{SIC} is shown in Fig.~\ref{fig.blockdiagram}.

\subsection{Passive Analog SI Cancellation}
In the propagation stage, an isolator can be used to separate the desired signal and the \ac{SI}. Passive analog cancellation, which does not require an active \ac{RF} component, is often employed in this stage. One approach is to use separate antennas for the transmit and receive chains, which can be described in terms of path loss between the two antennas as
\begin{equation}
	\label{eq.pathloss}
L =10n\log{10}{d}+C,
\end{equation}
where $n$, $d$, and $C$ denote the path loss exponent, the distance between the two antennas, and a system-dependent constant, respectively. Clearly, as~\eqref{eq.pathloss} indicates, the \ac{SIC} amount is determined by the size of the \ac{FD} transceiver (specified~by~$d$).  

Another analog \ac{SIC} solution relies on antenna cancellation via using an additional \ac{TX} (or \ac{RX} antenna), aiming to generate a $\pi$-phase rotated \ac{SI} signal at the \ac{RX} of the \ac{FD} transceiver \cite{Khan2}. Let us denote the wavelength of the transmitted signal by $\lambda$. Suppose also that the distance between a \ac{TX} antenna $1$ of the \ac{FD} radio and its \ac{RX} is $\lambda/2$ larger than the distance between its \ac{TX} $2$ and \ac{RX}. In this case, considering also line-of-sight signal propagation, the two transmit signals will arrive with a phase difference of $\pi$ at the \ac{RX}. To compensate for the different path loss, we have to allocate more power to the \ac{TX} $2$. This transmission scheme implies that the \ac{SI} signal will be perfectly mitigated, because the signal from the two transmit antennas will add destructively at the receive antenna.
The authors in~\cite{stan_conf1} adopted interference cancellation for this antenna-based \ac{SIC} scheme. Note that,  since it depends on wavelength $\lambda$, this asymmetric antenna cancellation method is inherently applicable in narrowband systems. 
\begin{figure}[!t]
	\begin{center}
   		\centerline{\resizebox{1.0\columnwidth}{!}{\includegraphics{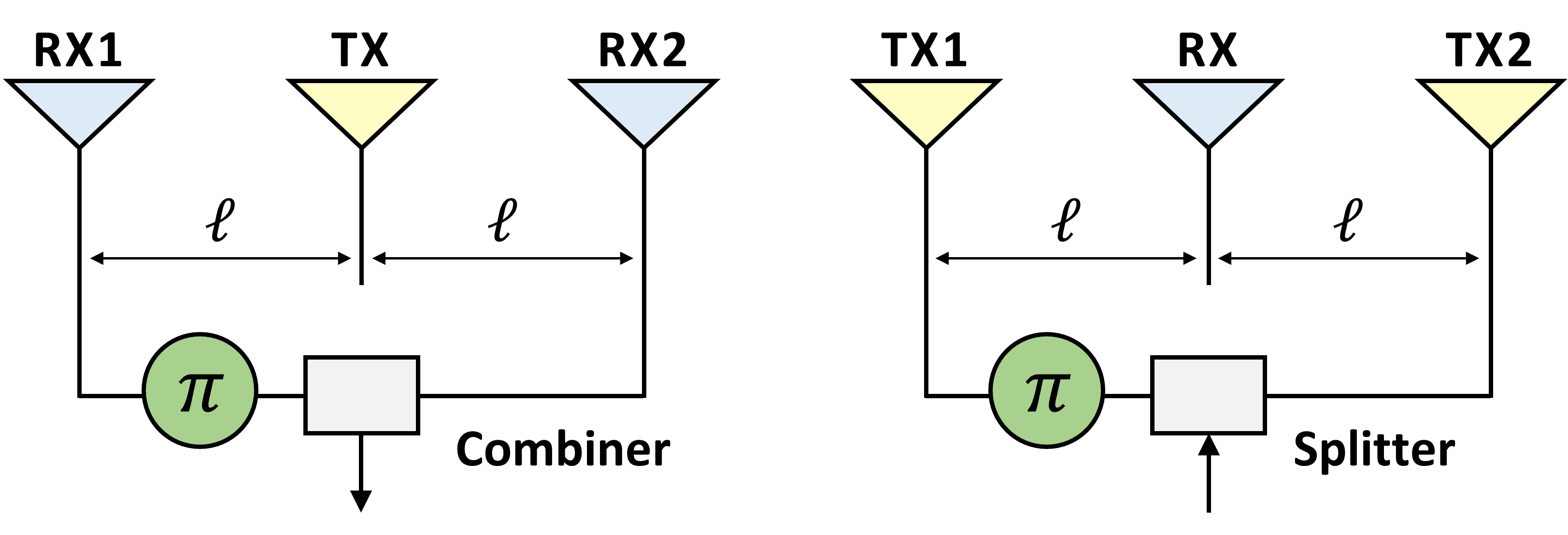}}}
		\caption{Antenna-based passive analog SI cancellation using a $\pi$-phase shifter either at the TX or the RX part of the FD system ~\cite{passive_symmetric}.} 
		\label{Fig.symmetric}
	\end{center}
\end{figure}

To alleviate the dependency on the narrow bandwidth assumption, a symmetric antenna cancellation scheme was presented in~\cite{passive_symmetric,Khan2}. The antenna system architecture for such an \ac{FD} system is demonstrated in Fig.~\ref{Fig.symmetric}. As shown, a $\pi$-phase shifter is employed instead of having space separate TXs. It should be emphasized here that we can mitigate \ac{SI} by using two RX antennas and one TX antenna. The $\pi$-phase shifter is placed after one of the RX antenna as illustrated in Fig.~\ref{Fig.symmetric}. 
In~\cite{passive_dualpol,realtime}, passive analog \ac{SIC} techniques using polarized antennas were proposed, while~\cite{realtime} experimented with an \ac{FD}-based \ac{LTE} prototype including a dual-polarized antenna.

\begin{figure}[!t]
	\begin{center}
		\centerline{\resizebox{0.9\columnwidth}{!}{\includegraphics{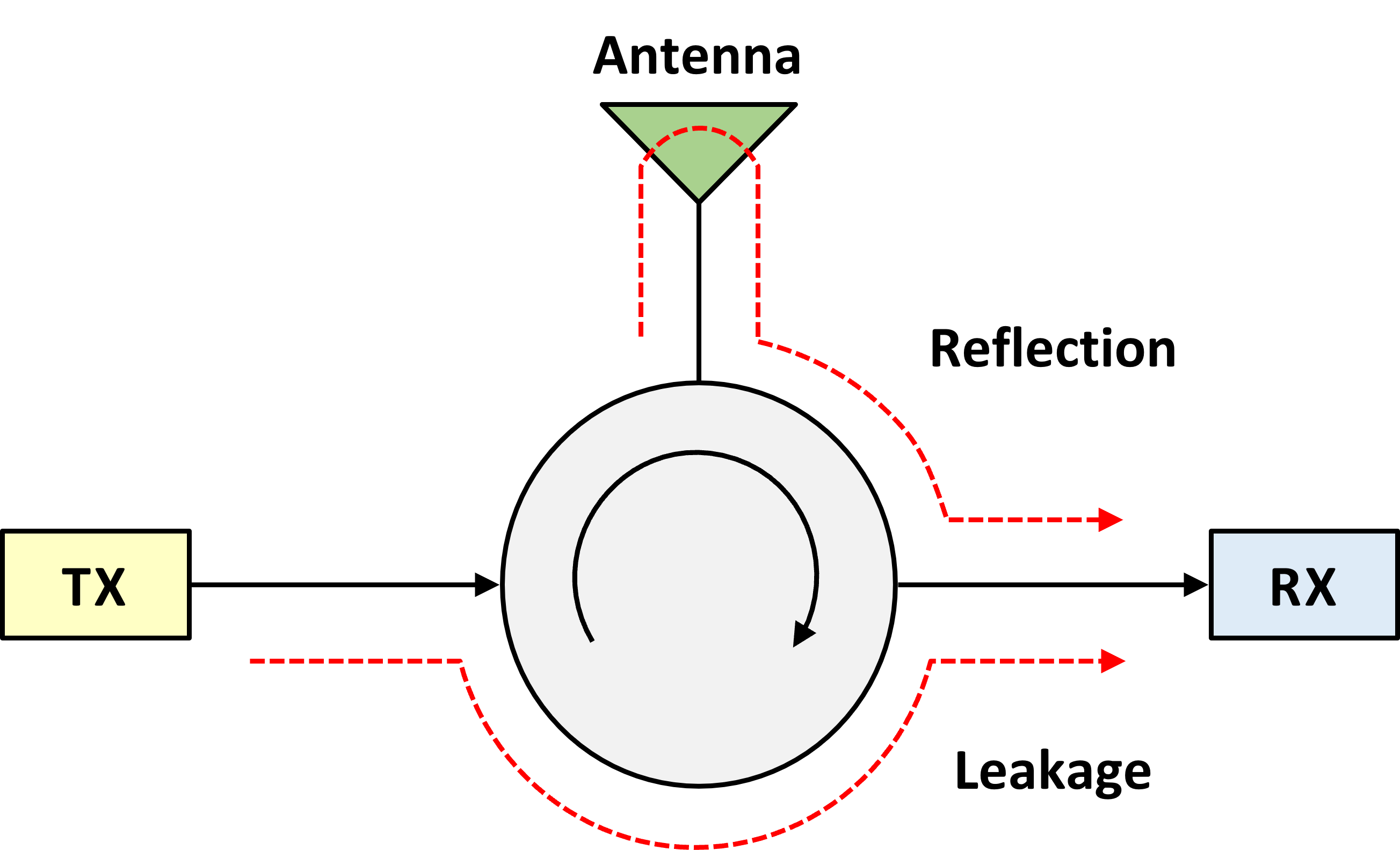}}}
		\caption{Schematic of a $3$-port circulator and the resulting signal components. }
		\label{Fig.circulator}
	\end{center}
\end{figure}

To achieve passive analog \ac{SIC} in \ac{FD} radios utilizing the same antenna for simultaneous transmission and reception, one needs a circulator. It is noted that circulators can provide isolation only for narrow bandwidths. As shown in Fig.~\ref{Fig.circulator}, this is a $3$-port device (TX, antenna, and RX ports) that steers the signal entering any port to only the next port; ferrite circulators are often used in communication systems. 
When a signal enters to Port $1$, the ferrite changes a magnetic resonance pattern to create a null at Port $3$. The resulting \ac{SI} components from a circulator are sketched in Fig.~\ref{Fig.circulator}. 
As shown, the strongest \ac{SI} component comes from the direct leakage, with a typical circulator being capable to provide $20$ and up to $30$ dB isolation for this signal. The remaining \ac{SI} components can mitigated using active analog \ac{SIC}, as will be described in the sequel. The second signal component contributed by the circulator is the signal reflected by the antenna. Its reflected power (a.k.a., return loss) can be computed in dB as follows:
\begin{equation}
\label{eq.returnloss}
L_{\text{ret}}=-20\log_{10}\left|\frac{Z_L-Z_S}{Z_L+Z_S}\right|,
\end{equation}
where $Z_L$ and $Z_S$ represent the load (i.e. receiver, transmitter and circulator) and the source (i.e. antenna) impedance, respectively. 
Finally, the third component resulting from the deployment of the circulator is a reflected signal from the environment that depends on the objects lying in the proximity of the \ac{FD} system. 
 

Passive analog \ac{SIC} has a major drawback in that it cannot eliminate the \ac{SI} signal resulting from reflections from the environment. In~\cite{passive_absorb}, the researchers compared the residual \ac{SI} after passive cancellation in both an anechoic chamber and a highly reflective room with metallic walls. To address this issue, they used a combination of directional isolation, absorptive shielding, and cross polarization. The absorptive shielding involved placing an RF-absorptive material between the TX and RX of the FD system to increase path loss. They used Eccosorb AN-79 as the absorber, which consists of discrete layers of lossy material and has an impedance designed to be similar to air. The impedance gradually increased from the incident layer to the rear layer, making the multi-layer structure more effective at absorbing electromagnetic waves over a broad frequency range than the single-layer structure used in~\cite{passive_absorb}. These findings suggest that active cancellation, in combination with passive suppression, may require higher-order filters or per-subcarrier cancellation.


\subsection{Active Analog SI Cancellation}
\begin{figure}[!t]
	\begin{center}
		\centerline{\resizebox{1.0\columnwidth}{!}{\includegraphics{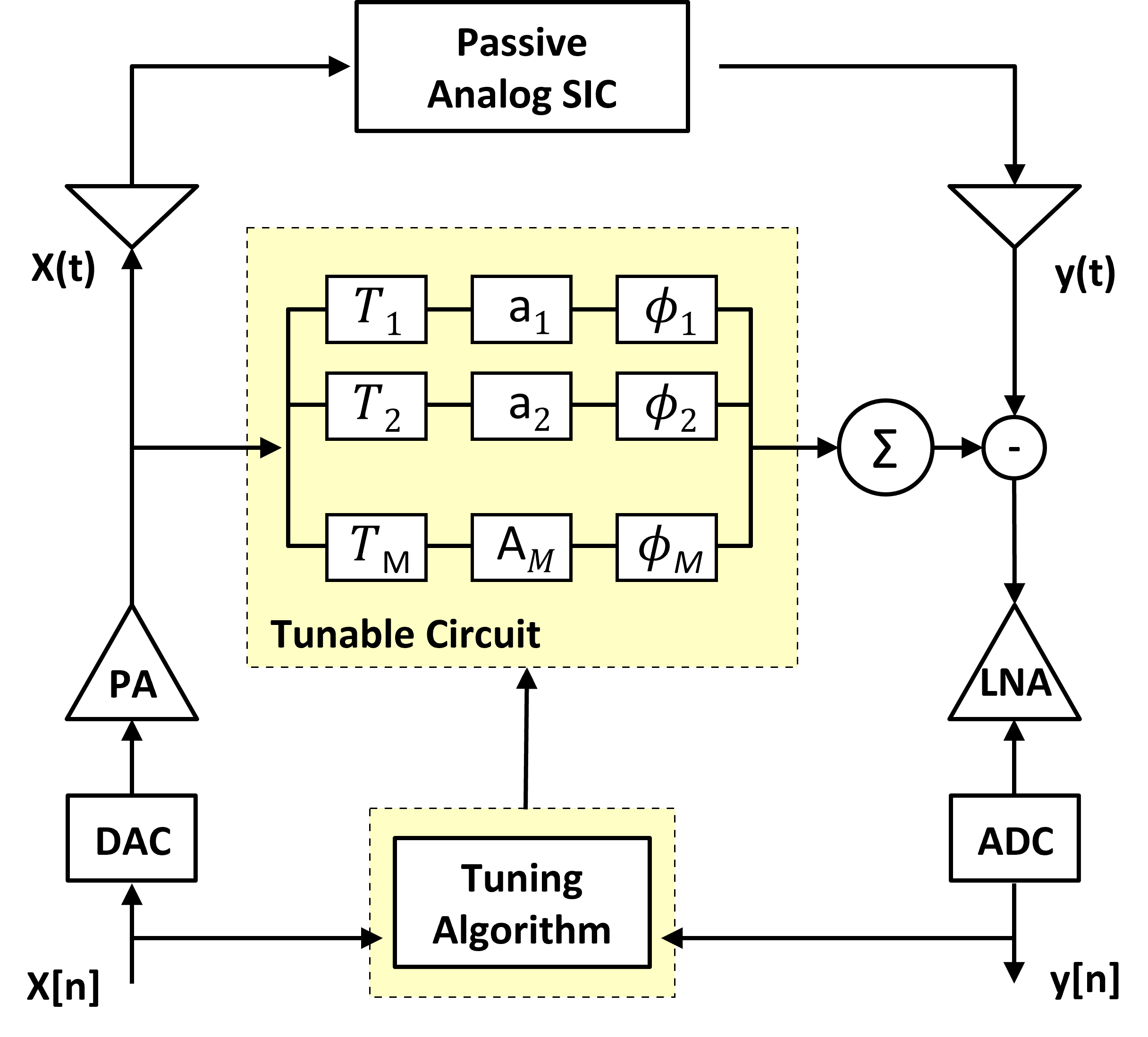}}}
  		\caption{Joint active and passive analog \ac{SIC}, emphasizing the $M$-tap tunable analog canceller and its optimization mechanism.}
  		\label{fig:adaptive circuit}       
	\end{center}
\end{figure}

Active analog \ac{SIC} targets at generating a signal that matches to the usually unavoidable leakage from the passive analog \ac{SIC}. This is done by using a dynamically tunable circuit mimicking an auxiliary transmit \ac{RF} chain \cite{rice_exp,9895328}, which is fed with a copy of the transmitted signal. A general multi-tap analog SI canceller consisting of $M$ delay line $\tau_1,\tau_2,\ldots,\tau_M$ is illustrated in Fig.~\ref{fig:adaptive circuit}. As shown, each $i$-th delay line, with $i=1,2,\ldots,M$, comprises an attenuator $a_i$ and a phase shifter $\phi_i$. The $M$ triplets of analog cancellation parameters are configured via solving the following optimization problem:
\begin{equation}
\label{eq.circuit}
\underset{\tau,a,\phi}{\text{min}}   \left(y(t)-\sum_{i=1}^{M}x_{\tau_i,a_i,\phi_i}(t)\right)^2,
\end{equation} 
where $x_{\tau_i,a_i,\phi_i}(t)$ denotes the output signal of each $i$-th delay line that will be a function of its tunable parameters, and $y(t)$ is the leaking signal at the output of the passive analog \ac{SIC}. In Fig.~\ref{fig:adaptive circuit}, $x(t)$ represents the transmitted signal, resulting from the baseband version $x[n]$ when it passes through the transmit RF chain. 
The paper \cite{9431091}  introduces the use of neural network machine learning to accelerate the tuning of multi-tap adaptive RF cancellers.
In an analogous way, $y_[n]$ is the baseband transformation of $y(t)$, when the latter feeds the receive RF chain. It is noted that the optimization problem \eqref{eq.circuit} is performed in the digital domain (i.e., the optimization algorithm receives as input variables the baseband signals $x[n]$ and $y[n]$). Recall that the aim of the analog \ac{SIC} is to avoid the \ac{ADC} saturation. Therefore, at an initial phase, the \ac{SI} signal is transmitted at a weak power to carry out the optimization (i.e., tune the active analog \ac{SIC} circuit), while avoiding \ac{ADC} the saturation. When the parameters are set, the \ac{FD} transmission is performed.


To eliminate the two main sources of interference in SI leakage, an adaptive circuit must be developed when using a circulator as a passive isolator. This circuit should incorporate delay lines with variable attenuators to fix the delays using the properties of the leakage \cite{7146163,maddio2013reconfigurable}. By selecting appropriate fixed delays, the original SI can be regenerated even if it has arbitrary delays within a particular range. Using fixed delays in this way eliminates the need for high-resolution delays, making the proposed approach more feasible to implement.

\begin{figure}[!t]
	\begin{center}
		\centerline{\resizebox{0.9\columnwidth}{!}{\includegraphics{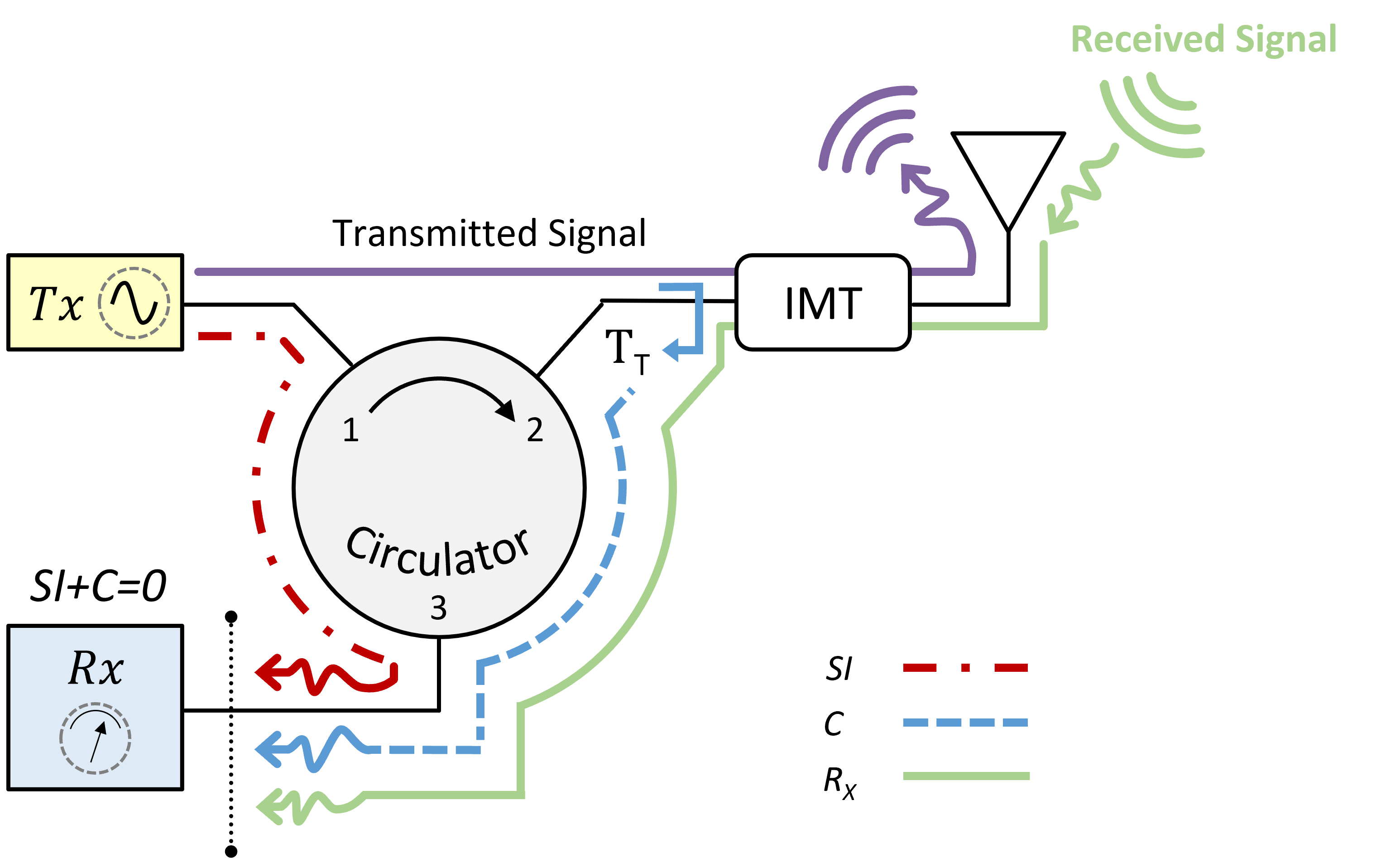}}}
  		\caption{Inherent active \ac{SIC} using a reconfigurable IMT circuit consisting of two varactor diodes at the antenna port.}
  		\label{fig:sic_imt}       
	\end{center}
\end{figure}

With a similar objective in mind, the authors of \cite{8363897,8335770} adopted a straightforward approach by utilizing an impedance mismatched terminal (IMT) circuit and the secondary \ac{SI} signals inherent at the circulator (i.e., those reflected by the antenna), to nullify the primary \ac{SI} signals that leaked from the \ac{TX} to the \ac{RX} port. They modified the frequency response of the secondary \ac{SI} signals, referred to as $C$, by employing a reconfigurable IMT circuit with two varactor diodes at the antenna port, as shown in Fig.~\ref{fig:sic_imt}. By controlling the bias voltages of those diodes, the frequency band and bandwidth were adjusted. The IMT adjustability made it resilient to fabrication errors and antenna input impedance variations. The proposed method achieved a cancellation of more than $60$ dB over $65$ MHz bandwidth at $2.4$ GHz.




 \subsection{Digital SI Cancellation} 
\subsubsection{Filtering} 
Digital-domain \ac{SIC} can be accomplished via by \ac{SI} regeneration using {\it digital} filters fitted to the detector input through the known transmitted data. A key challenge for this cancellation is the nonlinear distortion introduced at the \ac{TX} and \ac{RX}. The effect of the transmit and receive power amplifier nonlinearities on the digital \ac{SIC} techniques was analyzed in \cite{7051286}, whereas the impact of the \ac{I/Q} imbalances and the resulting image components were investigated in \cite{6832439}. Signal models including the phase noise at the \ac{TX} and \ac{RX} oscillators were presented in \cite{7815419}. It was observed that the phase noise is a bottleneck for perfect \ac{SIC} while using two different oscillators for transmission and reception. A scenario with a common oscillator for both the \ac{TX} and \ac{RX} sides was also analyzed in \cite{balatsoukas2015baseband}. A comprehensive \ac{SI} model for single-antenna \ac{FD} radios, considering all impairments including the \ac{TX} and \ac{RX} \ac{I/Q} imbalances, nonlinear distortions in all the transceiver components, \ac{RX}'s noise figure, and the phase noise effect at both \ac{TX} and \ac{RX} \ac{I/Q} mixers was provided in \cite{8761524}.
 
\subsubsection{Linearization via \acp{DNN}} 
All of the aforementioned digital \ac{SIC} techniques require knowledge of the nonlinear model prior to implementation, including the order of a polynomial model. In contrast, \ac{DNN}-based solutions, such as those proposed in \cite{Balatsoukas:2018,Guo:2019,Shi:2019,Kurzo:2018,DNN_ULDL2019,9195843}, do not require prior knowledge of the nonlinear distortion model. Additionally, it has been demonstrated in \cite{Balatsoukas:2018, 9195843,9736621} that the use of a \ac{DNN} can reduce computational complexity by up to $36\%$ when compared to non-\ac{DNN}-based solutions \cite{Korpi:2017}. Most current \ac{DNN}-based methods combine estimation of the typically time-invariant nonlinear distortion and the time-varying \ac{SI} propagation channel, which must be periodically re-acquired. As such, these approaches require \emph{online} training, since they do not make use of information from previous time intervals. This can be addressed by decoupling the time-varying \ac{SI} propagation channel from the estimation of the time-invariant nonlinear distortion and utilizing \emph{transfer learning} to combine elements of both off-line and on-line training while accumulating information across time intervals \cite{9552213,9732685}.
\begin{figure*}[!t]
	\begin{center}
	\includegraphics[width=0.8\textwidth]{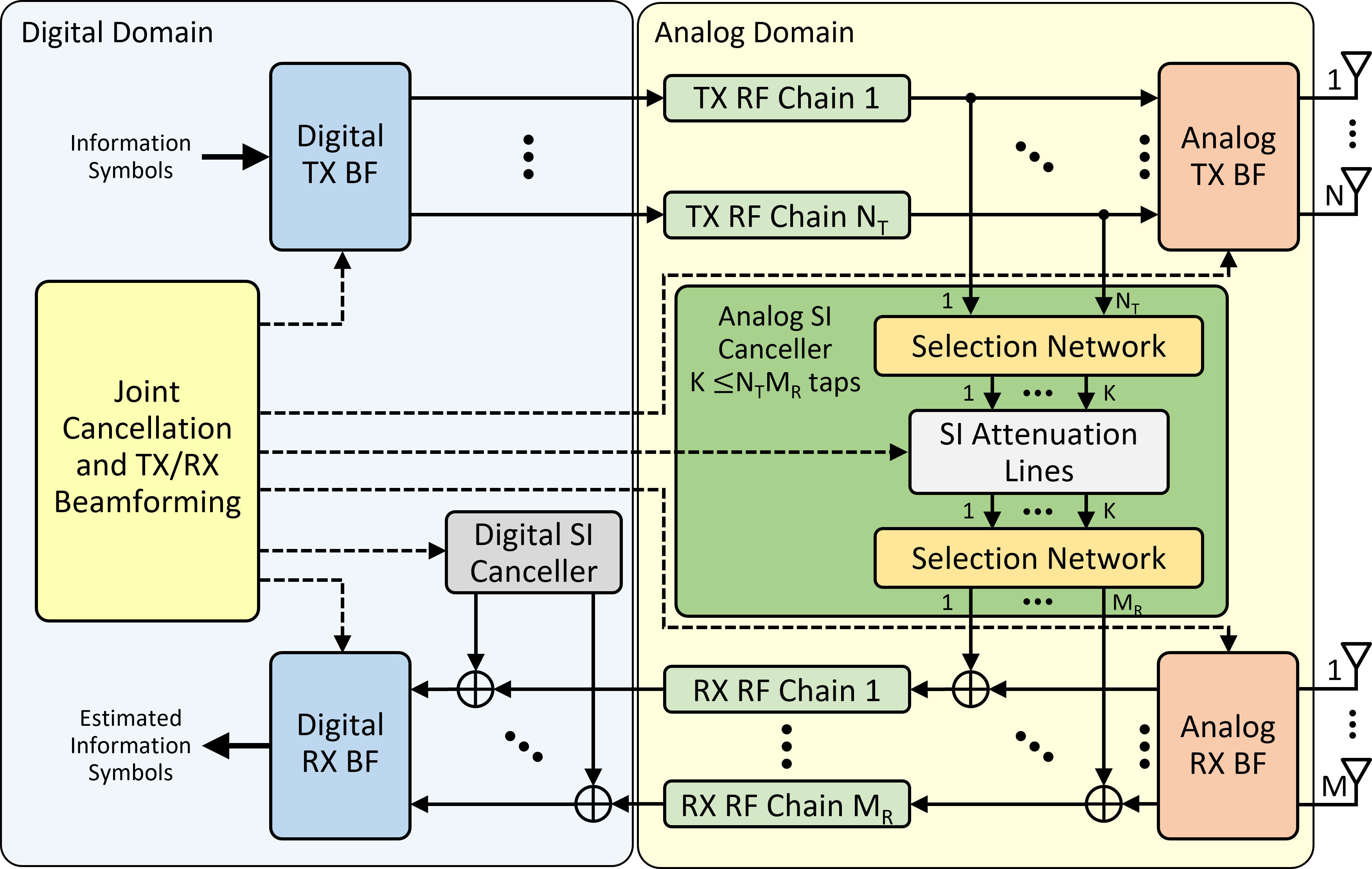}
	\caption{The generic \ac{FD} MIMO transceiver architecture of \cite{FD_MIMO_VTM2022} including most of the available multi-antenna \ac{STAR} modules as special cases. BF stands for beamforming.}
	\label{fig:FD_MIMO}
	\end{center}
\end{figure*} 

\subsection{MIMO Architectures and Spatial SI Cancellation} 
Spatial signal processing schemes, based on the \ac{MIMO} technology, can be used to suppress the \ac{SI} signal to desired levels, and have been usually deployed in conjunction with any of the previously described analog and digital SIC techniques \cite{full-tap_MIMO,Bharadia2014,rice_journal1,rice_conf3,Kolodziej2016}. In the available \ac{FD} \ac{MIMO} transceiver architectures, which intend to profit from the diversity or multiplexing gain offered by the efficient usage of the multiple antennas, the spatial degrees of freedom can be used to simultaneously handle the SI signal as well as the \ac{DL} and \ac{UL} signals of interest. It is noted, however, that more \ac{SI} signals are inevitably generated due to the larger number of antenna elements at an FD node's TX side, that need to be efficiently suppressed. To this end, FD MIMO targets to balance between the received signal quality and \ac{SI} suppression, and various techniques have been lately presented, namely, null-space projection or transmitting and receiving in orthogonal subspaces~\cite{full-tap_MIMO,Spatial_suppression,Gowda:18,Hien:2017,Lim_Kim}, antenna selection for choosing the best pair or subset of \ac{TX}/\ac{RX} antennas minimizing the \ac{SI}~\cite{Suraweera:2014,Zarifeh:2019,Mohammadi_Mobini_ICC18,Mobini:2022}, joint \ac{TX} and \ac{RX} beam selection~\cite{alexandropoulos2020full,IanRoberts:2022}, use of lenses~\cite{Chen:2023}, and \ac{ZF} beamforming for \ac{SI} elimination~\cite{Suraweera:2014,Mohammadi_CRAN:2018,Radwa:2021} or \ac{MMSE} filtering that maintains the useful signal's quality~\cite{alexandropoulos2017joint,Islam2019unified,FD_MIMO_Impairments,FD_MIMO_Arch,Chun-Tao:2017}.  

The generic unified \ac{FD} \ac{MIMO} transceiver architecture of Fig$.$~\ref{fig:FD_MIMO} was recently presented in \cite{alexandropoulos2020full,FD_MIMO_VTM2022} for the case of $N$ TX and $M$ RX antennas, where any of the $N$ and $M$ can be arbitrarily large. All \ac{TX} antenna elements are attached to $N_{\rm T}\leq N$ \ac{TX} \ac{RF} chains, and all RX antenna elements are connected to $M_{\rm R}\leq M$ \ac{RX} \ac{RF} chains. A \ac{TX} \ac{RF} chain consists of a \ac{DAC}, a mixer which upconverts the signal from baseband to \ac{RF}, and a power amplifier. An \ac{RX} \ac{RF} chain consists of a \ac{LNA}, a mixer which downconverts the signal from \ac{RF} to baseband, and an ADC converter. The \ac{RX} side of the architecture in Fig$.$~\ref{fig:FD_MIMO} is composed of analog \ac{RX} beamforming connecting the $M$ \ac{RX} antennas with the inputs of the $M_{\rm R}$ RX RF chains, and digital \ac{RX} beamforming that processes the outputs of the \ac{RX} \ac{RF} chains in baseband before symbol decoding. Similar to the \ac{TX} side, each of the \ac{RX} \ac{RF} chains is connected to a sub-array of $M/M_{\rm R}$ (again, assumed integer-valued) antenna elements via constant-magnitude phase shifters. The complex-valued analog \ac{RX} beamforming vectors belong in a predefined RX beam codebook, similar to the analog beamforming ones. The signals at the outputs of the receive \ac{RF} chains are being downconverted and fed to the digital RX beamformer to derive the estimated information symbols. As previously described, the objective of \ac{SIC} in \ac{FD} radios is to suppress the \ac{SI} signal below the RX noise floor. In practical wireless communication systems, each receive \ac{RF} chain is characterized by a maximum input signal power level, above which saturation happens. This means that when the \ac{SI} signal is larger than the chain's maximum allowable power level, this \ac{RX} \ac{RF} chain gets saturated. The role of the active analog \ac{SIC} is to suppress the \ac{SI} signal power level below the latter threshold in order to avoid saturation. 

Several variations of the MIMO architecture in Fig$.$~\ref{fig:FD_MIMO} were investigated over various hardware impairments in \cite{alexandropoulos2017joint,Islam2019unified,FD_MIMO_Impairments,FD_MIMO_Arch}. A active multi-tap wideband analog \ac{SIC} architecture, whose number of taps does not scale with the number of transceiver antennas and multipath SI components, was presented in \cite{WB_FD_MIMO_TWC2022}. An alternative architecture for the analog canceller, which is based on auxiliary \ac{TX}s for locally generating the \ac{SIC} signal, was presented in \cite{Duarte:20}. An extreme case of the unified architecture in Fig$.$~\ref{fig:FD_MIMO} relies solely on \ac{TX} digital beamforming to reduce SI at the RX antennas of the \ac{FD} node, without making use of any analog \ac{SIC} \cite{Spatial_suppression,spatial_suppression_chae}. In addition, \cite{Spatial_suppression2} studied two methods of partial active analog \ac{SIC}: the one, where the analog cancellers are assigned to a fixed set of antennas, and the other, where the cancellers are reconfigurable such that they can be dynamically assigned to any RX antennas based on the channel conditions.

As shown in Fig$.$~\ref{fig:FD_MIMO}, $K\leq N_{\rm T}M_{\rm R}$ \ac{SI} attenuation lines appear in the analog \ac{SIC}, which also includes two selection networks. The role of these networks is to decide which \ac{SI} signals will be mitigated from which inputs to the RX \ac{RF} chains. In \cite{alexandropoulos2017joint}, the selection network feeding the inputs to the \ac{SI} attenuation lines was implemented with $K$ MUltipleXers (MUXs) of $N_{\rm T}$-to-$1$, while $K$ DEMUltipleXers (DEMUXs) of $1$-to-$M_{\rm R}$ were considered for implementing the selection network at the outputs of the attenuation lines. Each attenuation line can be realized via a fixed delay, a variable phase shifter, and a variable attenuator, termed also compactly as an \textit{analog tap} \cite{alexandropoulos2017joint}, or with an auxiliary \ac{TX} for locally generating the \ac{SIC} signal \cite{Duarte:20}. The adders appearing before the \ac{RX} \ac{RF} chains in Fig$.$~\ref{fig:FD_MIMO} can be implemented via power combiners or directional couplers.

The conventional \ac{FD} \ac{MIMO} architectures of \cite{full-tap_MIMO,Bharadia2014,Kolodziej2016} deploy fully-connected analog \ac{SIC}, which interconnects all inputs to the \ac{TX} antennas to all outputs of the RX antennas in order to suppress all possible \ac{SI} signals. This cancellation approach requires $K=NM$ \ac{SI} attenuation lines. In the recent architecture of \cite{Vishwanath_2020}, the active analog \ac{SIC} connects all outputs of the \ac{TX} \ac{RF} chains with all inputs to the \ac{RX} \ac{RF} chains, which results in $K=N_{\rm T}M_{\rm R}$ attenuation lines. Clearly, when $N>N_{\rm T}$ or $M>M_{\rm R}$, the architecture of \cite{Vishwanath_2020} has the lowest complexity analog \ac{SIC}. However, an algorithmic design for the parameters of the attenuation lines as well as for \ac{TX}/\ac{RX} beamforming to justify improved performance with the fully-connected architectures of \cite{full-tap_MIMO} was not presented in \cite{Vishwanath_2020}.

The \ac{FD} \ac{MIMO} architectures in \cite{xiao2017full_all,satyanarayana2018hybrid,roberts2019beamforming,da20201} adopt also fully-connected analog \ac{TX}/\ac{RX} beamforming, which results in large numbers of phase shifters, as compared to the partially-connected architecture in Fig$.$~\ref{fig:FD_MIMO}. The architectures in \cite{xiao2017full_all,satyanarayana2018hybrid} include active analog \ac{SIC} with $K=NM$ taps, i.e., the canceller's hardware scales with the product of the \ac{TX} and \ac{RX} antennas. This complexity is prohibitive for \ac{FD} massive \ac{MIMO} systems. 
The paper \cite{9145036} presents a neural network structure that exploits the spatial correlation among the transmit and receive antennas to reduce the complexity of echo cancellation filters at the receivers. 
In \cite{roberts2019beamforming} and \cite{da20201}, passive \ac{SI} suppression was used instead of analog \ac{SIC}. However, it has been shown that only digital \ac{TX}/\ac{RX} beamforming falls short in efficiently handling nonlinear \ac{SI} resulting from \ac{TX} impairments. The architecture in Fig$.$~\ref{fig:FD_MIMO} deploys active analog \ac{SIC} with $K\leq N_{\rm T}M_{\rm R}$ taps (i.e., its complexity does not scale with the number of \ac{TX}/\ac{RX} antennas, hence, suitable for massive \ac{MIMO}), which is jointly designed with the analog and digital \ac{TX}/\ac{RF} beamformers.


\section{Current Applications and Trends}\label{sec:applications}
In this section, we present the latest applications and trends of the in-band \ac{FD} technology, highlighting the unique opportunities offered by its \ac{STAR} operation as well as the technical challenges that need to be addressed for each application's consideration for future wireless systems. We particularly discuss \ac{FD} for sensing, \ac{ISAC}, \ac{IAB}, latency reduction, its potential inclusion in the upcoming 5G-Advanced, interplay with the emerging wireless technology of reconfigurable metasurfaces and how \ac{NTNs} for ubiquitous connectivity could benefit from it.

\subsection{Wireless Networks That Sense} 
{\bf Opportunity}: In current and next generation wireless networks, higher spectral bands are of significant interest, e.g., \ac{mmWave} bands and above. Interestingly, those bands have been extensively used for sensing, e.g., automotive radars, airport security systems, and quality control in manufacturing. Thus, the use of such frequencies for wireless networks opens new possibilities for leveraging the same hardware and network infrastructure for both sensing and communications~\cite{Zhang:21b,Liu:22}. As a result, driven by efficient spectrum usage and enabled by the deployment of advanced antenna systems at cellular \acp{BS}, integrated or harmonized communication and sensing systems are gaining renewed interest both by the academic and industrial  research communities \cite{Liu:20,Liu:20b,S1-214242}. 

{\bf State-of-the-art}: In recent years, there has been significant activity in developing algorithms and establishing practical feasibility investigations in the broad area of wireless sensing. For example, several works on human sensing have demonstrated ability to sense heart rate, gait, or gestures using wireless signals~\cite{Behravan:22,Wu:22}. That body of work has also shed light on the differences between sensing and channel estimation; sensing appears to be a form of channel estimation that is commonly performed in all wireless systems. The difference between the latter two lies in the \emph{details} about the physical medium that is modeled to address communication and sensing objectives~\cite{Davis:11}. The aim of communication is maximal energy transfer from a source to a sink via the channel between them. Hence, the channel models need to capture how the energy transfer from an \ac{TX} to and \ac{RX} occurs, e.g., often modeled as a linear transfer function that does not distinguish between different objects in the scene. On the other hand, different sensing objectives explicitly model specific scene properties of interest. For example, radar sensing models the location and velocity of the targets of interest in the environment~\cite{Barneto:22}, \cite{Behravan:22} and human sensing models specify human-focused parameters, like heart rate and gait~\cite{Wu:22}.

\begin{figure}[t]
\begin{center}
\includegraphics[width=\hsize]{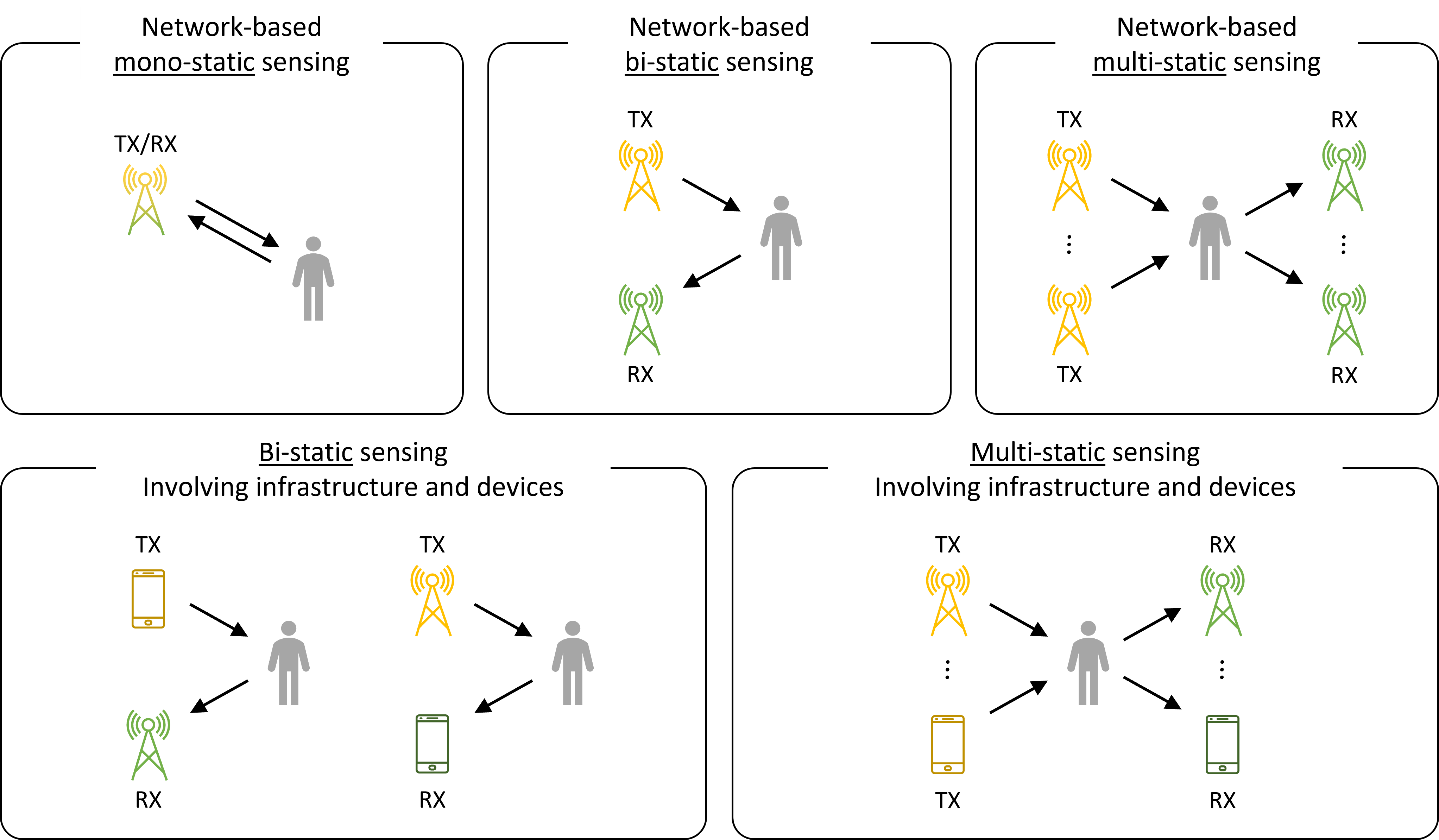}
\caption{
Mono-static, bi-static, and multi-static sensing scenarios in a cellular network. Mono-static sensing employing continuous-time signals require \ac{FD} capability at the \ac{BS}, while bi-static and multi-static sensing require accurate synchronization between the \acp{BS} and the \ac{UE} devices.}
\label{fig:scenarios}
\end{center}
\end{figure}

{\bf Challenges}: Three sensing architectures, namely mono-static, bi-static, and multi-static sensing, as illustrated in Fig.~\ref{fig:scenarios}, that are based on the location and number of \acp{TX} and \acp{RX}, are of practical interest. Two metrics of interest in sensing are that of the \emph{sensing accuracy} and \emph{sensing coverage}; these are equivalent to the communication systems' metrics of error probability and cell coverage. 

In the architecture of \emph{mono-static sensing}, the \ac{TX} and \ac{RX} are co-located. This architecture is commonly used in radar sensors. The co-location of a \ac{TX} and \ac{RX} allows time and frequency synchronization, enabling the highest sensing accuracy. For sensing applications where the distance to a target is of the order as that witnessed in wireless networks, mono-static sensing requires in-band \ac{FD} operation. However, when conventional \ac{SIC} is applied, all useful information for sensing is removed. To this end, a key challenge in developing mono-static wireless sensing is to retaining backscatter, while cancelling out \ac{SI} or reflection paths that do not provide any sensing information. Recent theoretical results~\cite{Mehrotra-Sabharwal:2022b} demonstrate that multipath could potentially be used to increase performance of monostatic sensing. 

The second architecture is \emph{bi-static sensing}, where the \ac{TX} and \ac{RX} cannot be time and frequency synchronized, which is often the case when the two communication ends are physically not co-located. The lack of synchronization creates a significant challenge, which has to be accounted for in the design of sensing algorithms. Overall, the sensing accuracy is lower in bi-static architectures than mono-static ones. However, the bi-static architectures do not require \ac{FD} operation, and hence, can leverage existing hardware capabilities of radio transceivers  

The last sensing architecture is \emph{multi-static sensing}, which is a generalization of bi-static sensing to allow arbitrary number of \acp{TX} and \acp{RX} to participate in sensing. The sensing coverage of multi-static sensing is (by definition) higher than mono- and bi-static systems, as it can illuminate more of the scene and capture larger portions of reflected energy. However, this gain in sensing coverage requires coordination in transmission and joint processing of received signals, which is a major challenge in mobile wireless networks. 

\subsection{Integrated Sensing and Communication (\ac{ISAC})}\label{sec:isac}

{\bf Opportunity}:
The emerging concept of \ac{ISAC}, according to which the previously competing sensing and communication functionalities are jointly optimized in the same hardware platform using a unified signal processing framework, is lately gaining increased attention from both academia and industry for future wireless networks \cite{liu2021integrated,ISAC_RIS_SPM,9099670}. In-band \ac{FD}  is being considered as a key enabling technology for \ac{ISAC} applications due to its \ac{STAR} capability, with promising theoretical results demonstrating advantages of joint design~\cite{Mehrotra-Sabharwal:2022}.

{\bf State-of-the-art}: In \cite{Barneto:19,liyanaarachchi2021optimized}, single-antenna \ac{FD} systems were considered for \ac{ISAC} operation, where both communication and radar waveforms were optimized for sensing performance. Frequency-domain radar processing building on the` \ac{LTE} or the 5G \ac{NR} time–frequency resource grid, and complemented with interpolation to account for missing samples due to the null subcarriers within the transmit waveform passband, was presented in \cite{Barneto:19}. It was shown that the large channel bandwidths supported by 5G NR provide good sensing performance already at below $6$ GHz frequencies. With
$100$ MHz channel bandwidth, range estimation accuracy in
the order of $1$ m and target detection probability exceeding
$90\%$ were shown to be feasible at \ac{SNR} values lower than $-30$ dB. To prevent \ac{RX} saturation in mono-static shared-antenna \ac{BS} deployments and reduce the inherent masking effect of the direct leakage sidelobes, efficient analog and digital \ac{SIC} solutions, tailored to the \ac{OFDM} radar use, were introduced in \cite{Barneto:19}. The various provided measurements verified successful sensing of static and moving targets, such as drones and cars, while evidencing measured \ac{TX}/\ac{RX} isolation at the \ac{FD} node of approximately $100$ dB.
\begin{figure}[!t]
	\centerline{\resizebox{0.9\columnwidth}{!}{\includegraphics{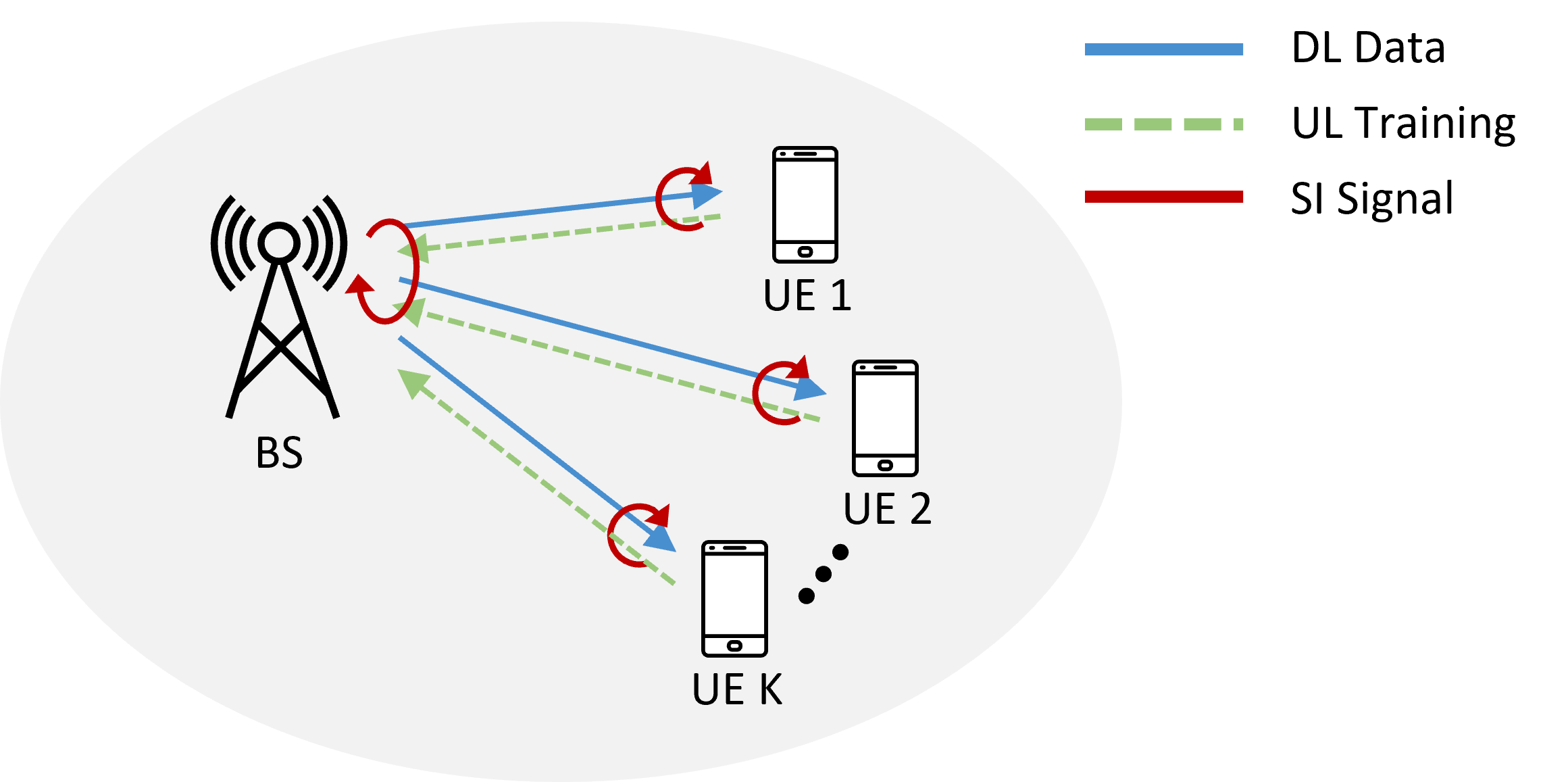}}}
	\caption{\ac{FD}-enabled simultaneous multi-user {DL} data transmission and channel estimation in the \ac{UL}, leveraging \ac{TDD}.}
	\label{Fig.SCDC}
\end{figure}

Apart from the typical optimization of the \ac{UL} and \ac{DL} sum rate, \ac{FD} radios have been considered to continuously update the \ac{CSIT}, which is required to compute the \ac{DL} precoding matrix in multi-user (massive) MIMO systems, as schematically shown in Fig.~\ref{Fig.SCDC}. By design principle, \ac{FD} operation leverages channel reciprocity for open-loop \ac{UL} training to estimate the \ac{DL} channels. However, the \ac{UL} transmission of training might create interference at the \ac{DL} receiving mobile nodes. Most predominantly, the \ac{DL} transmission contaminates the reception of the training symbols in the \ac{UL}. For this channel sounding process based on \ac{FD} systems, the optimal training resource allocation and its associated spectral efficiency was studied in \cite{du2015mu,du2016sequential,mirza2018performance,D2D_fd}. In particular, simultaneous \ac{DL} data transmission and \ac{UL} \ac{CSIR} has been considered for an \ac{FD} \ac{MIMO} \ac{BS} serving multiple \ac{HD} \ac{UE}. The latter were assigned to transmit training symbols through the \ac{UL} channel in a time division multiple access manner, which are utilized by the \ac{BS} to estimate the \ac{DL} channels, while at the same time transmitting the \ac{DL} payload to the user for whom the \ac{CSIT} is already available. The adopted multi-user \ac{FD} system models in \cite{du2015mu,du2016sequential,mirza2018performance} deployed ideal active analog \ac{SIC} that was based on \ac{MIMO} architectures with fully-connected analog \ac{SIC}, interconnecting all \ac{TX} antenna elements in the \ac{FD} node with all its \ac{RX} antennas. Recently, in \cite{Comms-CE2020,MultiuserComms-CE2020}, the reduced complexity acive analog \ac{SIC} of \cite{alexandropoulos2017joint} was adopted to jointly design the \ac{BS}’s transmit and receive BF matrices, as well as the settings for the multiple analog taps and the digital \ac{SI} canceller, with the objective to maximize the \ac{DL} sum rate. It was showcased that the proposed approach outperforms its conventional \ac{HD} counterpart with $50\%$ reduction in the hardware complexity of the analog canceller compared to the schemes in \cite{du2015mu,du2016sequential,mirza2018performance}.

\ac{FD} \ac{MIMO} radios have been also leveraged for efficient low-latency beam management in \ac{mmWave} \ac{MIMO} systems. In particular, \cite{Direction-Aided2020} presented a direction-assisted beam management framework, where the \ac{BS} was equipped with a large antenna array realizing \ac{DL} analog
beamforming and few digitally controlled reception antenna elements used for \ac{UL} estimation of the \ac{DoA} of the \ac{UL} signal from an intended \ac{UE}. A simultaneous \ac{DoA} estimation and data
transmission scheme for boosting beam management was presented, capitalizing on the \ac{FD} \ac{MIMO} architecture of \cite{alexandropoulos2017joint}. Leveraging channel reciprocity, the \ac{UL} dominant \ac{DoA} was estimated at the \ac{FD} \ac{BS}, which then transmits analog beamformed data in the estimated \ac{DL} direction. A joint design of the \ac{DoA}-assisted analog beamformer as well as the active analog and digital \ac{SIC} units was proposed with the objective to maximize the achievable \ac{DL} rate.

\begin{figure}[!t]
	\centerline{\resizebox{0.9\columnwidth}{!}{\includegraphics{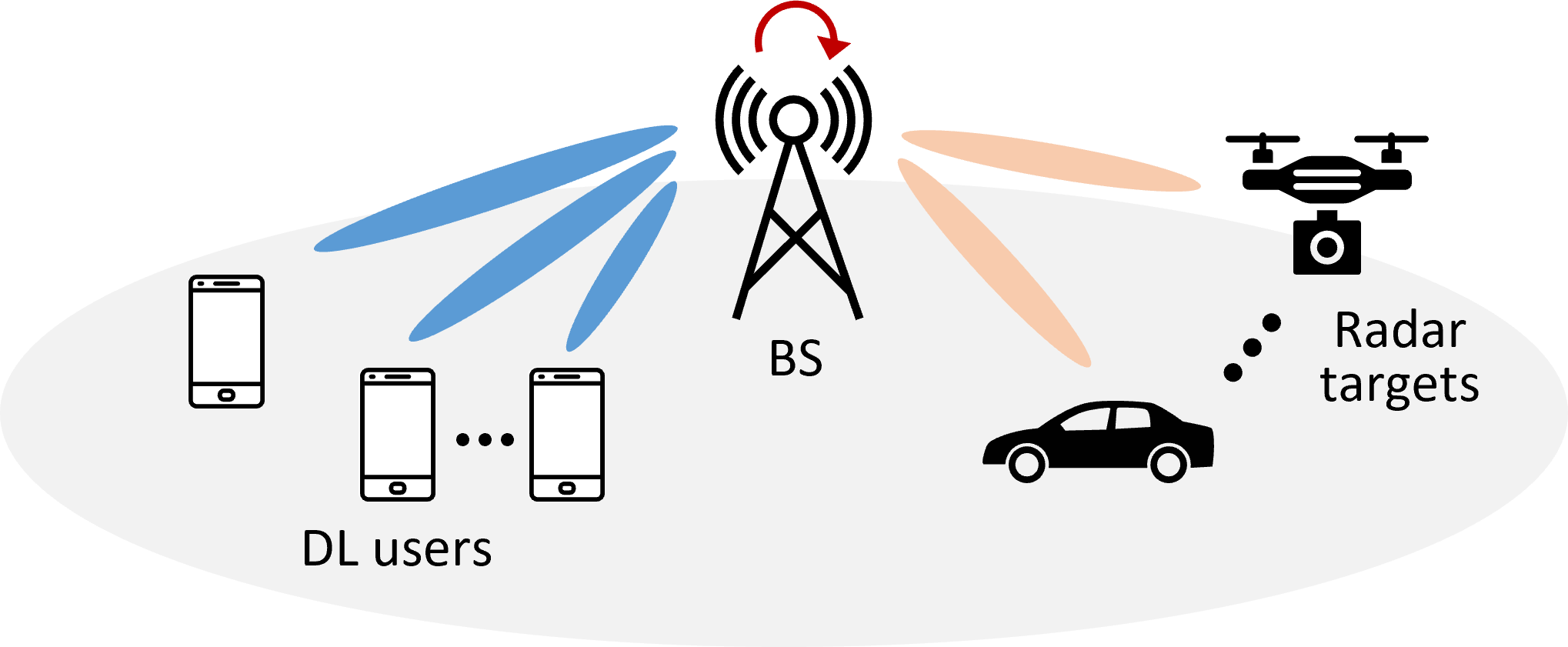}}}
	\caption{\ac{FD}-enabled simultaneous multi-user {DL} data transmission and reception in the \ac{UL} of signals reflected from targets in the propagation environment.}
	\label{Fig.ISAC}
\end{figure}
A multi-beam \ac{mmWave} \ac{FD} \ac{ISAC} system was presented \cite{barneto2020beamforming}, according to which the \ac{TX} and \ac{RX} beamformers were optimized to have multiple beams for both communications and sensing, as schematically presented in Fig.~\ref{Fig.ISAC}. The proposed optimization targeted at suppressing the SI signal arising from the \ac{FD} operation, and the sensing performance was further improved by minimizing the
reflections from the communication beams; the latter was accomplished by implementing
beamforming nulls in both the frequency and angular domains. In addition, an \ac{OFDM} waveform design, through
filling the unused subcarriers with optimized symbols, was designed, which was shown to minimize the delay estimation
error, while also minimizing the peak-to-average power
ratio of the waveform. In \cite{liyanaarachchi2021joint, https://doi.org/10.48550/arxiv.2211.00229}, the \ac{RX} spatial signal was further used to estimate the range and angle profiles corresponding to the targets. Very recently, in \cite{Comms-Target_Tracking2022}, the authors presented an \ac{ISAC} system where a massive \ac{MIMO} \ac{BS} equipped with hybrid analog and digital beamformers is communicating with multiple \ac{DL} users, and simultaneously estimates via the same signaling waveforms the \ac{DoA} as well as the range
of radar targets, which are randomly distributed within its coverage area. A joint radar target tracking and \ac{DL} data transmission protocol was presented together with an optimization framework for the joint design of the massive-antenna beamformers and the \ac{SIC} unit, with the dual objective of
maximizing the radar tracking accuracy and \ac{DL} communication performance. The provided simulation results at \ac{mmWave} frequencies, using 5G \ac{NR} wideband waveforms, showcased the accuracy of the radar target tracking performance of the proposed system, which simultaneously offered increased sum rate compared with benchmark \ac{HD} schemes. A more involved joint optimization framework for designing the analog and digital \ac{TX} and \ac{RX} beamformers, as well as the active analog and digital \ac{SIC} units, was presented in \cite{ISAC2022} with the objective to maximize the achievable \ac{DL} rate and the accuracy performance of the \ac{DoA}, range, as well as the relative velocity estimation of radar targets. The  simulation results, considering 5G \ac{NR} \ac{OFDM} waveforms, showcased high target parameters' estimation accuracy, while maximizing the \ac{DL} communication rate.

{\bf Challenges}: In multi-function networks enabling \ac{ISAC}, several important design aspects need to be considered, namely, the choice of the sensing architecture, the signal design for each functionality, and the resource sharing between the communications and sensing operations.  

\subsubsection{To Mono or Not} In a tightly \ac{ISAC} system, a key design question is whether
sensing should utilize a mono-static or bi-static setup, as illustrated in Fig.~\ref{fig:scenarios}. Mono-static operation together with continuous time radar signals imposes the requirement for sophisticated \ac{FD} operation with adequate \ac{SIC}. In some cases, the \ac{FD} requirement can be avoided by using pulsed radar signals, rather than continuous-wave signals, which use long silent periods of post-transmission to allow for the interference-free reception of the echo signals. Alternatively, when using continuous-wave signals, the known \ac{TX} signal can be used as ``template signal” to detect only the difference between the transmitted and received signals, and estimate the parameters of this ``beat” signal \cite{Zhang:21b}. In a bi-static radar setup, the \ac{FD} requirement is eliminated at the expense of a stringent synchronization problem, that applies to both the timing and frequency offsets between the oscillators of the \ac{TX} and \ac{RX} nodes. In practice, mono-static and bi-static sensing can be advantageously employed \emph{simultaneously} when sensing is integrated in the operation of cellular networks, depending on the duplexing scheme of the network. For example, in 5G \ac{NR} systems operating in \ac{TDD} mode, mobile terminals (i.e., \ac{UE} devices) obtain channel state information by means of synchronization signal blocks transmitted in a \ac{PBCH}, while the \ac{BS} may acquire information about the communication channel by decoding the sounding reference signals (pilots) sent periodically by the mobile terminals. Due to channel reciprocity, the channel state information obtained by the \ac{BS} serves as \ac{CSIR} and \ac{CSIT} facilitating both \ac{UL} data reception and \ac{DL} data transmission, respectively. In this operation, mono-static 
sensing is conveniently integrated in the \ac{DL} frames, while bi-static sensing can be optionally added to utilize the \ac{UL} pilot or data signals to enhance the sensing precision at the \ac{BS}.

\subsubsection{From \ac{SIC} to \ac{SI} Management} 
As discussed above, in mono-static \ac{DL} sensing, the \ac{BS} transmits the same signal to multiple \ac{UE} devices, while having its \ac{RX} active for backscattered 
signals to sense its surroundings. That is, sensing information is completely contained in the backscattered signal. However, in \ac{FD} communications, the backscattered signal is an SI signal that has to be eliminated to improve communications. Thus, \ac{SIC} developed for communications could potentially remove the key signals required in sensing. 
Thus, an open question for \ac{ISAC} networks is how to manage backscattering, i.e., it remains to understand which part of the backscattered signal to suppress and which to admit. 
To enable mono-static \ac{DL} sensing, it is necessary to admit reflections from the surrounding objects and not cancel useful backscattered signals, while ensuring sufficient \ac{TX}/\ac{RX} isolation \cite{Barneto:21, Barneto:19}. 
To enable \ac{ISAC} using mono-static radar, more than $100$ dB of total \ac{SI} suppression is required, which calls for multiple complementary methods, since no single technique can facilitate such high \ac{TX}/\ac{RX} isolation at \ac{FD} devices. Therefore, in addition to basic passive analog \ac{SIC} and radar-domain digital suppression methods, efﬁcient active analog and digital \ac{SIC} methods are needed, particularly from the static and slow-moving targets' viewpoint, as well as overall, to prevent \ac{RX} saturation \cite{Barneto:19}.
To this end, active analog and time-domain digital \ac{SIC} methods are important. Compared to the existing in-band \ac{FD} radio research, 
a specific radar-related aspect is that only the direct \ac{SI} should be canceled or suppressed, along with possible reﬂections from very close surfaces, while the echoes 
from the targets of interest must be properly detected. This aspect is an important difference to \ac{FD} radio works developed solely for communication networks, 
such as that in \cite{Korpi:17}, which does not distinguish between the direct and reﬂected \ac{SI} components. In addition, agnostic digital \ac{SIC} techniques, independent of the speciﬁc radar processing approach, are desirable, which can then be 
complemented with \ac{SIC} methods operating 
in the radar domain \cite{Mercier:19}.

\subsection{Integrated Access and Backhaul (IAB)} 
\begin{figure}[!t]
	\centerline{\resizebox{1\columnwidth}{!}{\includegraphics{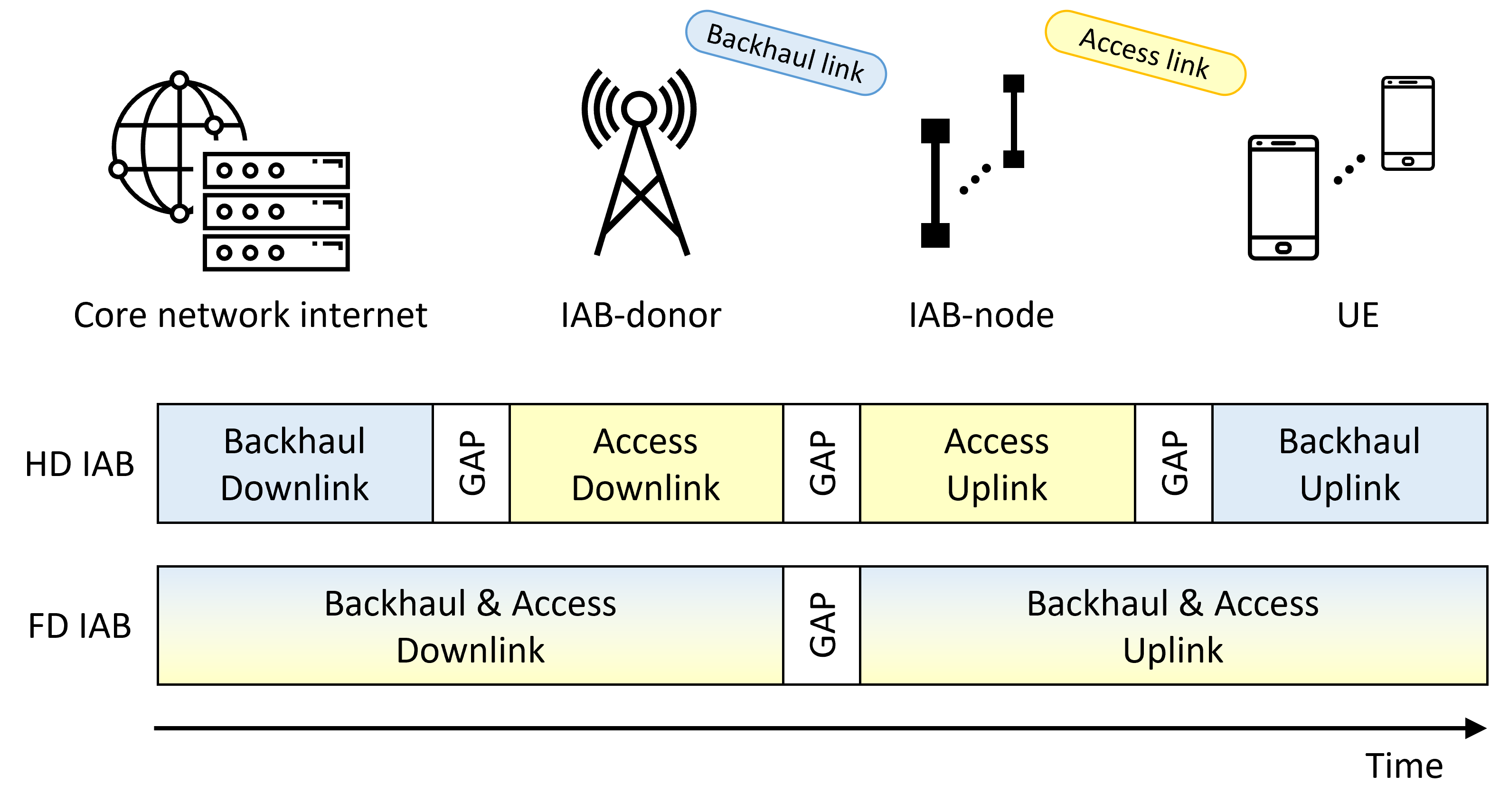}}}
        \caption{ A comparison of an operation protocol for \ac{IAB} based on \ac{HD} and \ac{FD} configurations.}
	\label{Fig.IAB}
\end{figure}
{\bf Opportunity}:
Future cellular networks are expected to be highly dense to support extended coverage and capacity expansion. However, traditional fiber-backhauling can be expensive. To address this, a new technology called Integrated Access and Backhaul (IAB) has been developed, which integrates access and backhaul functions using wireless links. IAB was included in the 3GPP NR Release 16, and the work on IAB was finalized in 2020~\cite{3GPP2018SI}.

{\bf State-of-the-art}:
A typical \ac{IAB} architecture, where only some of the gNBs (generation Node B) are connected to the traditional wired infrastructures (core network internet) via fiber as IAB-donors, while the rest of them, termed as \ac{IAB}-nodes, decode and forward the backhaul traffic wirelessly, is illustrated in Fig.~\ref{Fig.IAB}. It is noted that an gNB \ac{BS} provides NR protocol terminations to \ac{UE}. As defined in 3GPP technical specification $38.401$ \cite{3GPP38401}, the gNB is a logical node, which may be split into one central unit (CU) and one or more distributed units (DU). The CU hosts the higher layer protocols to the UE and terminates the control and user plane interfaces to the core network. The CU controls the DU nodes over the F1 interface(s), where the DU node hosts the lower layers for the NR Uu interface to the UE. The conventional scenario in Fig.~\ref{Fig.IAB}, which we refer to as \ac{HD} \ac{IAB}, involves an in-band system in which the backhaul and access links share the same frequency spectrum under the typical \ac{HD} constraints. In other words, the access and backhaul links must use the given radio resources in an orthogonal manner, be it time or frequency, to avoid collision. Thus, \ac{HD}-based \ac{IAB} clearly fails to exploit the full potential of the given radio resources. 

In contrast, researchers have proposed a smarter IAB framework, referred to as FD IAB, that eliminates the HD constraint and enables simultaneous transmission and reception of backhaul and access links. With advanced FD capability and proper SI mitigation, FD IAB can achieve a spectral efficiency more than double that of the conventional HD IAB scenario, as illustrated in Fig.~\ref{Fig.IAB}. Additionally, FD technology can reduce end-to-end and feedback delays and provide more degrees-of-freedom in wireless protocol design~\cite{kim2015survey}.

{\bf Challenges}: As previously discussed, to realize \ac{FD} \ac{IAB} in practice, and then include into the future \ac{3GPP} releases, it is necessary to verify the performance of the system.

\subsubsection{Link-Level Simulations}
Although FD IAB has a promising future, SI mitigation is a crucial challenge. Since SI increases with TX power, deploying IAB-nodes with high TX power infrastructures requires state-of-the-art SI reduction performance. In a macrocell scenario, the transmit power at the IAB-DU can be as high as SI 46 dBm, which means that more than 120 dB of aggregate SI reduction is necessary to reduce the SI close to the RX noise floor. This is a significant technical challenge that needs to be addressed for the successful deployment of FD IAB.

The strict performance requirements have led some to question the feasibility of implementing \ac{FD} in infrastructure systems. Nevertheless, exploiting highly directional beams in the millimeter wave band can lead to significant suppression of \ac{SI} in the propagation domain, making \ac{FD} feasible. In the subsection that follows, we provide link-level performance evaluations based on a realistic configuration of an IAB system.

\begin{figure}[!t]
	\centerline{\resizebox{1\columnwidth}{!}{\includegraphics{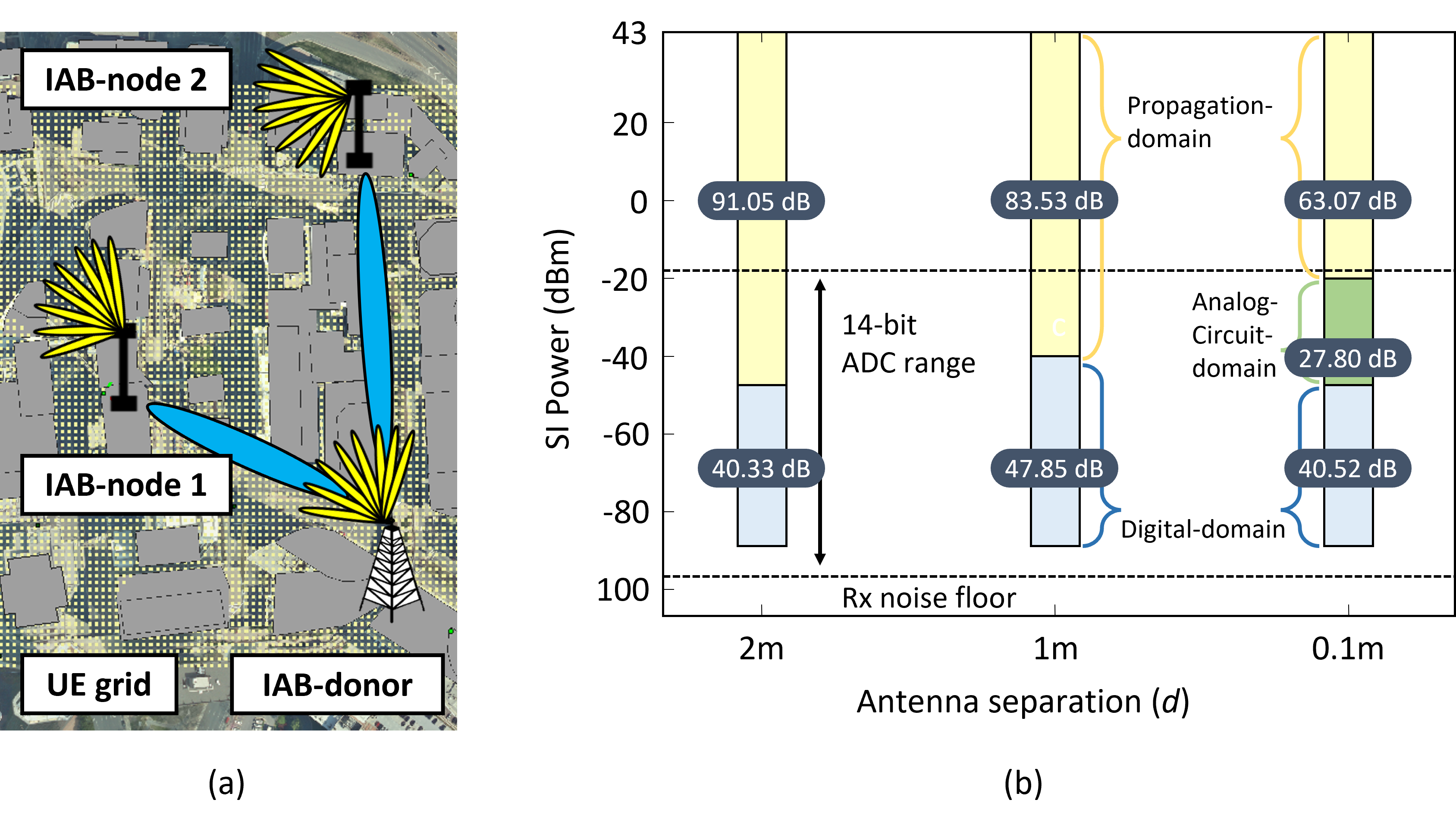}}}
	\caption{Link-level simulations for the \ac{SI} mitigation performance considering different antenna separations $d$ and \ac{SIC} mechanisms: (a) the virtual 3D environment of the downtown area in Rosslyn City, Virginia, USA; and (b) The \ac{SI} power in dB as a function of $d$ in meters.}
	\label{Fig.SIClink}
\end{figure}

Fig.~\ref{Fig.SIClink} presents the results of link-level simulations conducted in a virtual 3D environment modeling the downtown area of Rosslyn City, Virginia, USA, using realistic settings. The simulation considers an in-band \ac{FD} \ac{IAB} scenario, with physically fixed relays equipped with a narrow beam codebook and an \ac{RF} chain at both the DU and mobile termination (MT)\footnote{It should be noted that the backhaul links are terminated by an IAB-MT function \cite{3GPP38401}}. The IAB-MT could either use a separate antenna or share the access antenna of the \ac{BS} (virtual IAB-MT). The latter provides the ultimate level of integration, as well as utilizing the high-performance \ac{BS} antennas for backhaul over longer distances, operating over a 120 MHz bandwidth channel at the central frequency of $28$ GHz is considered.


The setting includes a \ac{IAB}-donor and two \ac{IAB}-nodes, all of which have \ac{FD} capabilities, to serve multiple UEs. Once the access link decisions are made, such as UE scheduling and beam alignment, the residual \ac{SI} channel after the propagation-domain \ac{SI} suppression can be calculated. The link-level simulation uses single-stream OFDM and includes a pilot-based, channel-aware, two-tap canceller circuit for analog-circuit-domain \ac{SIC}, as well as nonlinear digital \ac{SIC} with the fifth-order parallel Hammerstein model~\cite{kwack,kwack_j}.

Fig.~\ref{Fig.SIClink} illustrates the average reduction of \ac{SI} and its components across all \ac{FD} transmission scenarios with three distinct antenna separations. It is evident that the total average \ac{SI} suppression is adequate in all cases to decrease the residual \ac{SI} to a level that is similar to that of the \ac{RX} noise floor. For antenna separations of $d=1$ and $2$ meters, suppressing \ac{SI} in the propagation-domain alone was sufficient to achieve the effective \ac{ADC} range, which corresponds to $72.24$ dB for a $14$-bit resolution \ac{ADC} as reported in \cite{Sabharwal2014IBFD}, above the \ac{RX} noise floor. Thus, in these cases, using analog active-domain \ac{SIC} was unnecessary, leading to cost-effective systems. However, for the $d=0.1$ meter case, although the average suppression of \ac{SI} in the propagation-domain was adequate to stay within the effective \ac{ADC} range above the \ac{RX} noise floor, there were specific instances where the suppression was insufficient, causing significant performance degradation in the digital-domain \ac{SIC}. As a result, active analog \ac{SIC} was utilized to ensure that the residual \ac{SI} falls within the ADC range. Generally, the impact of antenna separation and analog beamforming in the propagation-domain is substantial in the considered \ac{mmWave} frequency band, accounting for a significant proportion of the total \ac{SI} reduction.

\begin{figure}[!t]
	\centerline{\resizebox{1\columnwidth}{!}{\includegraphics{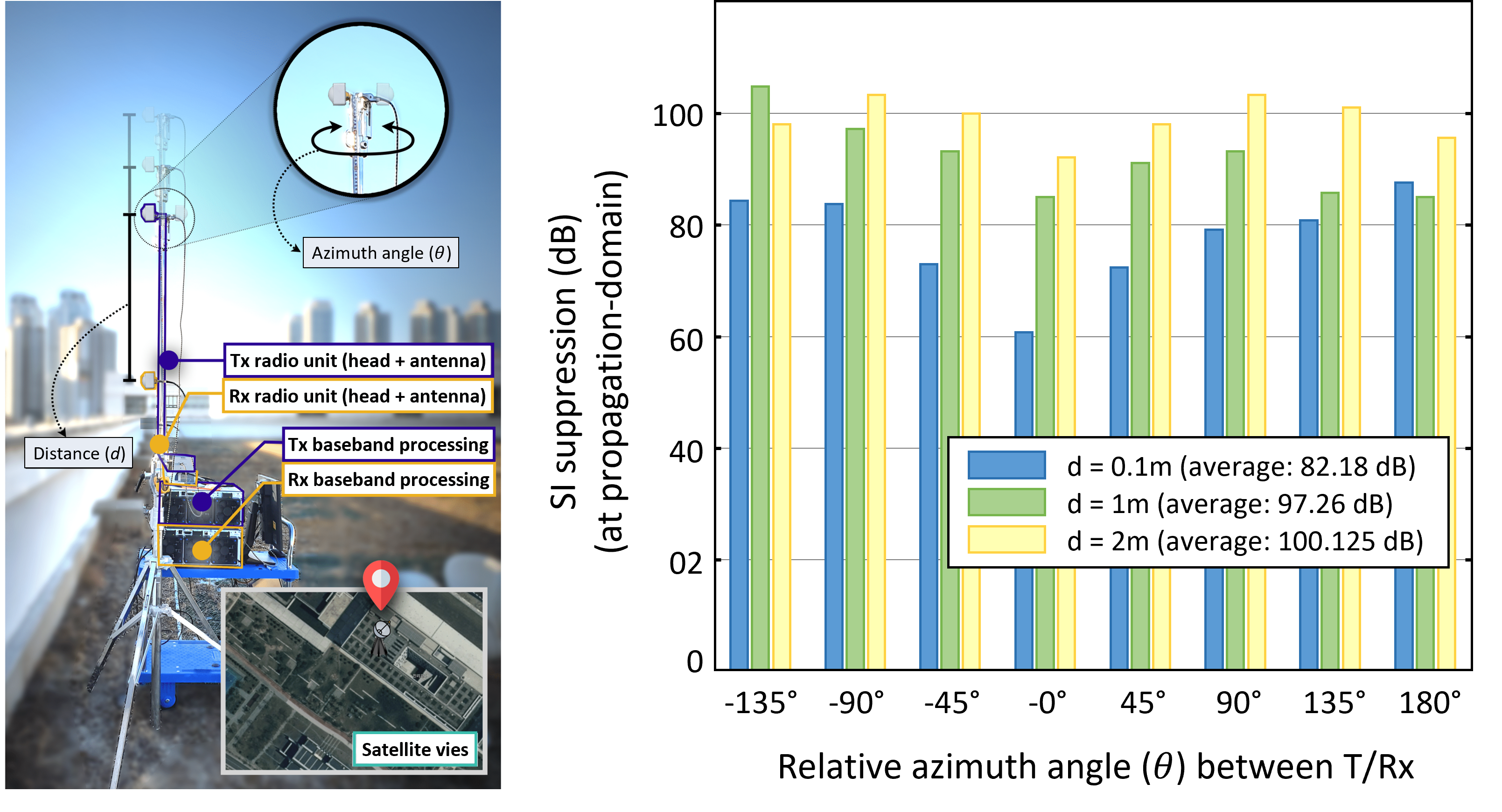}}}
	\caption{The set-up of the $28$~GHz real-time \ac{FD}-based \ac{IAB} hardware prototype in the considered outdoor scenario and its satellite view (left); and the measured SI suppression via propagation-domain passive analog \ac{SIC} in dB versus the relative azimuth angle ($\theta$) between TX and RX (right)~\cite{suk_IAB}.}	
	\label{Fig.prototype}
\end{figure}

\subsubsection{Prototyping at $28$ GHz}

Previous practical observations have shown that a significant part of the total suppression of \ac{SI} is due to passive analog \ac{SIC} in the propagation domain. Building on that practical evidence, researchers implemented a hardware prototype for an in-band \ac{FD}-based \ac{IAB} system at $28$ GHz in \cite{cho2018rf,park2020wcnc,park_squint}, and measured the propagation-domain \ac{SIC} with an $800$~MHz bandwidth. This bandwidth is twice the maximum channel bandwidth of the \ac{mmWave} band in 5G NR. The prototype was developed using the LabVIEW\texttrademark{} system design software and a PXIe software-defined radio platform based on a field-programmable-gate-array (FPGA). The measurement campaign was carried out on the rooftop of a 4-floor building, as shown in the left subfigure of Fig.~\ref{Fig.prototype}.


The goal of the prototype was to replicate an \ac{FD} \ac{IAB} node in a suburban \ac{DL} setting, where the DU and MT were positioned at a certain distance apart and had different orientations relative to one another. The \ac{RX} beneath the \ac{TX} represented the MT of an \ac{IAB} node, while the \ac{TX} acted as the DU. The \ac{SI} in this scenario was the signal on the access link that was not intended for the MT to receive. Nevertheless, we established a synchronized communication link between the DU and MT to measure the \ac{SI} as the received signal strength. While the experimentation setup provided a useful reference for evaluating the \ac{SI} suppression performance of \ac{IAB} by simulating an \ac{IAB}-node and its environment, improvements in the equipment's form factor and the surrounding reflectors have yet to be made. These enhancements would lead to a more refined and stable \ac{SI} mitigation performance. In the designed prototype, both the \ac{TX} and \ac{RX} were equipped with a single antenna and a corresponding \ac{RF} chain. To generate a sharp analog beam at the \ac{RX} \ac{RF} front-end, advanced lens antennas with a gain of $19.86$~dBi and an approximate 3 dB beamwidth of 13.4$^{\circ}$ were used. At the \ac{TX} end, an external \ac{PA} with 35 dB gain and a compression point of $30$ dBm P1dB was employed.

The measured \ac{SI} suppression via propagation-domain passive analog \ac{SIC} in dB as a function of the relative azimuth angle $\theta$ between the \ac{TX} and \ac{RX} antennas is illustrated in Fig.~\ref{Fig.prototype}(c) for various antenna separation distances $d$. The results show that the average suppression for $d=2$, $1$, and $0.1$ meters were $100.125$, $97.26$, and $82.18$ dB, respectively. These values were influenced by the antenna directionality and path loss.
This demonstrates that \ac{FD}-based \ac{IAB} systems operating at the \ac{mmWave} frequency band can effectively achieve a significant amount of \ac{SI} suppression, particularly when compared to the results obtained at the $2.4$ GHz band, as reported in~\cite{EEverett2014Passive,kim2015survey,suk2022}. The aforementioned study demonstrated a \ac{SI} suppression of $73.8$ dB for a separation distance of $0.5$ meters, utilizing absorbing shielding and cross-polarized antennas. In contrast, the $28$ GHz experiment showed an \ac{SI} suppression of $82.18$ {dB} for a separation distance of $0.1$ meters, highlighting the importance of using directional antennas~\cite{23_TAnt} (without absorbing shielding) that offer a sharp radiation pattern and high path loss.

\subsection{Full-Duplex in 5G-Advanced Systems \label{subsec:FD_in_5G}}

{\bf Opportunity}:
5G networks are primarily designed to operate in \ac{TDD} mode, which offers flexibility in resource allocation and can be deployed in unpaired spectrum bands. Since \ac{DL} traffic load is often much higher than the \ac{UL} traffic load, typical \ac{TDD} schedules assign more time slots to the former, for example, $3:1$ or $4:1$ \ac{DL}/\ac{UL} pattern. While this asymmetry is favorable to meet the \ac{UL}/\ac{DL} capacity demands, it may compromise the \ac{UL} performance in terms of latency, which may not be acceptable for applications requiring ultra-reliable low-latency communication (URLLC) links \cite{Guo:22}. Therefore, for \ac{FD}, there is an opportunity to ensure the required flexibility in terms of \ac{DL}/\ac{UL} traffic demands, while meeting latency requirements.

{\bf State-of-the-art}:
Recognizing that in multicell systems, different cells may benefit from differing \ac{DL}/\ac{UL} patterns, modern cellular networks support dynamic \ac{TDD}, which allows to dynamically configure of the \ac{DL}/\ac{UL} patterns. Although dynamic \ac{TDD} is attractive, since it can adjust the duplexing pattern to the prevailing
traffic load conditions in different cells, it suffers from \ac{CLI} \cite{Kim:2020, TR-38828}. Due to \ac{CLI}, which may severely degrade the \ac{UL} performance due to \ac{BS}-to-\ac{BS} interference, dynamic \ac{TDD} has so far not been successful in practice. However, as effective \ac{CLI} mitigation schemes emerge, dynamic \ac{TDD} systems may gain greater adoption in the future \cite{Silva:21, Clowdhury:2022, Razlighi:2020}.   

Therefore, meeting the throughput, capacity, reliability, and latency requirements 
motivates the adoption of \ac{FD} schemes in evolving 5G systems, since they allow to simultaneously serve \ac{UL} and \ac{DL} traffic irrespective of the traffic pattern and without imposing duplexing delay on the \ac{UL} traffic. However, due to the formidable hardware challenges of implementing high-power \ac{FD} radios in a cost-efficient fashion, the research and engineering communities have recently proposed 
the adoption of \ac{XDD} or subband \ac{FD}, as an intermediary step from dynamic \ac{TDD} systems towards true \ac{FD} communications \cite{Ji:21, Silva:21}. 
\begin{figure}[!t]
\begin{center}
\includegraphics[width=\hsize]{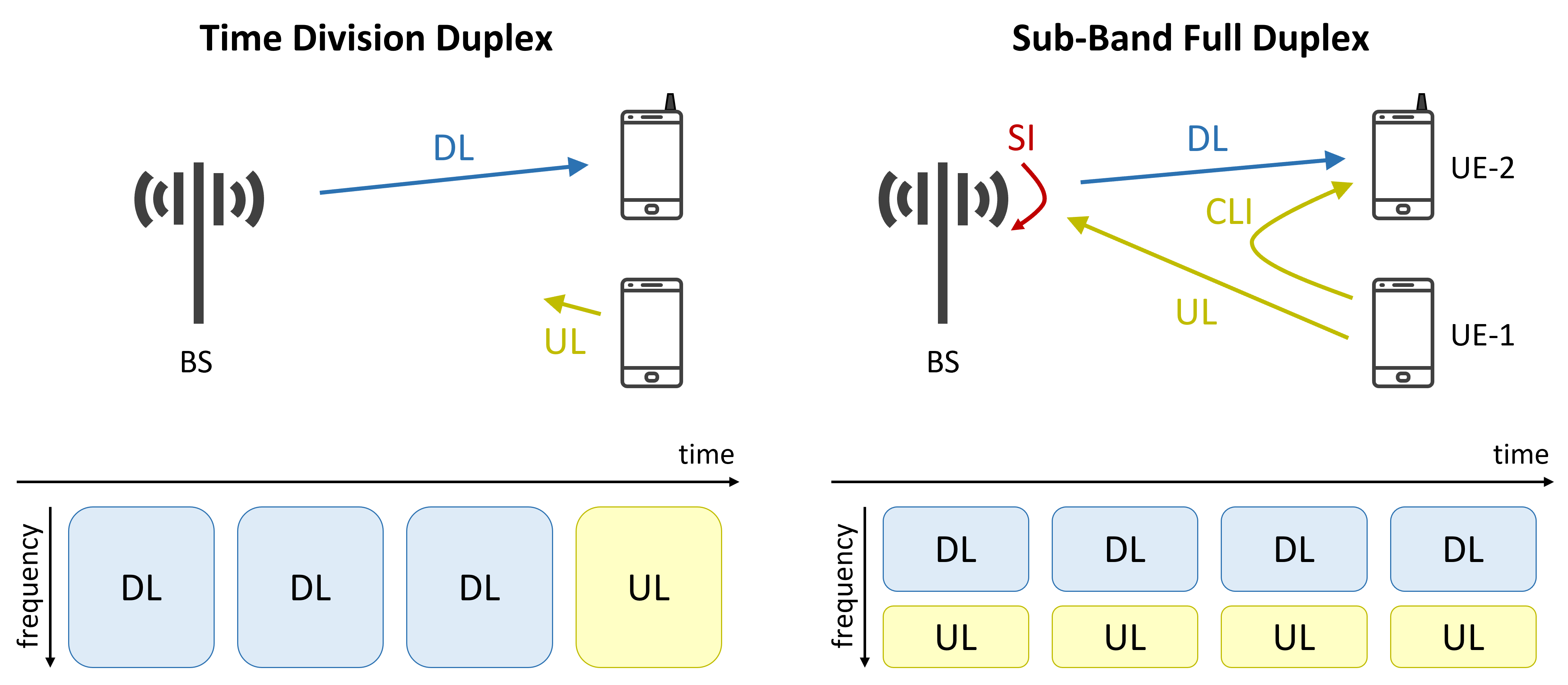}
\caption{
Comparing traditional \ac{TDD} transmissions with those employed in \ac{SBFD} -- also referred to as flexible duplex, \ac{XDD}, systems. Notice that in \ac{SBFD}, two separate \ac{TDD} schedules operate in adjacent frequency chunks, which gives rise to \ac{SI} at the \ac{BS} and cross-link interference between the \ac{UE} devices.
}
\label{fig:1}
\end{center}
\end{figure}

In subband \ac{FD}, each subband within the TDD carrier implements a specific \ac{DL}/\ac{UL} ratio, enabling the cellular \ac{BS} to operate \ac{UL} and \ac{DL} on the same carrier but on different frequency resources within the subband, see Fig. \ref{fig:1} \cite{R4-2212117}. Notice in Fig. \ref{fig:1} that even though separate chunks of the frequency band are allocated to \ac{DL} and \ac{UL}, the \ac{BS} simultaneously schedules \ac{UL} data from \ac{UE}-1 and \ac{DL} data to \ac{UE}-2, and thereby implements a (virtual) 3-node \ac{FD} schedule \cite{Mairton:21}.

{\bf Challenges}:
As discussed, with \ac{XDD}, two separate \ac{TDD} schedules operate in adjacent frequency chunks, which gives rise to two problems. 

The first problem (crosstalk) is caused by the -- typically high-power -- \ac{DL} \ac{TX} signals operating in close frequency channels to the those used by the \ac{UL} \ac{RX} channels, which may saturate the \ac{LNA} in the receiver chains \cite{R4-2212117}. 
The second problem is due to the \ac{ACL} power injected by the \ac{TX} signals into the \ac{UL} channels, which may desensitize (block) the receiver amplifier \cite{Rostomyan:19}. For cellular \acp{BS} employing advanced antenna systems, these problems can be to some extent mitigated by smart antenna assignments (referred to as smart antenna splitting) that assign some antenna elements to either \ac{DL} or \ac{UL} traffic, while the remaining antenna elements may act as integrated \ac{DL}/\ac{UL} antennas \cite{Mairton:21}. However, even with careful antenna splitting and antenna isolation techniques, additional isolation enhancements are necessary to reach the at least 100 dB of total isolation of the \ac{TX} and \ac{RX} signals. We list some promising techniques that are applicable in advanced antenna and massive \ac{MIMO} systems:

\subsubsection{Beam Nulling}
Multi-antenna algorithms aim to ensure high \ac{SINR} at the intended \ac{UE} devices using beamforming, while creating minimum interference to other \acp{UE} by creating nulls in the non-intended directions. In a massive \ac{MIMO} radio units that support subband or true \ac{FD} links, additional isolation between the \ac{TX} and \ac{RX} antennas may be achieved 
by creating nulls on the transmit side towards the co-located \ac{RX} antennas \cite{Hien:2017, Gowda:18, Mairton:21}. Such spatial domain interference suppression using beamforming and beam nulling algorithms may degrade the performance of the beamforming algorithm towards the intended \ac{UE} in terms of the achieved \ac{SINR}, since some degrees-of-freedom are used to form nulls towards the co-located \ac{RX} antennas \cite{R4-2212117}.

\subsubsection{Digital Cancellation and Linearization}
As mentioned, subband \ac{FD} requires to mitigate the impacts of both crosstalk and \ac{ACL}, while true \ac{FD} requires to minimize the \ac{SI}. In both cases, digital cancellation can increase the \ac{TX}/\ac{RX} isolation. For digital cancellation, a sufficient \ac{ADC}/\ac{DAC} dynamic range and resolution are required to capture both the \ac{SI} and the desired signals, with enough resolution to cancel the injected \ac{TX} signal into the \ac{RX} baseband \cite{Katanbaf:19}. In practice, the achievable digital cancellation is affected by hardware impairments and limitations, including nonlinearities, limited dynamic range and resolution of the \ac{ADC}/\ac{DAC} circuits, phase noise and multipath environmental reflections.

\subsubsection{\ac{RF} Cancellation}
\ac{RF} cancellation creates an intentional and controllable leakage path between the
\ac{TX} and \ac{RX} chains to provide cancellation at the analog stage of the receiver. 
It typically  provides much higher dynamic range path compared to digital cancellation, making it suitable for minimizing \ac{ACL}. However, the main challenge of \ac{RF} cancellation is to manage the associated hardware complexity due to the large number of \ac{TX} and \ac{RX} antennas \cite{Duarte:20,R4-2212117}. 

Similarly to the challenges faced by dynamic \ac{TDD} systems, a major challenge for introducing \ac{FD} technology in evolving 5G systems is the intra-cell and inter-cell \ac{CLI} \cite{Silva:21,R1-2207461}. As pointed out in \cite{Silva:21}, dynamic \ac{TDD} as well as \ac{FD} cellular networks suffer from two kinds of interference, which are not present in static \ac{TDD} and \ac{FDD} systems. The \ac{BS}-to-\ac{BS} interference is caused by a transmitting \ac{BS} to surrounding neighbour \acp{BS} that are currently receiving due to operating either in \ac{FD} or dynamic \ac{TDD} mode~\cite{Goyal:2015, Psomas:2017}. The \ac{UE}-to-{UE} interference is caused by a transmitting \ac{UE} -- either in a dynamic \ac{TDD} \ac{UL} slot or as part of an \ac{FD} schedule -- to a receiving {UE} either
in a neighboring dynamic \ac{TDD} cell or in the same \ac{FD} cell \cite{Mohammadi:TCOM2015}.  
These types of interference sources are referred to as intracell or intercell \ac{CLI} \cite{R1-2207461}. 
In practice, the \ac{BS}-to-\ac{BS} adjacent-channel \ac{CLI}, that is, \ac{CLI} caused by a \ac{DL} transmission of a \ac{BS} in a frequency channel to an \ac{UL} reception of a neighbouring \ac{BS} in an adjacent frequency channel also needs to be taken into account. This type of \ac{CLI} includes the adjacent-channel \ac{CLI} between \acp{BS} in the same and different sectors of the same or closely located sites. Finally, adjacent-channel \ac{CLI} also causes \ac{UE}-to-\ac{UE} interference, since an \ac{UL} transmission of a \ac{UE} may leak into the \ac{DL} reception of a \ac{DL} \ac{UE} operating in an adjacent frequency channel. To harvest the benefits of \ac{FD} systems, these additional interference sources need to be mitigated.  

As shown in \cite{Silva:21,R1-2207461,Goyal:2015}, for \ac{FD} networks to become feasible in multicell cellular networks, the \ac{SI}, \ac{CLI}, as well as the conventional intercell interference, must be properly mitigated, which is a difficult requirement in macro networks employing high-power \acp{BS}. 

Since combating \ac{CLI} is vital in both dynamic \ac{TDD} as well as in \ac{FD} multicell systems, recent research and standardization efforts proposed practical solutions to mitigate \ac{CLI} \cite{Huang:2019,Rachad:2020,Al-Saadeh:2017,Koh:2018,Wang:2019,Silva:21}.  
Specifically, a \ac{CLI} mitigation scheme utilizing a massive number of antennas and \ac{ZF}, termed \emph{BSint} was proposed in \cite{Silva:21}. This technique utilizes channel state information between neighbor \acp{BS} and additional information exchange among \acp{BS} through the standardized X2 interface. Building on these pieces of information, transmitting \acp{BS} are able to beamform away from neighbors, while using some of their spatial-degrees-of freedom to suppress interference that leaks into the useful signal. Using the \emph{BSint} technique, \acp{BS} can accurately cancel the BS-to-BS interference by using some of their spatial-degrees-of freedom to combat \ac{CLI} rather than scheduling multiple \acp{UE} in the spatial domain \cite{Silva:21}. It is important to notice that this scheme is applicable in both dynamic \ac{TDD} and \ac{FD} systems. 

\subsection{Low-Latency Networking}
{\bf Opportunity}:
Low-latency and highly reliable communications are increasingly important in private and public wireless networks that are deployed to provide mission critical and highly monetizable services, such as factory automation, advanced driver assistance, telemedicine, autonomous train control services and online gaming~\cite{Fodor:21}. Accordingly, one of the goals of 5G NR  is to achieve a consistent latency of 1 ms while \ac{6G} will further unlock services that demand extremely low latency with additional requirements such as improved spectral and energy efficiency, security and low age-of-information. Moreover, to enhance coverage and capacity, wireless backhauling and multi-hop relaying are emerging as natural components of contemporary and future wireless networks \cite{Zhang:21c,Malayter:22,Papadogiannis2012,Basak_Can_2007}. Cooperative schemes are also attractive for enhancing the coverage and capacity of 
random access \ac{MAC} protocols \cite{Vaezi:20}. However, cooperative schemes employing relaying nodes~\cite{relayselection2010,Alexandropoulos2011,HoVan_relay_selection,Peppas2013,Miridakis2016,Fozooni2019} in both scheduled and random access \ac{MAC} protocols may increase the end-to-end latency due to the need for processing at the relaying node. It has been observed in several recent works that \ac{FD}-capable relaying nodes may reduce the latency compared with their \ac{HD} counterparts.

{\bf State-of-the-art:}
Several academic and standardization contributions proposed \ac{FD} transmission schemes that reduce the end-to-end delay, while maintaining high spectrum utilization and simplifying the \ac{MAC} design \cite{Kwon:12, Zhang:16, FD_user_scheduling_2016, FD_UDN_2016, Gu:18, Airod:18, Malayter:22, Yuan:22, Vaezi:20}. As previously discussed, \ac{3GPP} proposed cost-effective dense deployments of wireless backhauling using \ac{IAB} nodes to enhance coverage and capacity with rather limited capital expenditures \cite{Zhang:21c}, as illustrated in Fig.~\ref{Fig.IAB}. In fact, \ac{FD}-based \ac{IAB} nodes can significantly reduce the delay of packet delivery in multi-hop deployments compared with \ac{HD} store-and-forward relay nodes \cite{Zhang:16, Gu:18}.

{\bf Challenges:}
As showcased via simulations in \cite{Kwon:12,FD_user_scheduling_2016, FD_UDN_2016}, \ac{FD} relays experience interference problems that can limit the relaying gains in the achievable rate. In addition, the \ac{FD} mode of operation imposes higher processing power at the \ac{IAB} node, since it has to process twice the number of packets per unit of time due to \ac{STAR}. An additional challenge for multi-antenna \ac{FD} \ac{IAB} nodes is to manage the trade-off between suppressing \ac{SI} caused by the \ac{TX} panel to the \ac{RX} antenna elements and the dynamic beamforming towards the intended \ac{RX} node, such as the served \ac{UE} device in the \ac{DL} direction \cite{Malayter:22}.   

As new use cases emerge, future network operations are set to take advantage of innovations made in the areas of computation, storage, and communications to improve user experience \cite{Hakkeon:22, Wen:22}. Additionally, features, such as wider bandwidth channels (e.g. in mmWave and sub-THz frequency bands) and strict quality-of-service requirements of applications, continue to demand reduced latency as well as distributed resources for computation and storage. \ac{MEC} has a significant role in 5G \ac{NR}, and is expected to do the same in 5G-Advanced and \ac{6G} networks, in terms of making use of cloud networking infrastructure to offload some of the user computational tasks to edge servers. However, as the number of \acp{UE} getting connected using limited spectrum resources and data generated increases, it becomes challenging for \ac{MEC} to guarantee low latency. Moreover, today's \ac{MEC} architectures are largely ineffective to cater emerging high-throughput and low-latency demands originating from applications such as mixed and extended reality, aiming to combine physical and virtual worlds with haptic interactions. Such applications are expected to gain from \ac{FD}, involving offloading and downloading made possible due to simultaneous \ac{UL} and \ac{DL} transmissions thus significantly reducing the associated latency. 

Connected and autonomous driving of the future is another example in which vehicle-to-vehicle communication would benefit from the low-latency capability of \ac{FD} radios. In those scenarios, the \ac{FD} mode of operation would enable communication among vehicles through simultaneous sensing and transmission, and without relying on cellular coverage, at low latency to ensure traffic safety \cite{Yuan_V2V:2020}. It is also argued in \cite{Campolo:2018} that transceivers capable of \ac{FD} would merge well with vehicular onboard units without causing power and processing issues. In addition, 5G-Advanced and beyond systems are expected to seamlessly integrate terrestrial, underwater, and air networks to deliver ubiquitous coverage and connect remote areas of the world. In this regard, ``connectivity from the sky'' is an innovative trend that would connect aerial vehicles such as drones, low- and high-altitude platforms, and satellites with terrestrial segments, bringing new communication opportunities and services. These aerial networks are characterized by frequently varying topologies and very long transmission distances, causing severe communication delays and, in certain cases, outages. To this end, \ac{FD} radios promise to reduce the involved high latency and improve the network throughput \cite{LZhang:2018,Quynh:2022}. However, under the \ac{FD} operation, long-distance satellite communications introduce a power imbalance between the \ac{TX} and \ac{RX} signals, rendering \ac{SIC} a significant challenge.   

\subsection{Interplay with Reconfigurable Intelligent Surfaces (RISs)}
{\bf Opportunity:}
A \ac{RIS} is a two-dimensional metasurface comprising metamaterials, whose responses on impinging electromagnetic signals can be programmed to offer several desired functionalities, like signal scattering, generalized reflection, and absorption \cite{huang2019reconfigurable,Marco2019}. The \ac{RIS}, commonly discrete, unit elements are usually implemented with basic and ultra-low-power integrated electronic circuits, which are orchestrated in real-time by a dedicated controller. This controller is usually a simple printed circuit board that manages the metasurface's reconfiguration and is responsible for its interface with the rest of the wireless network \cite{RIS_fundamentals}. It is envisioned that a \ac{RIS} can be used to shape the wireless environment for smart and sustainable connectivity~\cite{WavePropTCCN, Qingqing:2021}. Note that, up to date, wireless environments are not programmable, in the sense that, the propagation of electromagnetic signals is determined by the geometry of the wireless channel, which serves as an uncontrollable factor in a given deployment scenario. 

The operation of a \ac{RIS} closely mimics an amplify-and-forward \ac{FD} relay \cite{huang2019reconfigurable,Shaikh_2021}. Information-bearing electromagnetic fields are impinging upon an almost passive \ac{RIS} and get reflected with programmable phase shifts in order to reach the desired end-user location with minimal \ac{SI}. In such metasurfaces, signal amplification is not feasible, however, there are lately appearing various hardware architectures of \acp{RIS} with active elements \cite{Tsinghua_RIS_Tutorial}, enabling signal reception and sensing at the \ac{RIS} side \cite{hardware2020icassp,HRIS_Mag,HRIS_Nature}, as well as reflection amplification \cite{amplifying_RIS_2022}. The former category of architectures facilitates the \ac{RIS} reconfiguration/optimization with reduced overhead, while the latter confronts the inevitable dual path loss effect due to the dual-hop signal propagation. However, in general, a \ac{RIS} is not expected to be as costly, power-hungry, and computation-capable device as an amplify-and-forward relay, hence, it is envisioned to consist of hundreds of metamaterials. In this way, the RIS technology can become competitive with the relay and/or smart repeater deployments.
Very recently, relay-aided \ac{RIS} deployments have been proposed to exploit the advantages of both propagation extension technologies in order to offer increased network coverage \cite{yildirim_hybrid_2021}.    

It is already well known that simultaneously operated and overlapping channels in frequency, resulting from the \ac{FD} operation, can escalate both the inter- and intra-cell interference in cellular networks~\cite{Goyal:2015,FD_user_scheduling_2016,FD_UDN_2016,Psomas:2017}. In certain cellular applications, \ac{FD} \ac{BS}s and \ac{RIS} can be deployed in combination to improve the overall system performance. In such deployments, the \acp{RIS} can help to either enhance the received signals or to control interference at the \ac{UE}s. To this end, properly designed reflective beams to a particular direction through \ac{RIS} phase optimization \cite{Mustafa_ArbitraryBeam_6G2022} would be an effective solution to lower the number of \ac{UE}s affected by interfering signals.

{\bf State-of-the-art:}
Optimizing the \ac{RIS} placement and orientation has been shown to deliver higher gains \cite{RIS_placement}, while switching on/off the \ac{RIS} is effective to combat blockage-induced effects with high energy efficiency. Moreover, in some applications, such as wireless power transfer, sensor-like devices can harvest energy not only from power beacons but also from reflected signals by the \acp{RIS}~\cite{MengHua:2022}. Leveraging the high-aperture efficiency offered by metasurfaces with fine-grained adjustable phase profiles, passive beamforming gains can be achieved to further deal with the path loss-induced performance degradation!\cite{Chou:2021}, and harvesting a significant amount of energy to supplement the energy requirements of wireless devices stand out as a near-term possibility. The combination of \ac{FD} nodes and \acp{RIS} can also be used to deliver physical-layer security solutions in emerging networks \cite{Conf_all}. Such security techniques are well suited to enhance the security of Internet-of-Things devices that incorporate sensors \cite{WangNing:2019}, since such devices are unable to deploy sophisticated security countermeasures. As an example, an \ac{RIS} can be used to reflect jamming signals generated from \ac{FD} transceivers, and collectively, such transmissions can enhance the system's secrecy rate~\cite{YongJin:2022}. 

\begin{figure}[t]
\begin{center}
\includegraphics[width=\hsize]{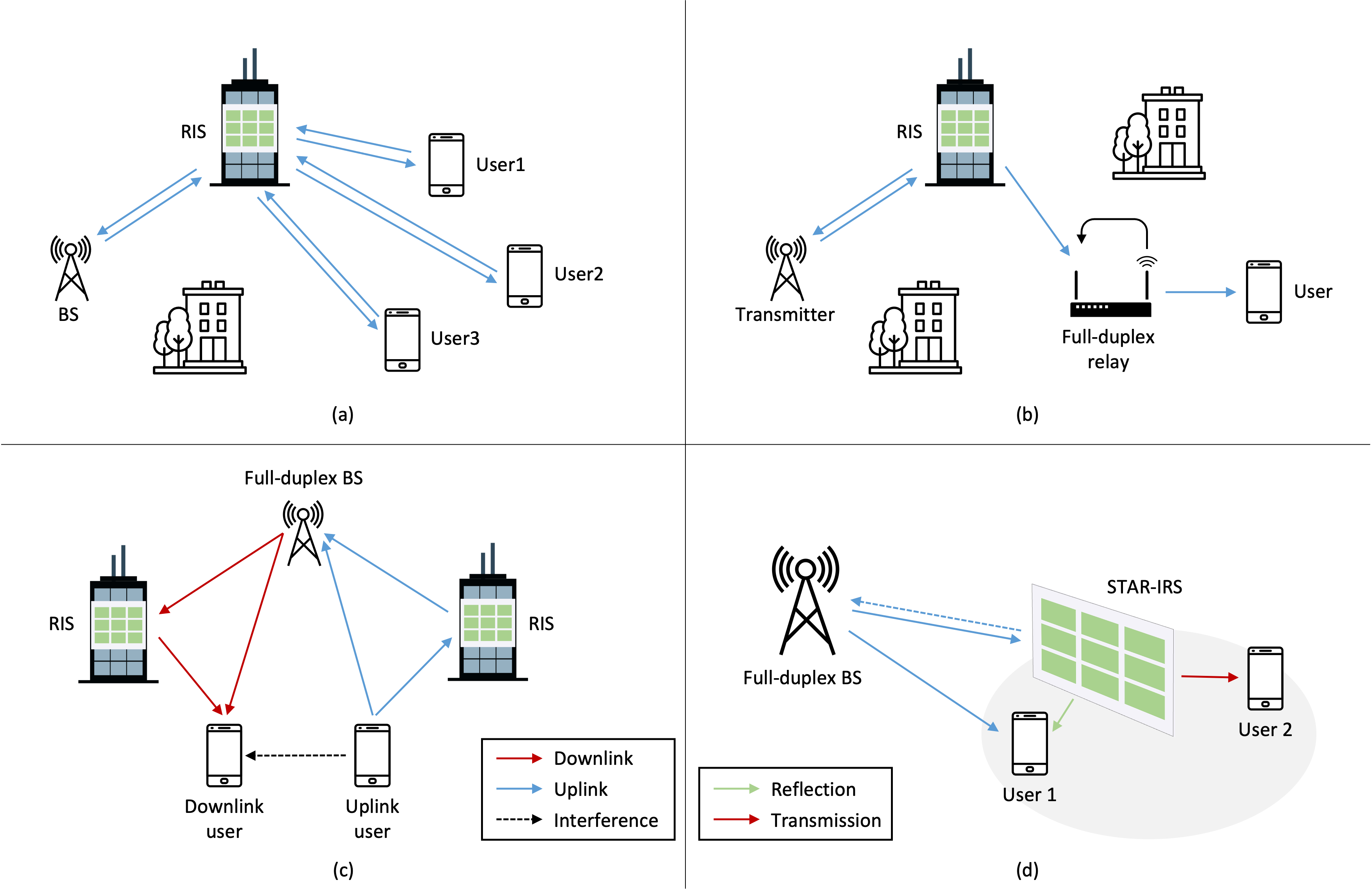}
\caption{\ac{RIS}-assisted \ac{FD} wireless systems: (a) Multi-user two-way communications; (b) relaying; (c) simultaneous \ac{UL} and \ac{DL} communications; and (d) communications with simultaneous transmitting and reflecting metasurfaces.}
\label{fig:RIS1}
\end{center}
\end{figure}

Potential deployment scenarios combining the FD and RIS technologies are illustrated in Fig.~\ref{fig:RIS1}. The deployment of an \ac{RIS} to cover the dead zones of cellular systems, enabling multi-user two-way \ac{FD} communications, is depicted in Fig.~\ref{fig:RIS1}(a). The benefits of \ac{FD} relays and \acp{RIS} can also be merged to deliver extended coverage and flexible deployment as illustrated in Fig.~ \ref{fig:RIS1}(b), where the orientation and placement of the metasurfaces, as well as the number of multiple signal propagation hops, can be optimized to provide large received \ac{SNR} within the cell coverage areas. Another example is shown in Fig.~\ref{fig:RIS1}(c) where an \ac{FD} BS serves both \ac{UL} and \ac{DL} users with the aid of an \ac{RIS}. The metasurface can dynamically adjust its reflection coefficients to boost the received signal power at the \ac{DL} receivers. This, in turn, helps to reduce transmit powers at the BS and \ac{UL} users, rendering the network design spectrally and energy efficient. 

In fact, several late works have investigated the combination of \ac{FD} transceivers with \ac{RIS}s for two-way communications \cite{Peng:TSP2021}, \ac{UL}/\ac{DL} transmission \cite{Cai:2022}, and relaying \cite{yildirim_hybrid_2021,Arzykulov:2022}. The findings of those papers revealed the benefits of using \ac{FD} communications over \ac{HD} counterparts in the presence of \acp{RIS}, in terms of sum-rate maximization and transmit power minimization~\cite{WangYiru:2022}. Interestingly, to perform \ac{SIC}, an \ac{RIS} can be utilized to form a suitable phase-reversed cancellation signal in the analog domain. This novel concept was introduced in \cite{tewes:2022} in which a greedy heuristic algorithm to converge to a locally optimal \ac{RIS} phase profile was developed. The $256$-element \ac{RIS} prototype designed in \cite{tewes:2022} has been shown to be capable of canceling the leaked signal by $59$ dB, achieving a total of $103$ dB \ac{SI} suppression. Clearly, \ac{RIS}-aided \ac{SIC} opens up new possibilities for reduced cost transceiver design, potentially allowing conventional transceivers to be operated in the \ac{FD} mode. However, for the latter framework to be applied with 5G \ac{mmWave} signals, experimental work to design new \ac{RIS}-assisted wideband \ac{SIC} techniques are required. Very recently in \cite{FD_HMIMO_2023}, an \ac{RIS} was deployed in the near-field region of an FD MIMO node to enable joint communications and sensing. In particular, the RIS phase configuration was jointly optimized with the MIMO digital beamformers to enable this simultaneous dual functionality, while efficiently handling SI. Near-field ISAC, but in sub-THz frequencies, was also lately investigated in \cite{FD_RIS_ISAC_2023} considering a novel FD architecture incorporating metasurface-based holographic MIMO transceivers \cite{HMIMO_survey}.

In the existing \ac{RIS}-empowered wireless commnunication systems, the TX and RX have to be on the same side, while this restriction can be removed with the use of the simultaneous transmitting and reflecting \ac{RIS} (STAR-RIS) concept~\cite{LiuYuanwei:2021, Xu_Jaiqi:2021}. Such hybrid metasurfaces are capable of reflecting and transmitting the incident electromagnetic waves into both of their sides. This allows full coverage since \ac{UL}/\ac{DL} users located in front and behind of the STAR-RIS can be served, enabling various applications (e.g., simultaneous indoor and outdoor 3D localization \cite{STARRIS_Loc2022}). In addition, the transmission and reflection coefficients of the STAR-\ac{RIS} unit elements can be configured according to a energy splitting, mode switching, and time switching protocols. Such flexible operation enables to optimize diverse design objectives in \ac{FD} wireless systems, e.g., the sum-rate maximization when users are concurrently operating in the \ac{UL} and \ac{DL} directions~\cite{Perera:2022, Papazafeiropoulos:2022}. As recently shown in \cite{Perera:2022}, all STAR-\ac{RIS} protocols outperform \ac{HD} transmissions and those based on the conventional \ac{RIS} operation. It was also found that, as the number of the STAR-\ac{RIS} unit elements increases, the average rates of the energy splitting and mode switching protocols increase, however with diminishing returns. The mode switching protocol and that based on the conventional \ac{RIS} exhibited similar performance for small metasurface elements. Those observations collectively revealed the importance of intelligently selecting the number of STAR-\ac{RIS} elements and optimizing the amplitudes of reflection and transmission coefficients to improve the system performance.


{\bf Challenges:}
Research on the efficient combination of \ac{FD}-capable nodes with the \ac{RIS} technology is still not mature. Acquisition of accurate channel information is a key challenge for active and passive beamforming, since large training overhead and channel estimation errors can limit the desired signal power at users in practice. This inspires the need to develop novel reflective coefficient configuring techniques with low complexity and low power consumption. Although progress has been made in this direction, based, e.g., on approaches such as random phase rotations \cite{RISrandom2022}, deep reinforcement learning \cite{IRS2022Pervasive}, and parallel factor decomposition \cite{PARAFAC2021}, the joint consideration with the FD technology is still absent. Hence, joint \ac{FD} transceiver and hybrid active/passive \ac{RIS} designs and respective algorithmic optimizations are notable research directions. Coordination of multiple \ac{RIS}s to control the interfering signals and enable their energy-efficient green operation serves as another key challenge in multi-cell mixed \ac{HD} and \ac{FD} networks. There are some promising directions to follow in this research direction, such as advanced network designs based on distributed machine learning, optimization and intelligent resource allocation, as well as research on novel signaling schemes and intelligent network orchestration. However, their effectiveness needs to be showcased through dedicated research.

\subsection{Empowering Non-Terrestrial Networks (NTNs)}

{\bf Opportunity:}
6G's vision represents a major departure from past wireless standards, which concentrated on providing coverage on the ground. Instead, the overarching goal is to integrate both ground and air by leveraging drones and high-altitude platforms (HAPs) to achieve worldwide connectivity~\cite{Geraci_2022,Karabulut_2021}. Additionally, due to advancements in materials development, inexpensive manufacturing, and satellite launches, \ac{NTNs} in the form of mega constellation cube satellites are anticipated to be a component of 6G. This new wave of innovation is highlighted in recent publications in 2022~\cite{Huawei_2022}.


In comparison to geostationary earth orbit satellites, which orbit at a distance of 35,800 km from Earth, drones, HAPs, and nanosatellites operate at a much lower altitude. Unmanned aerial vehicles (UAVs) typically fly at an altitude of a few hundred meters above the ground, while HAPs can be positioned in the stratosphere at an altitude of 20 km. Low-Earth-orbit (LEO) satellites, on the other hand, orbit at an altitude ranging between 200 and 2000 km~\cite{Giordani_2021}. These lower orbits cause LEO satellites to be ``non-stationary" in relation to the ground. Therefore, a constellation of LEO satellites is necessary to ensure service continuity. However, the shorter transmission distances of drones, HAPs, and LEO satellites result in less propagation loss, higher signal strength attenuation, and a lower transmission delay. These appealing features have prompted 3GPP to initiate study items through Rel. 14 to Rel. 20, identifying architectures, potential issues, and use cases such as roaming, backhauling, broadcast/multicast, and IoT, with the goal of integrating NTNs into modern cellular systems and beyond \cite{Huawei_2022,Xingqin_2021}.


The above-mentioned features bring forth fresh possibilities for FD operation in aerial networks of 6G. One such opportunity is the relaxation of the power imbalance between the transmitting and receiving signals, thereby reducing the burden of self-interference (SI) that can lead to performance deterioration due to shorter transmission distances. Furthermore, advancements in powerful on-board capabilities such as processors, high-gain antennas, RF hardware, and low complexity signal processing algorithms are making significant progress. This progress is expected to facilitate the implementation of feasible SI cancellation schemes that can leverage these developments for FD implementation.


One potential application of upcoming aerial networks is to enable Internet of Things (IoT) services in remote and unconnected locations, such as polar regions, oceans, and even space. Since many IoT applications are sensitive to latency, FD operation can help minimize delays in end-to-end data delivery. As \ac{NTNs} continue to evolve, new use cases requiring higher spectral efficiencies are expected to emerge. These applications would benefit from the increased spectral efficiencies offered by FD. Security is also a crucial consideration for communication in \ac{NTNs}. For example, drones are increasingly being used in the military for surveillance and localization purposes, making them susceptible to jamming attacks by external parties due to their relatively shorter distances. In response, FD-capable aerial vehicles can perform active jamming while simultaneously receiving information, thereby ensuring improved physical layer security.

{\bf State-of-the-art:}
Several recent studies have investigated the feasibility of using FD mode for communication in UAV and satellite networks, with a focus on performance analysis and optimization. For instance, \cite{Grayver_2015} aimed to achieve 130~dB of SI cancellation for an FD ground-LEO satellite link using expensive components. In \cite{BShankar_2015}, the benefits of using FD relaying on-board DVB-S2 satellite networks were demonstrated, including more efficient use of satellite spectrum. In \cite{Hua_Meng_2020}, the use of an FD UAV as an aerial base station for small cell wireless communication was studied, and system performance was optimized by optimizing the UAV trajectory, DL/UL user scheduling, and UL user transmit power to maximize the system's UL and DL sum rate. In \cite{Shi_Wenjuan_2021}, power allocation, user pairing, and subchannel assignment were optimized for a UL/DL communication system that used a UAV as a relay, and the authors demonstrated that a proposed non-orthogonal access method could provide better spectral and energy efficiencies than the orthogonal access method. In \cite{Araki_2022}, multiple UAVs simultaneously reused all available DL and UL channels, while using isolated DL and UL channels to avoid SI, but directional antennas and flight control of the UAVs were used to eliminate co-channel interference. In \cite{Nguyen_Tan_2022}, the outage probability of a satellite-ground network utilizing FD relays and relay selection was studied in the presence of interference, while \cite{Ngo_Quynh_2022} presented a two-tier cache-enabled satellite-terrestrial network that used FD transmission to reduce content delivery latency. However, the motion of aerial terminals can introduce significant challenges for FD implementation, including handover and interference control, and the limited onboard energy of aerial vehicles may limit the introduction of FD aerial terminals for certain use cases.

{\bf Challenges:}
The movement of aerial terminals poses considerable obstacles to implementing FD, particularly in terms of handover and interference control. For instance, the flight paths of drones may lead to time-varying interference patterns of varying strengths between air-to-air or air-to-ground signals. Additionally, dynamic and challenging-to-model shadowing conditions in certain scenarios could further complicate how interference affects performance. Thus, integrating air or space segments with terrestrial networks would require meticulous planning. Furthermore, for certain use cases, the limited on-board energy available could significantly constrain the adoption of FD aerial terminals, unless lightweight and cost-effective energy storage technologies are developed in the near future.

\section{Future Outlook and Concluding Remarks}


Wireless networks are poised for a significant transformation, offering new degrees of design freedom in the form of fresh spectrum, advanced sensing capabilities, and novel use cases. Additionally, learning-based methods will play a pivotal role in this evolution, particularly in 5G-Advanced Rel-18 and beyond, where artificial intelligence and machine learning techniques will enhance the air interface by enabling adaptive waveform design, energy-efficient sustainable operation, and predictive user mobility patterns and behavior. Devices and systems that support \ac{FD} communication will complement these new design degrees, pushing the limits of throughput and latency. Moreover, learning-based methods hold promise for eliminating the need for hand-tuning and effectively canceling self-interference in the presence of system nonlinearities and challenging wireless propagation conditions. As in-band full-duplex communication gains wider acceptance in standards, it is not only becoming a well-established concept but also an active area of research. This overview highlighted some of the emerging research and usage directions for in-band \ac{FD} communication, with many new opportunities and challenges remaining, as evidenced by the strong set of papers in this special issue.

\bibliographystyle{IEEEtran}
\bibliography{Bib1}

\begin{thebibliography}{100}
\providecommand{\url}[1]{#1}
\csname url@samestyle\endcsname
\providecommand{\newblock}{\relax}
\providecommand{\bibinfo}[2]{#2}
\providecommand{\BIBentrySTDinterwordspacing}{\spaceskip=0pt\relax}
\providecommand{\BIBentryALTinterwordstretchfactor}{4}
\providecommand{\BIBentryALTinterwordspacing}{\spaceskip=\fontdimen2\font plus
\BIBentryALTinterwordstretchfactor\fontdimen3\font minus
  \fontdimen4\font\relax}
\providecommand{\BIBforeignlanguage}[2]{{%
\expandafter\ifx\csname l@#1\endcsname\relax
\typeout{** WARNING: IEEEtran.bst: No hyphenation pattern has been}%
\typeout{** loaded for the language `#1'. Using the pattern for}%
\typeout{** the default language instead.}%
\else
\language=\csname l@#1\endcsname
\fi
#2}}
\providecommand{\BIBdecl}{\relax}
\BIBdecl

\bibitem{Ericsson:2022}
{Ericsson}, ``Ericsson {M}obility {R}eport,'' Nov. 2022 [Online],
  https://www.ericsson.com/en/reports-and-papers/mobility-report.

\bibitem{Chowdhury_2020}
M.~Z. Chowdhury, M.~Shahjalal, S.~Ahmed, and Y.~M. Jang, ``6{G} wireless
  communication systems: Applications, requirements, technologies, challenges,
  and research directions,'' \emph{IEEE Open J. Commun. Soc.}, vol.~1, pp.
  957--975, 2020.

\bibitem{Tataria_2021}
H.~Tataria, M.~Shafi, A.~F. Molisch, M.~Dohler, H.~Sj{\"o}land, and
  F.~Tufvesson, ``6{G} wireless systems: Vision, requirements, challenges,
  insights, and opportunities,'' \emph{Proc. IEEE}, vol. 109, no.~7, pp.
  1166--1199, 2021.

\bibitem{TFaisal_2020}
F.~Tariq, M.~R.~A. Khandaker, K.-K. Wong, M.~A. Imran, M.~Bennis, and
  M.~Debbah, ``A speculative study on 6{G},'' \emph{IEEE Wireless Commun.},
  vol.~27, no.~4, pp. 118--125, 2020.

\bibitem{Wanshi_2023}
W.~Chen, X.~Li, J.~Lee, A.~Toskala, S.~Sun, C.~F. Chiasserini, and L.~Liu,
  ``5{G}-{A}dvanced towards 6{G}: {P}ast, present, and future,'' \emph{arXiv
  preprint arXiv:2303.07456}, 2023.

\bibitem{Fodor:2021}
G.~Fodor, C.-B. Chae, R.~Wichman, A.~Sabharwal, H.~A. Suraweera, R.~Rao, and
  H.~Alves, ``Guest editorial: Full duplex communications theory,
  standardization, and practice,'' \emph{{IEEE} Wireless Commun. Mag.},
  vol.~28, no.~1, pp. 10--11, 2021.

\bibitem{chae_mole}
J.~Kwak, H.~B. Yilmaz, N.~Farsad, C.-B. Chae, and A.~Goldsmith, ``Two-way
  molecular communications,'' \emph{IEEE Trans. Commun.}, vol.~6, no.~68, pp.
  3550--3563, June 2020.

\bibitem{rice_conf1}
M.~Duarte and A.~Sabharwal, ``Full-duplex wireless communications using
  off-the-shelf radios: Feasibility and first results,'' in \emph{Proc.
  Asilomar Conference on Signals, Systems and Computers}, 2010, pp. 1558--1562.

\bibitem{stan_conf1}
J.~I. Choi, M.~Jain, K.~Srinivasan, P.~Levis, and S.~Katti, ``Achieving single
  channel, full duplex wireless communication,'' in \emph{Proc. Annual
  International Conference on Mobile Computing and Networking (MobiCom)}, 2010,
  pp. 1--12.

\bibitem{old_fd_mit}
D.~W. Bliss, P.~A. Parker, and A.~R. Margetts, ``Simultaneous transmission and
  reception for improved wireless network performance,'' in \emph{Proc. IEEE/SP
  Workshop on Statistical Signal Processing}, 2007, pp. 478--482.

\bibitem{Khan2}
A.~K. Khandani, ``Methods for spatial multiplexing of wireless two-way
  channels,'' in \emph{US Patent US7817641}, Oct. 2010.

\bibitem{rice_conf2}
E.~Everett, M.~Duarte, C.~Dick, and A.~Sabharwal, ``Empowering full-duplex
  wireless communication by exploiting directional diversity,'' in \emph{Proc.
  Asilomar Conference on Signals, Systems and Computers (ASILOMAR)}, 2011, pp.
  2002--2006.

\bibitem{stan_conf2}
M.~Jain, J.~I. Choi, T.~Kim, D.~Bharadia, S.~Seth, K.~Srinivasan, P.~Levis,
  S.~Katti, and P.~Sinha, ``Practical, real-time, full duplex wireless,'' in
  \emph{Proc. Annual International Conference on Mobile Computing and
  Networking (MobiCom)}, 2011, pp. 301--312.

\bibitem{Khan1}
A.~K. Khandani, ``Shaping the future of wireless: Two-way connectivity,''
  http://www.nortel-institute.uwaterloo.ca/content/Shaping Future of Wireless
  Two-way Connectivity 18ne2012.pdf, June 2012.

\bibitem{rice_exp}
M.~Duarte, C.~Dick, and A.~Sabharwal, ``Experiment-driven characterization of
  full-duplex wireless systems,'' \emph{IEEE Trans. Wireless Commun.}, vol.~11,
  no.~12, pp. 4296--4307, 2012.

\bibitem{passive_symmetric}
E.~Aryafar, M.~A. Khojastepour, K.~Sundaresan, S.~Rangarajan, and M.~Chiang,
  ``{MIDU}: Enabling {MIMO} full duplex,'' in \emph{Proc. {ACM} Mobile Comp.
  and Net. (MobiCOM)}, Aug. 2012, pp. 257--268.

\bibitem{rice_journal1}
M.~Duarte, A.~Sabharwal, V.~Aggarwal, R.~Jana, K.~K. Ramakrishnan, C.~W. Rice,
  and N.~K. Shankaranarayanan, ``Design and characterization of a full-duplex
  multiantenna system for {W}i{F}i networks,'' \emph{IEEE Trans. Veh.
  Technol.}, vol.~63, no.~3, pp. 1160--1177, 2014.

\bibitem{Righini_2022}
D.~Righini and A.~M. Tonello, ``{MIMO} in-band-full-duplex {PLC}: {D}esign,
  analysis and first hardware realization of the analog self-interference
  cancellation stage,'' \emph{IEEE Open J. Commun. Soc.}, vol.~2, pp.
  1344--1357, 2021.

\bibitem{Riihonen_2011}
T.~Riihonen, S.~Werner, and R.~Wichman, ``Hybrid full-duplex/half-duplex
  relaying with transmit power adaptation,'' \emph{IEEE Trans. Wireless
  Commun.}, vol.~10, no.~9, pp. 3074--3085, 2011.

\bibitem{full-tap_MIMO}
------, ``Mitigation of loopback self-interference in full-duplex {MIMO}
  relays,'' \emph{IEEE Trans. Signal Process.}, vol.~59, no.~12, pp.
  5983--5993, 2011.

\bibitem{Krikidis_TWC_2012}
I.~Krikidis, H.~A. Suraweera, S.~Yang, and K.~Berberidis, ``Full-duplex
  relaying over block fading channel: {A} diversity perspective,'' \emph{IEEE
  Trans. Wireless Commun.}, vol.~11, no.~12, pp. 4524--4535, 2012.

\bibitem{Hong_2014}
S.~Hong, J.~Brand, J.~I. Choi, M.~Jain, J.~Mehlman, S.~Katti, and P.~Levis,
  ``Applications of self-interference cancellation in 5{G} and beyond,''
  \emph{IEEE Commun. Mag.}, vol.~52, no.~2, pp. 114--121, 2014.

\bibitem{Sabharwal2014IBFD}
A.~Sabharwal, P.~Schniter, D.~Guo, D.~W. Bliss, S.~Rangarajan, and R.~Wichman,
  ``In-band full-duplex wireless: Challenges and opportunities,'' \emph{{IEEE}
  J. Sel. Areas Commun.}, vol.~32, no.~9, pp. 1637--1652, Jun. 2014.

\bibitem{Heino_2015}
M.~Heino, D.~Korpi, T.~Huusari, E.~Antonio-Rodriguez, S.~Venkatasubramanian,
  T.~Riihonen, L.~Anttila, C.~Icheln, K.~Haneda, R.~Wichman, and M.~Valkama,
  ``Recent advances in antenna design and interference cancellation algorithms
  for in-band full duplex relays,'' \emph{IEEE Commun. Mag.}, vol.~53, no.~5,
  pp. 91--101, 2015.

\bibitem{Zhang:16}
Z.~Zhang, K.~Long, A.~V. Vasilakos, and L.~Hanzo, ``Full-duplex wireless
  communications: {C}hallenges, solutions, and future research directions,''
  \emph{Proc. IEEE}, vol. 104, no.~7, pp. 1369--1409, 2016.

\bibitem{Shende_2018}
N.~V. Shende, O.~G{\"u}rb{\"u}z, and E.~Erkip, ``Half-duplex or full-duplex
  communications: {D}egrees of freedom analysis under self-interference,''
  \emph{IEEE Trans. Wireless Commun.}, vol.~17, no.~2, pp. 1081--1093, 2018.

\bibitem{M_Mohammadi_2019}
M.~Mohammadi, X.~Shi, B.~K. Chalise, Z.~Ding, H.~A. Suraweera, C.~Zhong, and
  J.~S. Thompson, ``Full-duplex non-orthogonal multiple access for next
  generation wireless systems,'' \emph{IEEE Commun. Mag.}, vol.~57, no.~5, pp.
  110--116, 2019.

\bibitem{Kolodziej:2019}
K.~E. Kolodziej, B.~T. Perry, and J.~S. Herd, ``In-band full-duplex technology:
  {T}echniques and systems survey,'' \emph{IEEE Trans. Microw. Theory Tech.},
  vol.~67, no.~7, pp. 3025--3041, 2019.

\bibitem{Vaibhav:2020}
S.~Vaibhav, M.~Susnata, G.~Akshay, S.~Milind, P.~Jeyanandh, and K.~Swarun,
  ``Millimeter-wave full duplex radios,'' in \emph{Proc. 26th Annual
  International Conference on Mobile Computing and Networking (MobiCom '20)},
  2020, pp. 1--14.

\bibitem{Roberts_BookCh_2022}
I.~P. Roberts and H.~A. Suraweera, ``Full-duplex transceivers for
  next-generation wireless communication systems,'' \emph{arXiv preprint
  arXiv:2210.08094}, 2022.

\bibitem{Lingyang_Song_Book}
L.~Song, R.~Wichman, Y.~Li, and Z.~Han, \emph{Full-Duplex Communications and
  Networks}.\hskip 1em plus 0.5em minus 0.4em\relax Cambridge University Press,
  2017.

\bibitem{Alves_FD_Book}
H.~Alves, T.~Riihonen, and H.~A. Suraweera, \emph{Full-Duplex Communications
  for Future Wireless Networks}.\hskip 1em plus 0.5em minus 0.4em\relax
  Springer, 2020.

\bibitem{Kolodziej_Book}
K.~Kolodziej, \emph{In-Band Full-Duplex Wireless Systems Handbook}.\hskip 1em
  plus 0.5em minus 0.4em\relax Artech House, 2021.

\bibitem{Berscheid:2019}
B.~Berscheid and C.~Howlett, ``Full duplex {DOCSIS}: {O}pportunities and
  challenges,'' \emph{IEEE Commun. Mag.}, vol.~57, no.~8, pp. 28--33, 2019.

\bibitem{nonlinear_digital}
L.~Anttila, D.~Korpi, V.~Syrj{\"a}l{\"a}, and M.~Valkama, ``Cancellation of
  power amplifier induced nonlinear self-interference in full-duplex
  transceivers,'' in \emph{Proc. Asilomar Conference on Signals, Systems and
  Computers}, 2013, pp. 1193--1198.

\bibitem{nonlinear_digital2}
A.~Sahai, G.~Patel, C.~Dick, and A.~Sabharwal, ``On the impact of phase noise
  on active cancelation in wireless full-duplex,'' \emph{{IEEE} Trans. Veh.
  Technol.}, vol.~62, no.~9, pp. 4494--4510, Nov. 2013.

\bibitem{passive_dualpol}
T.~Oh, Y.-G. Lim, C.-B. Chae, and Y.~Lee, ``Dual-polarization slot antenna with
  high cross-polarization discrimination for indoor small-cell {MIMO}
  systems,'' \emph{IEEE Antennas Wirel. Propag. Lett.}, vol.~14, no. Feb., pp.
  374--377, 2015.

\bibitem{realtime}
M.~Chung, M.~S. Sim, J.~Kim, D.~K. Kim, and C.-B. Chae, ``Prototyping real-time
  full duplex radios,'' \emph{IEEE Commun. Mag.}, vol.~53, no.~9, pp. 56--63,
  Sep. 2015.

\bibitem{passive_absorb}
E.~Everett, A.~Sahai, and A.~Sabharwal, ``Passive self-interference suppression
  for full-duplex infrastructure nodes,'' \emph{{IEEE} Trans. Wireless
  Commun.}, vol.~13, no.~2, pp. 680--694, Jan. 2014.

\bibitem{9895328}
Y.~Xing, S.~Li, X.~Chen, X.~Xue, and X.~Zheng, ``Optical multi-tap rf canceller
  for in-band full-duplex wireless communication systems,'' \emph{IEEE
  Photonics J.}, vol.~14, no.~5, pp. 1--7, 2022.

\bibitem{9431091}
K.~E. Kolodziej, A.~U. Cookson, and B.~T. Perry, ``{RF} canceller tuning
  acceleration using neural network machine learning for in-band full-duplex
  systems,'' \emph{IEEE Open J. Commun. Soc.}, vol.~2, pp. 1158--1170, 2021.

\bibitem{7146163}
T.~Huusari, Y.-S. Choi, P.~Liikkanen, D.~Korpi, S.~Talwar, and M.~Valkama,
  ``Wideband self-adaptive {RF} cancellation circuit for full-duplex radio:
  Operating principle and measurements,'' in \emph{IEEE Vehicular Technology
  Conference (VTC Spring)}, 2015, pp. 1--7.

\bibitem{maddio2013reconfigurable}
S.~Maddio, A.~Cidronali, A.~Palonghi, and G.~Manes, ``A reconfigurable leakage
  canceler at 5.8 {GH}z for {DSRC} applications,'' in \emph{Proc. IEEE MTT-S
  International Microwave Symposium Digest (MTT 2013)}, June 2013, pp. 1--3.

\bibitem{8363897}
S.~Khaledian, F.~Farzami, B.~Smida, and D.~Erricolo, ``Inherent
  self-interference cancellation at 900 {MH}z for in-band full-duplex
  applications,'' in \emph{Proc. IEEE Wireless and Microwave Technology
  Conference (WAMICON 2018)}, 2018, pp. 1--4.

\bibitem{8335770}
------, ``Inherent self-interference cancellation for in-band full-duplex
  single-antenna systems,'' \emph{IEEE Trans. Microw. Theory Tech.}, vol.~66,
  no.~6, pp. 2842--2850, 2018.

\bibitem{7051286}
E.~Ahmed and A.~M. Eltawil, ``All-digital self-interference cancellation
  technique for full-duplex systems,'' \emph{IEEE Trans. Wireless Commun.},
  vol.~14, no.~7, pp. 3519--3532, 2015.

\bibitem{6832439}
D.~Korpi, L.~Anttila, V.~Syrj{\"a}l{\"a}, and M.~Valkama, ``Widely linear
  digital self-interference cancellation in direct-conversion full-duplex
  transceiver,'' \emph{IEEE J. Sel. Areas Commun.}, vol.~32, no.~9, pp.
  1674--1687, 2014.

\bibitem{7815419}
X.~Quan, Y.~Liu, S.~Shao, C.~Huang, and Y.~Tang, ``Impacts of phase noise on
  digital self-interference cancellation in full-duplex communications,''
  \emph{IEEE Trans. Signal Process.}, vol.~65, no.~7, pp. 1881--1893, 2017.

\bibitem{balatsoukas2015baseband}
A.~Balatsoukas-Stimming, A.~Austin, P.~Belanovic, and A.~Burg, ``Baseband and
  rf hardware impairments in full-duplex wireless systems: experimental
  characterisation and suppression,'' \emph{EURASIP J. Wirel. Commun. Netw.},
  vol. 2015, no.~1, pp. 1--11, 2015.

\bibitem{8761524}
M.~A. Islam and B.~Smida, ``A comprehensive self-interference model for
  single-antenna full-duplex communication systems,'' in \emph{Proc. IEEE
  International Conference on Communications (ICC)}, May 2019, pp. 1--7.

\bibitem{Balatsoukas:2018}
A.~{Balatsoukas-Stimming}, ``Non-linear digital self-interference cancellation
  for in-band full-duplex radios using neural networks,'' in \emph{Proc. IEEE
  SPAWC 2018}, Kalamata, Greece, June 2018, pp. 1--5.

\bibitem{Guo:2019}
H.~{Guo}, S.~{Wu}, H.~{Wang}, and M.~{Daneshmand}, ``Deep learning based
  self-interference cancellation for in-band full duplex wireless,'' in
  \emph{Proc. IEEE Global Commun. Conf. (GLOBECOM)}, Dec. 2019, pp. 1--6.

\bibitem{Shi:2019}
C.~{Shi}, Y.~{Hao}, Y.~{Liu}, and S.~{Shao}, ``Digital self-interference
  cancellation for full duplex wireless communication based on neural
  networks,'' in \emph{Proc. Int. Conf. Commun. and Inform. Syst.}, Wuhan,
  China, Dec. 2019, pp. 53--57.

\bibitem{Kurzo:2018}
Y.~{Kurzo}, A.~{Burg}, and A.~{Balatsoukas-Stimming}, ``Design and
  implementation of a neural network aided self-interference cancellation
  scheme for full-duplex radios,'' in \emph{Proc. Asilomar Conf. Signals, Syst.
  and Comput.}, Oct. 2018, pp. 589--593.

\bibitem{DNN_ULDL2019}
C.~Huang, G.~C. Alexandropoulos, A.~Zappone, C.~Yuen, and M.~Debbah, ``Deep
  learning for {UL/DL} channel calibration in generic massive {MIMO} systems,''
  in \emph{Proc. IEEE Intl. Conf. Commun. (ICC)}, Dec. 2019, pp. 1--6.

\bibitem{9195843}
M.~Elsayed, A.~A.~A. El-Banna, O.~A. Dobre, W.~Shiu, and P.~Wang, ``Low
  complexity neural network structures for self-interference cancellation in
  full-duplex radio,'' \emph{IEEE Commun. Lett.}, vol.~25, no.~1, pp. 181--185,
  2021.

\bibitem{9736621}
------, ``Hybrid-layers neural network architectures for modeling the
  self-interference in full-duplex systems,'' \emph{IEEE Trans. Veh. Technol.},
  vol.~71, no.~6, pp. 6291--6307, 2022.

\bibitem{Korpi:2017}
D.~Korpi, L.~Anttila, and M.~Valkama, ``Nonlinear self-interference
  cancellation in {MIMO} full-duplex transceivers under crosstalk,''
  \emph{EURASIP J. Wireless Commun. Netw.}, no.~24, 2017.

\bibitem{9552213}
K.~Muranov, M.~A. Islam, B.~Smida, and N.~Devroye, ``On deep learning assisted
  self-interference estimation in a full-duplex relay link,'' \emph{IEEE
  Wireless Commun. Lett.}, vol.~10, no.~12, pp. 2762--2766, 2021.

\bibitem{9732685}
D.~H. Kong, Y.-S. Kil, and S.-H. Kim, ``Neural network aided digital
  self-interference cancellation for full-duplex communication over
  time-varying channels,'' \emph{IEEE Trans. Veh. Technol.}, vol.~71, no.~6,
  pp. 6201--6213, 2022.

\bibitem{FD_MIMO_VTM2022}
G.~C. Alexandropoulos, M.~A. Islam, and B.~Smida, ``Full duplex massive {MIMO}
  architectures: {R}ecent advances, applications, and future directions,''
  \emph{IEEE Veh. Technol. Mag.}, vol.~17, no.~4, pp. 83--91, Dec. 2022.

\bibitem{Bharadia2014}
D.~Bharadia and S.~Katti, ``Full duplex {MIMO} radios,'' in \emph{Proc. {IEEE
  USENIX NSDI}}, Apr. 2014, pp. 359--372.

\bibitem{rice_conf3}
A.~Sahai, G.~Patel, and A.~Sabharwal, ``Asynchronous full-duplex wireless,'' in
  \emph{"Proc. Int. Conf. on Commun. Systems and Networks"}, 2012, pp. 1 -- 9.

\bibitem{Kolodziej2016}
K.~E. Kolodziej, J.~G. McMichael, and B.~T. Perry, ``Multitap {RF} canceller
  for in-band full-duplex wireless communications,'' \emph{IEEE Trans. Wireless
  Commun.}, vol.~15, no.~6, pp. 4321--4334, Jun. 2016.

\bibitem{Spatial_suppression}
E.~Everett, C.~Shepard, L.~Zhong, and A.~Sabharwal, ``{SoftNull: M}any-antenna
  full-duplex wireless via digital beamforming,'' \emph{IEEE Trans. Wireless
  Commun.}, vol.~12, no.~15, pp. 80\,770--8092, Dec. 2016.

\bibitem{Gowda:18}
N.~M. Gowda and A.~Sabharwal, ``{JointNull}: Combining partial analog
  cancellation with transmit beamforming for large-antenna full-duplex wireless
  systems,'' \emph{IEEE Trans. Wireless Commun.}, vol.~17, no.~3, pp.
  2094--2108, 2018.

\bibitem{Hien:2017}
H.~Q. Ngo, H.~A. Suraweera, M.~Matthaiou, and E.~G. Larsson, ``Multipair
  full-duplex relaying with massive arrays and linear processing,'' \emph{IEEE
  J. Sel. Areas Commun.}, vol.~32, no.~9, pp. 1721--1737, 2014.

\bibitem{Lim_Kim}
S.-M. Kim, Y.-G. Lim, L.~Dai, and C.-B. Chae, ``Performance analysis of
  self-interference cancellation in full-duplex massive {MIMO} systems:
  {S}ubtraction versus spatial suppression,'' \emph{{IEEE} Trans. Wireless
  Commun.}, vol.~22, no.~1, pp. 642--657, Jan. 2023.

\bibitem{Suraweera:2014}
H.~A. Suraweera, I.~Krikidis, G.~Zheng, C.~Yuen, and P.~J. Smith,
  ``Low-complexity end-to-end performance optimization in {MIMO} full-duplex
  relay systems,'' \emph{IEEE Trans. Wireless Commun.}, vol.~13, no.~2, pp.
  913--927, 2014.

\bibitem{Zarifeh:2019}
N.~Zarifeh, Y.~Zantah, Y.~Gao, and T.~Kaiser, ``Full-duplex femto base-station
  with antenna selection: Experimental validation,'' \emph{IEEE Access},
  vol.~7, pp. 108\,781--108\,794, 2019.

\bibitem{Mohammadi_Mobini_ICC18}
M.~Mohammadi, Z.~Mobini, H.~A. Suraweera, and Z.~Ding, ``Antenna selection in
  full-duplex cooperative {NOMA} systems,'' in \emph{Proc. IEEE Intl. Conf.
  Commun. (ICC)}, May 2018, pp. 1--6.

\bibitem{Mobini:2022}
Z.~Mobini, M.~Mohammadi, T.~A. Tsiftsis, Z.~Ding, and C.~Tellambura, ``New
  antenna selection schemes for full-duplex cooperative {MIMO}-{NOMA}
  systems,'' \emph{IEEE Trans. Commun.}, vol.~70, no.~7, pp. 4343--4358, 2022.

\bibitem{alexandropoulos2020full}
G.~C. Alexandropoulos, M.~A. Islam, and B.~Smida, ``Full duplex hybrid {A/D}
  beamforming with reduced complexity multi-tap analog cancellation,'' in
  \emph{Proc. {IEEE SPAWC}}, May 2020, pp. 1--6.

\bibitem{IanRoberts:2022}
I.~P. Roberts, A.~Chopra, T.~Novlan, S.~Vishwanath, and J.~G. Andrews, ``Steer:
  Beam selection for full-duplex millimeter wave communication systems,''
  \emph{IEEE Trans. Commun.}, vol.~70, no.~10, pp. 6902--6917, 2022.

\bibitem{Chen:2023}
Z.~Chen, C.~B. Barati, J.~Veihl, C.~Shepard, and A.~Sabharwal, ``Lensfd: Using
  lenses for improved sub-6 ghz massive mimo full-duplex,'' \emph{IEEE
  Transactions on Vehicular Technology}, 2023.

\bibitem{Mohammadi_CRAN:2018}
M.~Mohammadi, H.~A. Suraweera, and C.~Tellambura, ``Uplink/downlink rate
  analysis and impact of power allocation for full-duplex cloud-{RAN}s,''
  \emph{IEEE Trans. Wireless Commun.}, vol.~17, no.~9, pp. 5774--5788, 2018.

\bibitem{Radwa:2021}
R.~Sultan, K.~G. Seddik, Z.~Han, and B.~Aazhang, ``Joint transmitter-receiver
  optimization and self-interference suppression in full-duplex {MIMO}
  systems,'' \emph{IEEE Trans. Veh. Technol.}, vol.~70, no.~7, pp. 6913--6929,
  2021.

\bibitem{alexandropoulos2017joint}
G.~C. Alexandropoulos and M.~Duarte, ``Joint design of multi-tap analog
  cancellation and digital beamforming for reduced complexity full duplex
  {MIMO} systems,'' in \emph{Proc. IEEE Intl. Conf. Commun. (ICC)}, May 2017.

\bibitem{Islam2019unified}
M.~A. Islam, G.~C. Alexandropoulos, and B.~Smida, ``A unified beamforming and
  {A/D} self-interference cancellation design for full duplex {MIMO} radios,''
  in \emph{Proc. {IEEE PIMRC}}, Sep. 2019.

\bibitem{FD_MIMO_Impairments}
H.~Iimori, G.~Abreu, and G.~C. Alexandropoulos, ``{MIMO} beamforming schemes
  for hybrid {SIC FD} radios with imperfect hardware and {CSI},'' \emph{IEEE
  Trans. Wireless Commun.}, vol.~18, no.~10, pp. 4816--4830, Oct. 2019.

\bibitem{FD_MIMO_Arch}
G.~C. Alexandropoulos, ``Low complexity full duplex {MIMO} systems: {A}nalog
  canceler architectures, beamforming design, and future directions,''
  \emph{ITU J. Future Evolving Technol.}, vol.~2, no.~2, pp. 1--19, Dec. 2021.

\bibitem{Chun-Tao:2017}
C.-T. Lin, F.-S. Tseng, and W.-R. Wu, ``{MMSE} transceiver design for
  full-duplex {MIMO} relay systems,'' \emph{IEEE Trans. Veh. Technol.},
  vol.~66, no.~8, pp. 6849--6861, 2017.

\bibitem{WB_FD_MIMO_TWC2022}
M.~A. Islam, G.~C. Alexandropoulos, and B.~Smida, ``Joint analog and digital
  transceiver design for wideband full duplex {MIMO} systems,'' \emph{IEEE
  Trans. Wireless Commun.}, early access, 2022.

\bibitem{Duarte:20}
M.~Duarte and G.~C. Alexandropoulos, ``Full duplex {MIMO} digital beamforming
  with reduced complexity {AUXTX} analog cancellation,'' in \emph{Proc. IEEE
  Intl. Conf. Commun. (ICC)}, Jun. 2020, pp. 1--6.

\bibitem{spatial_suppression_chae}
S.-M. Kim, Y.-G. Lim, L.~Dai, and C.-B. Chae, ``Performance analysis of
  self-interference cancellation in full-duplex massive {MIMO} systems:
  Subtraction versus spatial suppression,'' \emph{IEEE Trans. Wireless
  Commun.}, vol.~22, no.~1, pp. 642--657, 2023.

\bibitem{Spatial_suppression2}
N.~M. Gowda and A.~Sabharwal, ``Jointnull: Combining partial analog
  cancellation with transmit beamforming for large-antenna full-duplex wireless
  systems,'' \emph{IEEE Trans. Wireless Commun.}, vol.~17, no.~3, pp.
  2094--2108, Mar. 2018.

\bibitem{Vishwanath_2020}
I.~P. Roberts, H.~B. Jain, and S.~Vishwanath, ``Equipping millimeter-wave
  full-duplex with analog self-interference cancellation,'' in \emph{Proc.
  {IEEE Intl. Conf. Commun. (ICC)}}, Jun. 2020, pp. 1--6.

\bibitem{xiao2017full_all}
Z.~Xiao, P.~Xia, and X.-G. Xia, ``Full-duplex millimeter-wave communication,''
  \emph{{IEEE} {W}ireless {C}ommun.}, vol.~24, no.~6, pp. 136--143, Dec. 2017.

\bibitem{satyanarayana2018hybrid}
K.~Satyanarayana, M.~El-Hajjar, P.-H. Kuo, A.~Mourad, and L.~Hanzo, ``Hybrid
  beamforming design for full-duplex millimeter wave communication,''
  \emph{{IEEE} {T}rans. {V}eh. {T}echnol.}, vol.~68, no.~2, pp. 1394--1404,
  Dec. 2018.

\bibitem{roberts2019beamforming}
I.~P. Roberts and S.~Vishwanath, ``Beamforming cancellation design for
  millimeter-wave full-duplex,'' in \emph{Proc. IEEE Global Commun. Conf.
  (GLOBECOM)}, Dec. 2019, pp. 1--6.

\bibitem{da20201}
J.~M.~B. Da~Silva, A.~Sabharwal, G.~Fodor, and C.~Fischione, ``1-bit phase
  shifters for large-antenna full-duplex mmwave communications,'' \emph{{IEEE}
  {T}rans. {W}ireless {C}ommun.}, vol.~19, no.~10, pp. 6916--6931, Jul. 2020.

\bibitem{9145036}
Y.~Chen, R.~K. Mishra, D.~Schwartz, and S.~Vishwanath, ``{MIMO} full duplex
  radios with deep learning,'' in \emph{Proc. 2020 IEEE Intl. Conf. Commun.
  Workshops (ICC Workshops)}, 2020, pp. 1--6.

\bibitem{Zhang:21b}
J.~A. Zhang, F.~Liu, C.~Masouros, R.~W. Heath, Z.~Feng, L.~Zheng, and
  A.~Petropulu, ``An overview of signal processing techniques for joint
  communication and radar sensing,'' \emph{IEEE J. Sel. Topics Signal
  Process.}, vol.~15, no.~6, pp. 1295--1315, 2021.

\bibitem{Liu:22}
F.~Liu, Y.~Cui, C.~Masouros, J.~Xu, T.~X. Han, Y.~C. Eldar, and S.~Buzzi,
  ``Integrated sensing and communications: Toward dual-functional wireless
  networks for {6G} and beyond,'' \emph{IEEE J. Sel. Areas Commun.}, vol.~40,
  no.~6, pp. 1728--1767, 2022.

\bibitem{Liu:20}
F.~Liu, C.~Masouros, A.~P. Petropulu, H.~Griffiths, and L.~Hanzo, ``Joint radar
  and communication design: Applications, state-of-the-art, and the road
  ahead,'' \emph{IEEE Trans. Commun.}, vol.~68, no.~6, pp. 3834--3862, 2020.

\bibitem{Liu:20b}
X.~Liu, T.~Huang, N.~Shlezinger, Y.~Liu, J.~Zhou, and Y.~C. Eldar, ``Joint
  transmit beamforming for multiuser {MIMO} communications and {MIMO} radar,''
  \emph{IEEE Trans. Signal Process.}, vol.~68, pp. 3929--3944, 2020.

\bibitem{S1-214242}
``{3GPP S1-214242} {New SID} on study on integrated sensing and communication,
  {3GPP System Architecture Group SA1},'' Nov. 2021.

\bibitem{Behravan:22}
A.~Behravan, R.~Baldemair, S.~Parkvall, E.~Dahlman, V.~Yajnanarayana,
  H.~Bj\"{o}rkegren, and D.~Shrestha, ``Introducing sensing into future
  wireless communication systems,'' in \emph{Proc. 2nd IEEE International
  Symposium on Joint Communications \& Sensing {(JC\&S 2022)}}, Seefeld,
  Austria, 2022, pp. 1--5.

\bibitem{Wu:22}
C.~Wu, B.~Wang, O.~C. Au, and K.~R. Liu, ``Wi-fi can do more: Toward ubiquitous
  wireless sensing,'' \emph{IEEE Commun. Stand. Mag.}, vol.~6, no.~2, pp.
  42--49, 2022.

\bibitem{Davis:11}
M.~Davis, ``Key differences between radar and communications systems,'' in
  Proc. Annu. Int. Symp. Adv. Radio Technol., Jul. 2011, {Available:}
  https://www.its.bldrdoc.gov/media/31075/DavisRadar\_vs\_comms.pdf.

\bibitem{Barneto:22}
C.~B. Barneto, E.~Rastorgueva-Foi, M.~F. Keskin, T.~Riihonen, M.~Turunen,
  J.~Talvitie, H.~Wymeersch, and M.~Valkama, ``Millimeter-wave mobile sensing
  and environment mapping: {M}odels, algorithms and validation,'' \emph{IEEE
  Trans. Veh. Technol.}, vol.~71, no.~4, pp. 3900--3916, 2022.

\bibitem{Mehrotra-Sabharwal:2022b}
N.~Mehrotra and A.~Sabharwal, ``When does multipath improve imaging
  resolution?'' \emph{IEEE Journal on Selected Areas in Information Theory},
  vol.~3, no.~1, pp. 135 -- 146, March 2022.

\bibitem{liu2021integrated}
F.~Liu, Y.~Cui, C.~Masouros, J.~Xu, T.~X. Han, Y.~C. Eldar, and S.~Buzzi,
  ``Integrated sensing and communications: Towards dual-functional wireless
  networks for {6G} and beyond,'' \emph{{IEEE} J. Sel. Areas Commun.}, vol.~40,
  no.~6, pp. 1728--1767, 2022.

\bibitem{ISAC_RIS_SPM}
S.~Prabhakar~Chepuri, N.~Shlezinger, F.~Liu, G.~C. Alexandropoulos, S.~Buzzi,
  and Y.~C. Eldar, ``Integrated sensing and communications with reconfigurable
  intelligent surfaces,'' https://arxiv.org/abs/2211.01003.

\bibitem{9099670}
S.~A. Hassani, V.~Lampu, K.~Parashar, L.~Anttila, A.~Bourdoux, B.~v. Liempd,
  M.~Valkama, F.~Horlin, and S.~Pollin, ``In-band full-duplex
  radar-communication system,'' \emph{IEEE Syst. J.}, vol.~15, no.~1, pp.
  1086--1097, 2021.

\bibitem{Mehrotra-Sabharwal:2022}
N.~Mehrotra and A.~Sabharwal, ``On the degrees of freedom region for
  simultaneous imaging \& uplink communication,'' \emph{{IEEE} {T}rans.
  {W}ireless {C}ommun.}, vol.~40, no.~6, pp. 1768 -- 1779, June 2022.

\bibitem{Barneto:19}
C.~B. Barneto, T.~Riihonen, M.~Turunen, L.~Anttila, M.~Fleischer, K.~Stadius,
  J.~Ryyn{\"a}nen, and M.~Valkama, ``Full-duplex {OFDM} radar with {LTE} and
  {5G NR} waveforms: Challenges, solutions, and measurements,'' \emph{IEEE
  Trans. Microw. Theory Techn.}, vol.~67, no.~10, pp. 4042--4054, Oct. 2019.

\bibitem{liyanaarachchi2021optimized}
S.~D. Liyanaarachchi, T.~Riihonen, M.~Turunen, A.~Lauri, M.~Fleischer,
  K.~Stadius, and M.~Valkama, ``Optimized waveforms for {5G--6G} communication
  with sensing: Theory, simulations and experiments,'' \emph{IEEE Trans.
  Wireless Commun.}, vol.~20, no.~12, Dec. 2021.

\bibitem{du2015mu}
X.~Du, J.~Tadrous, C.~Dick, and A.~Sabharwal, ``{MU-MIMO} beamforming with
  full-duplex open-loop training,'' in \emph{Proc. IEEE SPAWC 2015}, Jun. 2015,
  pp. 301--305.

\bibitem{du2016sequential}
X.~Du, J.~Tadrous, and A.~Sabharwal, ``Sequential beamforming for multi-user
  {MIMO} with full-duplex training,'' \emph{IEEE Trans. Wireless Commun.},
  vol.~15, no.~12, pp. 8551--8564, Dec. 2016.

\bibitem{mirza2018performance}
J.~Mirza, G.~Zheng, K.-K. Wong, S.~Lambotharan, and L.~Hanzo, ``On the
  performance of multi-user {MIMO} systems relying on full-duplex {CSI}
  acquisition,'' \emph{IEEE Trans. Wireless Commun.}, vol.~66, no.~10, pp.
  4563--4577, Oct. 2018.

\bibitem{D2D_fd}
H.-B. Jeon, B.-H. Koo, S.-H. Park, J.~Park, and C.-B. Chae,
  ``Graph-theory-based resource allocation and mode selection in {D2D}
  communication systems: The role of full-duplex,'' \emph{{IEEE} Wireless
  Commun. Lett.}, vol.~10, no.~2, pp. 236--240, Feb. 2021.

\bibitem{Comms-CE2020}
M.~A. Islam, G.~C. Alexandropoulos, and B.~Smida, ``Simultaneous data
  communication and channel estimation in multiuser full duplex {MIMO}
  systems,'' in \emph{Proc. IEEE Asilomar Signals Sys. Comp. Conf.}, Nov. 2020,
  pp. 1--5.

\bibitem{MultiuserComms-CE2020}
------, ``Simultaneous downlink data transmission and uplink channel estimation
  with reduced complexity full duplex {MIMO} radios,'' in \emph{Proc. IEEE
  Intl. Conf. Commun. (ICC)}, Jun. 2020, pp. 1--6.

\bibitem{Direction-Aided2020}
------, ``Direction-assisted beam management in full duplex millimeter wave
  massive {MIMO} systems,'' in \emph{Proc. IEEE Global Commun. Conf.
  (GLOBECOM)}, Dec. 2021, pp. 1--6.

\bibitem{barneto2020beamforming}
C.~B. Barneto, S.~D. Liyanaarachchi, T.~Riihonen, M.~Heino, L.~Anttila, and
  M.~Valkama, ``Beamforming and waveform optimization for {OFDM}-based joint
  communications and sensing at mm-waves,'' in \emph{Proc. IEEE Asilomar
  Signals Sys. Comp. Conf.}, Nov. 2020, pp. 895--899.

\bibitem{liyanaarachchi2021joint}
S.~D. Liyanaarachchi, C.~B. Barneto, T.~Riihonen, M.~Heino, and M.~Valkama,
  ``Joint multi-user communication and {MIMO} radar through full-duplex hybrid
  beamforming,'' in \emph{Proc. IEEE Int. Symp. Joint Commun. \& Sensing
  (JC\&S)}, Feb. 2021, pp. 1--5.

\bibitem{https://doi.org/10.48550/arxiv.2211.00229}
\BIBentryALTinterwordspacing
Z.~He, W.~Xu, H.~Shen, D.~W.~K. Ng, Y.~C. Eldar, and X.~You, ``Full-duplex
  communication for isac: Joint beamforming and power optimization,'' 2022.
  [Online]. Available: \url{https://arxiv.org/abs/2211.00229}
\BIBentrySTDinterwordspacing

\bibitem{Comms-Target_Tracking2022}
M.~A. Islam, G.~C. Alexandropoulos, and B.~Smida, ``Simultaneous multi-user
  {MIMO} communications and multi-target tracking with full duplex radios,'' in
  \emph{Proc. IEEE Global Commun. Conf. (GLOBECOM)}, Dec. 2022, pp. 1--6.

\bibitem{ISAC2022}
------, ``Integrated sensing and communication with millimeter wave full duplex
  hybrid beamforming,'' in \emph{Proc. IEEE Intl. Conf. Commun. (ICC)}, May
  2022, pp. 1--6.

\bibitem{Barneto:21}
C.~B. Barneto, S.~D. Liyanaarachchi, M.~Heino, T.~Riihonen, and M.~Valkama,
  ``Full duplex radio/radar technology: The enabler for advanced joint
  communication and sensing,'' \emph{IEEE Wireless Commun.}, vol.~28, no.~1,
  pp. 82--88, Feb. 2021.

\bibitem{Korpi:17}
D.~Korpi, M.~Heino, C.~Icheln, K.~Haneda, and M.~Valkama, ``Compact inband
  full-duplex relays with beyond {100 dB} self-interference suppression:
  Enabling techniques and field measurements,'' \emph{IEEE Trans. Antennas
  Propag.}, vol.~65, no.~2, pp. 960--965, Feb. 2017.

\bibitem{Mercier:19}
S.~Mercier, D.~Roque, and S.~Bidon, ``Study of the target self-interference in
  a low-complexity {OFDM}-based radar receiver,'' \emph{IEEE Trans. Aerosp.
  Electron. Syst.}, vol.~55, no.~3, pp. 1200--1212, 2019.

\bibitem{3GPP2018SI}
\emph{{TS} 38.174, Integrated access and backhaul radio transmission and
  reception, Release 16}, 3GPP Std.

\bibitem{3GPP38401}
\emph{{TS} 38.401, NG-RAN; Architecture description}, 3GPP Std.

\bibitem{kim2015survey}
D.~Kim, H.~Lee, and D.~Hong, ``A survey of in-band full-duplex transmission:
  From the perspective of {PHY} and {MAC} layers,'' \emph{{IEEE} Commun.
  Surveys Tuts.}, vol.~17, no.~4, pp. 2017--2046, Feb. 2015.

\bibitem{kwack}
J.~W. Kwak, M.~S. Sim, I.-W. Kang, J.~S. Park, J.~Park, and C.-B. Chae, ``A
  comparative study of analog/digital self-interference cancellation for full
  duplex radios,'' in \emph{Proc. Asilomar Conf. on Signal, Syst. and Comput.},
  Nov. 2019, pp. 1114--1119.

\bibitem{kwack_j}
J.~W. Kwak, M.~S. Sim, and C.-B. Chae, ``Analog self-interference cancellation
  with practical {RF} components for full-duplex radios,'' \emph{{IEEE} Trans.
  Wireless Commun.}, 2023.

\bibitem{suk_IAB}
G.~Y. Suk, S.-M. Kim, J.~Kwak, S.~Hur, E.~Kim, and C.-B. Chae, ``Full duplex
  integrated access and backhaul for {5G NR}: Analyses and prototype
  measurements,'' \emph{{IEEE} Wireless Commun. Mag.}, vol.~29, no.~4, pp.
  40--46, Aug. 2022.

\bibitem{cho2018rf}
Y.~J. Cho, G.-Y. Suk, B.~Kim, D.~K. Kim, and C.-B. Chae, ``{RF} lens-embedded
  antenna array for mm{W}ave {MIMO}: Design and performance,'' \emph{{IEEE}
  Commun. Mag.}, vol.~56, no.~7, pp. 42--48, July 2018.

\bibitem{park2020wcnc}
S.-H. Park, D.~Jun, B.~Kim, D.~K. Kim, and C.-B. Chae, ``Demo: Mmwave lens
  {MIMO},'' in \emph{Proc. IEEE Wireless Commun. and Netw. Conf. (WCNC)}, April
  2020, pp. 1--2.

\bibitem{park_squint}
S.-H. Park, B.~Kim, D.~Ku~Kim, L.~Dai, K.-K. Wong, and C.-B. Chae, ``Beam
  squint in {mmWave} systems: {RF} lens array vs. phase-shifter-based array,''
  \emph{{IEEE} Wireless Commun. Mag.}, vol.~30, no.~4, pp. 1--8, Aug. 2023.

\bibitem{EEverett2014Passive}
E.~Everett, A.~Sahai, and A.~Sabharwal, ``Passive self-interference suppression
  for full-duplex infrastructure nodes,'' \emph{{IEEE} Trans. Wireless
  Commun.}, vol.~13, no.~2, pp. 680--694, Jan. 2014.

\bibitem{suk2022}
G.-Y. Suk, J.~Kwak, J.~Choi, and C.-B. Chae, ``Dynamic {RF} beam codebook
  reduction for cost-efficient mmwave full-duplex systems,'' in \emph{Proc.
  IEEE Global Commun. Conf. (GLOBECOM)}, Dec. 2022, pp. 1--5.

\bibitem{23_TAnt}
S.-H. Park, C.-K. Park, H.~Yoo, B.~Kim, and C.-B. Chae, ``Window-type and {AR}
  glass-type transparent antenna systems for {B5G/6G},'' in \emph{Proc. IEEE
  Consumer Commun. and Net. Conf. (CCNC)}, Jan. 2023, pp. 937--938.

\bibitem{Guo:22}
S.~Guo, B.~Lu, M.~Wen, S.~Dang, and N.~Saeed, ``Customized 5{G} and beyond
  private networks with integrated {URLLC}, e{MBB}, m{MTC}, and positioning for
  industrial verticals,'' \emph{IEEE Commun. Stand. Mag.}, vol.~6, no.~1, pp.
  52--57, 2022.

\bibitem{Kim:2020}
H.~Kim, J.~Kim, and D.~Hong, ``Dynamic {TDD} systems for {5G} and beyond: A
  survey of cross-link interference mitigation,'' \emph{IEEE Commun. Surv.
  Tutor.}, vol.~22, no.~4, pp. 2315--2348, 2020.

\bibitem{TR-38828}
``{3GPP TR 38.828} cross link interference {(CLI)} handling and remote
  interference management {(RIM)} for {NR},'' Sep. 2019.

\bibitem{Silva:21}
J.~M.~B. da~Silva, G.~Wikstr{\"o}m, R.~K. Mungara, and C.~Fischione, ``Full
  duplex and dynamic {TDD}: Pushing the limits of spectrum reuse in multi-cell
  communications,'' \emph{IEEE Wirel. Commun.}, vol.~28, no.~1, pp. 44--50,
  2021.

\bibitem{Clowdhury:2022}
A.~Chowdhury, R.~Chopra, and C.~R. Murthy, ``Can dynamic {TDD} enabled
  half-duplex cell-free massive {MIMO} outperform full-duplex cellular massive
  {MIMO}?'' \emph{IEEE Trans. Commun.}, vol.~70, no.~7, pp. 4867--4883, 2022.

\bibitem{Razlighi:2020}
M.~M. Razlighi, N.~Zlatanov, and P.~Popovski, ``Dynamic time-frequency division
  duplex,'' \emph{IEEE Trans. Wireless Commun.}, vol.~19, no.~5, pp.
  3118--3132, 2020.

\bibitem{Ji:21}
H.~Ji, Y.~Kim, K.~Muhammad, C.~Tarver, M.~Tonnemacher, S.~Lim, J.~Shim, J.~Kim,
  B.~Yu, and G.~Xu, ``{XDD}: Cross division duplex in {5G-A}dvanced,'' in
  \emph{Proc. IEEE Vehicular Technology Conference (VTC2021-Fall)}, 2021, pp.
  1--5.

\bibitem{R4-2212117}
``{3GPP R4-2212117} discussion on {RF} requirements for massive {MIMO} antenna
  for subband non-overlapping full duplex, {3GPP Radio Access Network Group
  RAN4},'' Aug. 2022.

\bibitem{Mairton:21}
J.~M. B.~d. Silva, H.~Ghauch, G.~Fodor, M.~Skoglund, and C.~Fischione, ``Smart
  antenna assignment is essential in full-duplex communications,'' \emph{IEEE
  Trans. Commun.}, vol.~69, no.~5, pp. 3450--3466, 2021.

\bibitem{Rostomyan:19}
N.~Rostomyan, V.~Diddi, P.~Gudem, and P.~Asbeck, ``Adaptive cancellation of
  digital power amplifier receive band noise for {FDD} transceivers,''
  \emph{IEEE Microw. Wireless Compon. Lett.}, vol.~29, no.~1, pp. 59--61, 2019.

\bibitem{Katanbaf:19}
M.~Katanbaf, K.-D. Chu, T.~Zhang, C.~Su, and J.~C. Rudell, ``Two-way traffic
  ahead: {RF} analog self-interference cancellation techniques and the
  challenges for future integrated full-duplex transceivers,'' \emph{IEEE
  Microw. Mag.}, vol.~20, no.~2, pp. 22--35, 2019.

\bibitem{R1-2207461}
``{3GPP R1-2207461} {Evaluation of NR} duplex evolution, {3GPP Radio Access
  Network Group RAN1},'' Aug. 2022.

\bibitem{Goyal:2015}
S.~Goyal, P.~Liu, S.~S. Panwar, R.~A. Difazio, R.~Yang, and E.~Bala, ``Full
  duplex cellular systems: will doubling interference prevent doubling
  capacity?'' \emph{IEEE Commun. Mag.}, vol.~53, no.~5, pp. 121--127, 2015.

\bibitem{Psomas:2017}
C.~Psomas, M.~Mohammadi, I.~Krikidis, and H.~A. Suraweera, ``Impact of
  directionality on interference mitigation in full-duplex cellular networks,''
  \emph{IEEE Trans. Wireless Commun.}, vol.~16, no.~1, pp. 487--502, 2017.

\bibitem{Mohammadi:TCOM2015}
M.~Mohammadi, H.~A. Suraweera, Y.~Cao, I.~Krikidis, and C.~Tellambura,
  ``Full-duplex radio for uplink/downlink wireless access with spatially random
  nodes,'' \emph{IEEE Trans. Commun.}, vol.~63, no.~12, pp. 5250--5266, 2015.

\bibitem{Huang:2019}
Y.~Huang, B.~Jalaian, S.~Russell, and H.~Samani, ``Reaping the benefits of
  dynamic {TDD} in massive {MIMO},'' \emph{IEEE Syst. J.}, vol.~13, no.~1, pp.
  117--124, 2019.

\bibitem{Rachad:2020}
J.~Rachad, R.~Nasri, and L.~Decreusefond, ``3{D} beamforming based dynamic
  {TDD} interference mitigation scheme,'' in \emph{Proc. IEEE Vehicular
  Technology Conference (VTC2020-Spring)}, 2020, pp. 1--7.

\bibitem{Al-Saadeh:2017}
O.~Al-Saadeh and K.~W. Sung, ``A performance comparison of in-band full duplex
  and dynamic {TDD} for {5G} indoor wireless networks,'' \emph{{EURASIP} J.
  Wirel. Commun. Netw.}, vol. 2017, no.~50, pp. 1--14, 2017.

\bibitem{Koh:2018}
J.~Koh, Y.-G. Lim, C.-B. Chae, and J.~Kang, ``On the feasibility of full-duplex
  large-scale {MIMO} cellular systems,'' \emph{IEEE Trans. Wireless Commun.},
  vol.~17, no.~9, pp. 6231--6250, 2018.

\bibitem{Wang:2019}
K.~Wang, F.~R. Yu, L.~Wang, J.~Li, N.~Zhao, Q.~Guan, B.~Li, and Q.~Wu,
  ``Interference alignment with adaptive power allocation in
  full-duplex-enabled small cell networks,'' \emph{IEEE Trans. Veh. Technol.},
  vol.~68, no.~3, pp. 3010--3015, 2019.

\bibitem{Fodor:21}
G.~Fodor, J.~Vinogradova, P.~Hammarberg, K.~K. Nagalapur, Z.~T. Qi, H.~Do,
  R.~Blasco, and M.~U. Baig, ``5{G} new radio for automotive, rail, and air
  transport,'' \emph{IEEE Commun. Mag.}, vol.~59, no.~7, pp. 22--28, 2021.

\bibitem{Zhang:21c}
J.~Zhang, N.~Garg, M.~Holm, and T.~Ratnarajah, ``Design of full duplex
  millimeter-wave integrated access and backhaul networks,'' \emph{IEEE Wirel.
  Commun.}, vol.~28, no.~1, pp. 60--67, 2021.

\bibitem{Malayter:22}
J.~R. Malayter and D.~J. Love, ``Precoding for low-latency full-duplex {MIMO}
  relays: {A} dynamic approach,'' in \emph{Proc. IEEE Wireless Commun. and
  Networking Conf. (WCNC)}, Apr. 2022, pp. 2417--2422.

\bibitem{Papadogiannis2012}
A.~Papadogiannis, G.~C. Alexandropoulos, A.~G. Burr, and D.~Grace, ``Bringing
  mobile relays for wireless access networks into practice--learning when to
  relay,'' \emph{IET Commun.}, vol.~60, no.~6, pp. 618--627, Jun. 2012.

\bibitem{Basak_Can_2007}
B.~Can, M.~Portalski, H.~S.~D. Lebreton, S.~Frattasi, and H.~A. Suraweera,
  ``Implementation issues for {OFDM}-based multihop cellular networks,''
  \emph{IEEE Commun. Mag.}, vol.~45, no.~9, pp. 74--81, 2007.

\bibitem{Vaezi:20}
K.~Vaezi and F.~Ashtiani, ``Delay-optimal cooperation policy in a slotted aloha
  full-duplex wireless network: Static approach,'' \emph{IEEE Syst. J.},
  vol.~14, no.~2, pp. 2257--2268, 2020.

\bibitem{relayselection2010}
G.~C. Alexandropoulos, A.~Papadogiannis, and K.~Berberidis, ``Performance
  analysis of cooperative networks with relay selection over {N}akagami-$m$
  fading channels,'' \emph{IEEE Signal Process. Lett.}, vol.~17, no.~5, pp.
  441--444, May 2010.

\bibitem{Alexandropoulos2011}
G.~C. Alexandropoulos, A.~Papadogiannis, and P.~C. Sofotasios, ``A comparative
  study of relaying schemes with decode-and-forward over {N}akagami-$m$ fading
  channels,'' \emph{HINDAWI J. Comp. Netw. Commun.}, vol. 2011, no. 560528,
  2011.

\bibitem{HoVan_relay_selection}
K.~Ho-Van, P.~C. Sofotasios, G.~C. Alexandropoulos, and S.~Freear, ``Bit error
  rate of underlay decode-and-forward cognitive networks with best relay
  selection,'' \emph{J. Commun. Netw.}, vol.~17, no.~2, pp. 162--171, Apr.
  2015.

\bibitem{Peppas2013}
K.~P. Peppas, G.~C. Alexandropoulos, and P.~T. Mathiopoulos, ``Performance
  analysis of dual-hop {AF} relaying systems over mixed $\eta$-$\mu$ and
  $\kappa$-$\mu$ fading channels,'' \emph{IEEE Trans. Veh. Tehncol.}, vol.~62,
  no.~7, pp. 3149--3163, Sep. 2013.

\bibitem{Miridakis2016}
N.~I. Miridakis, T.~A. Tsiftsis, G.~C. Alexandropoulos, and M.~Debbah, ``Green
  cognitive relaying: {O}pportunistically switching between data transmission
  and energy harvesting,'' \emph{{IEEE} J. Sel. Areas Commun.}, vol.~34,
  no.~12, pp. 3725--3738, Dec. 2016.

\bibitem{Fozooni2019}
M.~Fozooni, H.~Q. Ngo, M.~Matthaiou, S.~Jin, and G.~C. Alexandropoulos,
  ``Hybrid processing design for multipair massive {MIMO} relaying with channel
  spatial correlation,'' \emph{IEEE Trans. Commun.}, vol.~67, no.~1, pp.
  107--123, Jan. 2019.

\bibitem{Kwon:12}
T.~Kwon, Y.~Kim, and D.~Hong, ``Comparison of {FDR} and {HDR} under adaptive
  modulation with finite-length queues,'' \emph{IEEE Trans. Veh. Technol.},
  vol.~61, no.~2, pp. 838--843, 2012.

\bibitem{FD_user_scheduling_2016}
G.~C. Alexandropoulos, M.~Kountouris, and I.~Atzeni, ``User scheduling and
  optimal power allocation for full-duplex cellular networks,'' in \emph{Proc.
  IEEE SPAWC}, Jul. 2016, pp. 1--6.

\bibitem{FD_UDN_2016}
I.~Atzeni, M.~Kountouris, and G.~C. Alexandropoulos, ``Performance evaluation
  of user scheduling for full-duplex small cells in ultra-dense networks,'' in
  \emph{Proc. European Wireless Conference}, May 2016, pp. 1--6.

\bibitem{Gu:18}
Y.~Gu, H.~Chen, Y.~Li, and B.~Vucetic, ``Ultra-reliable short-packet
  communications: Half-duplex or full-duplex relaying?'' \emph{IEEE Wireless
  Commun. Lett.}, vol.~7, no.~3, pp. 348--351, 2018.

\bibitem{Airod:18}
F.~E. Airod, H.~Chafnaji, and H.~Yanikomeroglu, ``Performance analysis of low
  latency multiple full-duplex selective decode and forward relays,'' in
  \emph{Proc. IEEE Wireless Commun. and Networking Conf. (WCNC)}, 2018, pp.
  1--6.

\bibitem{Yuan:22}
L.~Yuan, Q.~Du, and F.~Fang, ``Performance analysis of full-duplex cooperative
  {NOMA} short-packet communications,'' \emph{IEEE Trans. Veh. Technol.}, pp.
  1--6, 2022.

\bibitem{Hakkeon:22}
H.~Lee, J.~Choi, and D.~Hong, ``Resource configuration for full-duplex-aided
  multiple-access edge computation offloading,'' \emph{IEEE Trans. Wireless
  Commun.}, vol.~21, no.~4, pp. 2799--2812, 2022.

\bibitem{Wen:22}
W.~He, Y.~Zhang, Y.~Huang, D.~He, Y.~Xu, Y.~Guan, and W.~Zhang, ``Integrated
  resource allocation and task scheduling for full-duplex mobile edge
  computing,'' \emph{IEEE Trans. Veh. Technol.}, vol.~71, no.~6, pp.
  6488--6502, 2022.

\bibitem{Yuan_V2V:2020}
Z.~Yuan, Y.~Ma, Y.~Hu, and W.~Li, ``High-efficiency full-duplex {V}2{V}
  communication,'' in \emph{Proc. 6{G} Wireless Summit (6{G SUMMIT})}, 2020,
  pp. 1--5.

\bibitem{Campolo:2018}
C.~Campolo, A.~Molinaro, A.~O. Berthet, and A.~Vinel, ``Full-duplex radios for
  vehicular communications,'' \emph{IEEE Commun. Mag.}, vol.~55, no.~6, pp.
  182--189, 2017.

\bibitem{LZhang:2018}
L.~Zhang, Q.~Fan, and N.~Ansari, ``3-{D} drone-base-station placement with
  in-band full-duplex communications,'' \emph{IEEE Commun. Lett.}, vol.~22,
  no.~9, pp. 1902--1905, 2018.

\bibitem{Quynh:2022}
Q.~T. Ngo, K.~T. Phan, W.~Xiang, A.~Mahmood, and J.~Slay, ``Two-tier
  cache-aided full-duplex hybrid satellite--terrestrial communication
  networks,'' \emph{IEEE Trans. Aerosp. Electron. Syst.}, vol.~58, no.~3, pp.
  1753--1765, March 2022.

\bibitem{huang2019reconfigurable}
C.~Huang, A.~Zappone, G.~C. Alexandropoulos, M.~Debbah, and C.~Yuen,
  ``Reconfigurable intelligent surfaces for energy efficiency in wireless
  communication,'' \emph{IEEE Trans. Wireless Commun.}, vol.~18, no.~8, pp.
  4157--4170, Aug. 2019.

\bibitem{Marco2019}
M.~Di~Renzo, M.~Debbah, D.-T. Phan-Huy, A.~Zappone, M.-S. Alouini, C.~Yuen,
  V.~Sciancalepore, G.~C. Alexandropoulos, J.~Hoydis, H.~Gacanin, J.~de~Rosny,
  A.~Bounceu, G.~Lerosey, and M.~Fink, ``Smart radio environments empowered by
  reconfigurable {AI} meta-surfaces: {A}n idea whose time has come,''
  \emph{{EURASIP} J. Wirel. Commun. Netw.}, vol. 2019, no.~1, pp. 1--20, May
  2019.

\bibitem{RIS_fundamentals}
G.~C. Alexandropoulos, N.~Shlezinger, and P.~del Hougne, ``Reconfigurable
  intelligent surfaces for rich scattering wireless communications: {R}ecent
  experiments, challenges, and opportunities,'' \emph{IEEE Commun. Mag.},
  vol.~59, no.~6, pp. 28--34, Jun. 2021.

\bibitem{WavePropTCCN}
G.~C. Alexandropoulos, G.~Lerosey, M.~Debbah, and M.~Fink, ``Reconfigurable
  intelligent surfaces and metamaterials: {T}he potential of wave propagation
  control for {6G} wireless communications,'' \emph{IEEE ComSoc TCCN
  Newslett.}, vol.~6, no.~1, pp. 25--37, Jun. 2020.

\bibitem{Qingqing:2021}
Q.~Wu, S.~Zhang, B.~Zheng, C.~You, and R.~Zhang, ``Intelligent reflecting
  surface-aided wireless communications: A tutorial,'' \emph{IEEE Trans.
  Commun.}, vol.~69, no.~5, pp. 3313--3351, 2021.

\bibitem{Shaikh_2021}
M.~H.~N. Shaikh, V.~A. Bohara, A.~Srivastava, and G.~Ghatak, ``Intelligent
  reflecting surfaces versus full-duplex relaying: {P}erformance comparison for
  non-ideal transmitter case,'' in \emph{Proc. 2021 IEEE 32nd Annual Intl.
  Symp. Personal, Indoor and Mobile Radio Commun. (PIMRC)}, Helsinki, Finland,
  2021, pp. 513--518.

\bibitem{Tsinghua_RIS_Tutorial}
M.~Jian, G.~C. Alexandropoulos, E.~Basar, C.~Huang, R.~Liu, Y.~Liu, and
  C.~Yuen, ``Reconfigurable intelligent surfaces for wireless communications:
  {O}verview of hardware designs, channel models, and estimation techniques,''
  \emph{Intell. Converged Netw.}, vol.~3, no.~1, pp. 1--32, Mar. 2022.

\bibitem{hardware2020icassp}
G.~C. Alexandropoulos and E.~Vlachos, ``A hardware architecture for
  reconfigurable intelligent surfaces with minimal active elements for explicit
  channel estimation,'' in \emph{Proc. IEEE Int. Conf. Acoustics, Speech and
  Signal Process. (ICASSP)}, May 2020, pp. 9175--9179.

\bibitem{HRIS_Mag}
G.~C. Alexandropoulos, N.~Shlezinger, I.~Alamzadeh, M.~F. Imani, H.~Zhang, and
  Y.~C. Eldar, ``Hybrid reconfigurable intelligent metasurfaces: {E}nabling
  simultaneous tunable reflections and sensing for {6G} wireless
  communications,'' \emph{arXiv preprint arXiv:2104.04690}, 2021.

\bibitem{HRIS_Nature}
I.~Alamzadeh, G.~C. Alexandropoulos, N.~Shlezinger, and M.~F. Imani, ``A
  reconfigurable intelligent surface with integrated sensing capability,''
  \emph{Scientific Reports}, vol.~11, no. 20737, pp. 1--10, Oct. 2021.

\bibitem{amplifying_RIS_2022}
R.~A. Tasci, F.~Kilinc, E.~Basar, and G.~C. Alexandropoulos, ``A new {RIS}
  architecture with a single power amplifier: {E}nergy efficiency and error
  performance analysis,'' \emph{IEEE Access}, vol.~10, pp. 44\,804--44\,815,
  May 2022.

\bibitem{yildirim_hybrid_2021}
I.~Yildirim, F.~Kilinc, E.~Basar, and G.~C. Alexandropoulos, ``Hybrid
  {RIS}-empowered reflection and decode-and-forward relaying for coverage
  extension,'' \emph{IEEE Commun. Lett.}, vol.~25, no.~5, pp. 1692--1696, May
  2021.

\bibitem{Mustafa_ArbitraryBeam_6G2022}
M.~Rahal, B.~Denis, K.~Keykhosravi, M.~F. Keskin, B.~Uguen, G.~C.
  Alexandropoulos, and H.~Wymeersch, ``Arbitrary beam pattern approximation via
  {RISs} with measured element responses,'' in \emph{Proc. EuCNC \& 6G Summit},
  Jun. 2022, pp. 506--511.

\bibitem{RIS_placement}
A.~L. Moustakas, G.~C. Alexandropoulos, and M.~Debbah, ``Reconfigurable
  intelligent surfaces and capacity optimization: {A} large system analysis,''
  \emph{IEEE Trans. Wireless Commun.}, to appear, 2023.

\bibitem{MengHua:2022}
M.~Hua and Q.~Wu, ``Joint dynamic passive beamforming and resource allocation
  for {IRS}-aided full-duplex {WPCN},'' \emph{IEEE Trans. Wireless Commun.},
  vol.~21, no.~7, pp. 4829--4843, 2022.

\bibitem{Chou:2021}
S.-K. Chou, O.~Yurduseven, H.~Q. Ngo, and M.~Matthaiou, ``On the aperture
  efficiency of intelligent reflecting surfaces,'' \emph{IEEE Wireless Commun.
  Lett.}, vol.~10, no.~3, pp. 599--603, March 2021.

\bibitem{Conf_all}
G.~C. Alexandropoulos, K.~Katsanos, M.~Wen, and D.~B. da~Costa, ``Safeguarding
  {MIMO} communications with reconfigurable metasurfaces and artificial
  noise,'' in \emph{Proc. IEEE Intl. Conf. Commun. (ICC)}, Jun. 2021, pp. 1--6.

\bibitem{WangNing:2019}
N.~Wang, P.~Wang, A.~Alipour-Fanid, L.~Jiao, and K.~Zeng, ``Physical-layer
  security of 5{G} wireless networks for {I}o{T}: {C}hallenges and
  opportunities,'' \emph{IEEE Internet Things J.}, vol.~6, no.~5, pp.
  8169--8181, 2019.

\bibitem{YongJin:2022}
Y.~Jin, R.~Guo, L.~Zhou, and Z.~Hu, ``Secure beamforming for {IRS}-assisted
  nonlinear {SWIPT} systems with full-duplex user,'' \emph{IEEE Commun. Lett.},
  vol.~26, no.~7, pp. 1494--1498, 2022.

\bibitem{Peng:TSP2021}
Z.~Peng, Z.~Zhang, C.~Pan, L.~Li, and A.~L. Swindlehurst, ``Multiuser
  full-duplex two-way communications via intelligent reflecting surface,''
  \emph{IEEE Trans. Signal Process.}, vol.~69, pp. 837--851, 2021.

\bibitem{Cai:2022}
Y.~Cai, M.-M. Zhao, K.~Xu, and R.~Zhang, ``Intelligent reflecting surface aided
  full-duplex communication: {P}assive beamforming and deployment design,''
  \emph{IEEE Trans. Wireless Commun.}, vol.~21, no.~1, pp. 383--397, 2022.

\bibitem{Arzykulov:2022}
S.~Arzykulov, G.~Nauryzbayev, A.~Celik, and A.~M. Eltawil, ``Ris-assisted
  full-duplex relay systems,'' \emph{IEEE Syst. J.}, pp. 1--12, 2022.

\bibitem{WangYiru:2022}
Y.~Wang, P.~Guan, H.~Yu, and Y.~Zhao, ``Transmit power optimization of
  simultaneous transmission and reflection {RIS} assisted full-duplex
  communications,'' \emph{IEEE Access}, vol.~10, pp. 61\,192--61\,200, 2022.

\bibitem{tewes:2022}
S.~Tewes, M.~Heinrichs, P.~Staat, R.~Kronberger, and A.~Sezgin, ``Full-duplex
  meets reconfigurable surfaces: {RIS}-assisted {SIC} for full-duplex radios,''
  in \emph{Proc. IEEE Intl. Conf. Commun. (ICC)}, 2022, pp. 1106--1111.

\bibitem{FD_HMIMO_2023}
I.~Gavras, M.~A. Islam, B.~Smida, and G.~C. Alexandropoulos, ``Full duplex
  holographic {MIMO} for near-field integrated sensing and communications,'' in
  \emph{European Signal Proces. Conf.}, Helsinki, Finland, Sep. under review,
  2023.

\bibitem{FD_RIS_ISAC_2023}
C.~K. Sheemar, G.~C. Alexandropoulos, D.~Slock, J.~Querol, and S.~Chatzinotas,
  ``Full-duplex-enabled joint communications and sensing with reconfigurable
  intelligent surfaces,'' in \emph{European Signal Proces. Conf.}, Helsinki,
  Finland, Sep. under review, 2023.

\bibitem{HMIMO_survey}
T.~Gong, I.~Vinieratou, R.~Ji, C.~Huang, G.~C. Alexandropoulos, L.~Wei,
  M.~Debbah, H.~V. Poor, and C.~Yuen, ``Holographic {MIMO} communications:
  {T}heoretical foundations, enabling technologies, and future directions,''
  \emph{arXiv preprint arXiv:2212.01257}, 2022.

\bibitem{LiuYuanwei:2021}
Y.~Liu, X.~Mu, J.~Xu, R.~Schober, Y.~Hao, H.~V. Poor, and L.~Hanzo, ``{STAR}:
  Simultaneous transmission and reflection for 360$\,^{\circ}$ coverage by
  intelligent surfaces,'' \emph{IEEE Wirel. Commun.}, vol.~28, no.~6, pp.
  102--109, 2021.

\bibitem{Xu_Jaiqi:2021}
J.~Xu, Y.~Liu, X.~Mu, and O.~A. Dobre, ``{STAR-RIS}s: {S}imultaneous
  transmitting and reflecting reconfigurable intelligent surfaces,'' \emph{IEEE
  Commun. Lett.}, vol.~25, no.~9, pp. 3134--3138, 2021.

\bibitem{STARRIS_Loc2022}
J.~He, A.~Fakhreddine, and G.~C. Alexandropoulos, ``Simultaneous indoor and
  outdoor {3D} localization with {STAR-RIS}-assisted millimeter wave systems,''
  in \emph{Proc. IEEE VTC-Fall}, Sep. 2022, pp. 1--6.

\bibitem{Perera:2022}
P.~P. Perera, V.~G. Warnasooriya, D.~Kudathanthirige, and H.~A. Suraweera,
  ``Sum rate maximization in {STAR-RIS} assisted full-duplex communication
  systems,'' in \emph{Proc. IEEE Intl. Conf. Commun. (ICC)}, 2022, pp.
  3281--3286.

\bibitem{Papazafeiropoulos:2022}
A.~Papazafeiropoulos, P.~Kourtessis, and I.~Krikidis, ``{STAR-RIS} assisted
  full-duplex systems: {I}mpact of correlation and maximization,'' \emph{IEEE
  Commun. Lett.}, pp. 1--1, 2022.

\bibitem{RISrandom2022}
K.~Chen-Hu, G.~C. Alexandropoulos, and A.~G. Armada, ``Simultaneous {RIS}
  tuning and differential data transmission for {MISO OFDM} wireless systems,''
  in \emph{Proc. IEEE Global Conf. Commun. (GLOBECOM)}, Dec. 2022.

\bibitem{IRS2022Pervasive}
G.~C. Alexandropoulos, K.~Stylianopoulos, C.~Huang, C.~Yuen, M.~Bennis, and
  M.~Debbah, ``Pervasive machine learning for smart radio environments enabled
  by reconfigurable intelligent surfaces,'' \emph{Proc, {IEEE}}, pp. 1--32,
  2022.

\bibitem{PARAFAC2021}
L.~Wei, C.~Huang, G.~C. Alexandropoulos, C.~Yuen, and Z.~Zhang, ``Channel
  estimation for {RIS}-empowered multi-user {MISO} wireless communications,''
  \emph{IEEE Trans. Commun.}, vol.~69, no.~6, pp. 4144--4157, Jun. 2021.

\bibitem{Geraci_2022}
G.~Geraci, A.~Garcia-Rodriguez, M.~M. Azari, A.~Lozano, M.~Mezzavilla,
  S.~Chatzinotas, Y.~Chen, S.~Rangan, and M.~D. Renzo, ``What will the future
  of {UAV} cellular communications be? {A} flight from 5{G} to 6{G},''
  \emph{IEEE Commun. Surv. Tutor.}, vol.~24, no.~3, pp. 1304--1335, 2022.

\bibitem{Karabulut_2021}
G.~Karabulut~Kurt, M.~G. Khoshkholgh, S.~Alfattani, A.~Ibrahim, T.~S.~J.
  Darwish, M.~S. Alam, H.~Yanikomeroglu, and A.~Yongacoglu, ``A vision and
  framework for the high altitude platform station ({HAPS}) networks of the
  future,'' \emph{IEEE Commun. Surv. Tutor.}, vol.~23, no.~2, pp. 729--779,
  2021.

\bibitem{Huawei_2022}
H.~Luo, X.~Shi, Y.~Chen, X.~Meng, F.~Zhao, M.~Mayer, P.~A. Smith, B.~McCormick,
  A.~Akhavain, D.~Liu, H.~Wen, Y.~Wang, X.~Wang, R.~Yang, R.~Li, B.~Wang,
  J.~Wang, and W.~Tong, ``Very-low-earth-orbit satellite networks for 6{G},''
  [Online], Available:
  \url{https://www.huawei.com/en/huaweitech/future-technologies/very-low-earth-orbit-satellite-networks-6g},
  Huawei, Dec. 2022.

\bibitem{Giordani_2021}
M.~Giordani and M.~Zorzi, ``Non-terrestrial networks in the 6{G} era:
  {C}hallenges and opportunities,'' \emph{IEEE Netw.}, vol.~35, no.~2, pp.
  244--251, 2021.

\bibitem{Xingqin_2021}
X.~Lin, S.~Cioni, G.~Charbit, N.~Chuberre, S.~Hellsten, and J.-F. Boutillon,
  ``On the path to {6G}: Embracing the next wave of low earth orbit satellite
  access,'' \emph{IEEE Commun. Mag.}, vol.~59, no.~12, pp. 36--42, Decc. 2021.

\bibitem{Grayver_2015}
E.~Grayver, R.~Keating, and A.~Parower, ``Feasibility of full duplex
  communications for {LEO} satellite,'' in \emph{Proc. IEEE Aerospace
  Conference}, 2015, pp. 1--8.

\bibitem{BShankar_2015}
M.~R. Bhavani~Shankar, G.~Zheng, S.~Maleki, and B.~Ottersten, ``Feasibility
  study of full-duplex relaying in satellite networks,'' in \emph{Proc. IEEE
  International Workshop on Signal Processing Advances in Wireless
  Communications (SPAWC)}, 2015, pp. 560--564.

\bibitem{Hua_Meng_2020}
M.~Hua, L.~Yang, C.~Pan, and A.~Nallanathan, ``Throughput maximization for
  full-duplex {UAV} aided small cell wireless systems,'' \emph{IEEE Wireless
  Commun. Lett.}, vol.~9, no.~4, pp. 475--479, 2020.

\bibitem{Shi_Wenjuan_2021}
W.~Shi, Y.~Sun, M.~Liu, H.~Xu, G.~Gui, T.~Ohtsuki, B.~Adebisi, H.~Gacanin, and
  F.~Adachi, ``Joint {UL}/{DL} resource allocation for {UAV}-aided full-duplex
  {NOMA} communications,'' \emph{IEEE Trans. Commun.}, vol.~69, no.~12, pp.
  8474--8487, 2021.

\bibitem{Araki_2022}
T.~Yu, K.~Araki, and K.~Sakaguchi, ``Ground experiment of full-duplex
  multi-{UAV} system enabled by directional antennas,'' in \emph{Proc. IEEE
  Annual Computing and Communication Workshop and Conference (CCWC)}, 2022, pp.
  1092--1097.

\bibitem{Nguyen_Tan_2022}
T.~N. Nguyen, L.-T. Tu, D.-H. Tran, V.-D. Phan, M.~Voznak, S.~Chatzinotas, and
  Z.~Ding, ``Outage performance of satellite terrestrial full-duplex relaying
  networks with co-channel interference,'' \emph{IEEE Wireless Commun. Lett.},
  vol.~11, no.~7, pp. 1478--1482, July 2022.

\bibitem{Ngo_Quynh_2022}
Q.~T. Ngo, K.~T. Phan, W.~Xiang, A.~Mahmood, and J.~Slay, ``Two-tier
  cache-aided full-duplex hybrid satellite--terrestrial communication
  networks,'' \emph{IEEE Trans. Aerosp. Electron. Syst.}, vol.~58, no.~3, pp.
  1753--1765, March 2022.

\end{thebibliography}
\end{document}